\documentclass[12pt, a4paper]{article}
\pdfoutput=1
 \usepackage[font=small,format=plain,labelfont=bf,up,textfont=normal,up,justification=justified,singlelinecheck=false]{caption}
\usepackage{subcaption}
\usepackage[a4paper, left=2cm,right=2cm]{geometry}
\usepackage[colorlinks=true,linkcolor=black,citecolor=teal,urlcolor=MidnightBlue,filecolor=black]{hyperref}
\usepackage{amsfonts}
\usepackage{amsmath,amssymb}
\usepackage{setspace,braket}
\usepackage[dvipsnames]{xcolor}
\definecolor{SchoolColor}{rgb}{0.6471, 0.1098, 0.1882} 
\usepackage{algorithmic}

\usepackage[utf8,applemac]{inputenc}
\usepackage{tensor}
\usepackage{cite}
\usepackage{tikz}
\usepackage{graphicx}
\usepackage{bm} 

\usepackage[utf8,applemac]{inputenc}
\usepackage{tensor}
\usepackage{cite}
\usepackage{tikz}
\usepackage{graphicx}
\graphicspath{{figure/}}
\usepackage{array}
\usepackage{booktabs} 
\usepackage{multirow}

\usepackage{dcolumn}
\usepackage{bm}

\usepackage{verbatim}

\usepackage{textcomp} 
\usepackage{graphicx} 

\numberwithin{equation}{section}
\newcommand{\bea}{\begin{eqnarray}}
\newcommand{\eea}{\end{eqnarray}}
\newcommand{\be}{\begin{equation}}
\newcommand{\ee}{\end{equation}}
\newcommand{\bs}{\begin{subequations}}
\newcommand{\es}{\end{subequations}}
\def\nn{\nonumber}

\def\p{\partial}

\newcommand{\beqs}{\begin{eqnarray}}
\newcommand{\eeqs}{\end{eqnarray}}
\numberwithin{equation}{section}

\newcommand{\Rmnum}[1]{\uppercase\expandafter{\romannumeral #1\relax}}

\newcommand{\liu}{\color{red}}
\newcommand{\yrz}{\color{purple}}
\newcommand{\li}{\color{brown}}

\begin{document}
\begin{titlepage}

\begin{flushright}\vspace{-3cm}
{\small
\today }\end{flushright}
\vspace{0.5cm}
\begin{center}
	{{ \LARGE{\bf{Quantum flux	operators for Carrollian 
	
	diffeomorphism in general dimensions}}}}\vspace{5mm}

	\centerline{\large{Ang Li\footnote{liang121@hust.edu.cn}, Wen-Bin Liu\footnote{liuwenbin0036@hust.edu.cn}, Jiang Long\footnote{longjiang@hust.edu.cn} \&  Run-Ze Yu\footnote{m202270239@hust.edu.cn}}}
	\vspace{2mm}
	\normalsize
	\bigskip\medskip

	\textit{School of Physics, Huazhong University of Science and Technology, \\ Luoyu Road 1037, Wuhan, Hubei 430074, China
	}
	\vspace{25mm}
	
	\begin{abstract}
		\noindent
		We construct Carrollian scalar field theories in general dimensions, mainly focusing on the boundaries of Minkowski and  Rindler spacetime, whose quantum flux operators  
		form a faithful representation of Carrollian diffeomorphism up to a central charge, respectively. At future/past null infinity, the fluxes are physically observable and encode rich information of the radiation. The central charge may be regularized to be finite by the spectral zeta function  or heat kernel method on the unit sphere. 
		For the theory at the  Rindler horizon, the effective central charge is proportional to the area of the bifurcation surface after regularization. Moreover, the zero mode of supertranslation is identified as the modular Hamiltonian, linking Carrollian diffeomorphism to quantum information  theory. Our results may hold for general null hypersurfaces and provide new insight in the study of the Carrollian field theory, asymptotic symmetry group and entanglement entropy.
		\end{abstract}
	

\end{center}

\end{titlepage}
\tableofcontents
\section{Introduction}
Null hypersurfaces play a central role in the causal structure of spacetime. They are important in many physical systems,  such as future/past null infinity of asymptotically flat spacetime and black hole horizons.  The notable feature of a null hypersurface is that its metric is degenerate. Recently, the null geometry has been studied in the context of Carrollian manifolds \cite{Une, Gupta1966OnAA,Duval_2014a,Duval_2014b,Duval:2014uoa}.

The Carrollian diffeomorphism preserving the Carroll structure \cite{Liu:2022mne} could provide a natural separation between space and time \cite{Ciambelli:2019lap}. More interestingly, it has been shown in a series of papers \cite{Liu:2022mne, Liu:2023qtr,Liu:2023gwa} that one may construct quantum flux operators that generate Carrollian diffeomorphism up to anomalous terms at future null infinity. The operators are
defined through  the flux densities which are radiated to future null infinity and they encode rich information of  the time and angle dependencies about the radiation.  The operators are totally defined at future null infinity and could form a representation of Carrollian diffeomorphism.  Rather interestingly, the radiation fluxes are physical observables. Therefore, it indicates that all the Carrollian diffeomorphism should be regarded as large gauge transformations, extending the BMS groups \cite{Bondi:1962px,Sachs:1962wk,Barnich:2009se,Barnich:2010eb,Campiglia:2014yka,Campiglia:2015yka,Freidel:2021fxf}.

However, there are still two remaining problems. In higher dimensions, there are consistent fall-off conditions that  prevent the supertranslations  from being  asymptotic symmetries \cite{Hollands:2003ie, Hollands:2004ac}. On the other hand, the Weinberg soft theorem is believed to be valid in all dimensions \cite{1965PhRv..140..516W,Strominger:2013jfa,He:2014laa}. There exist interesting discussions on the extended symmetry in higher dimensions \cite{Kapec:2015vwa,Pate:2017fgt,Campoleoni:2020ejn}. Moreover, there are various null hypersurfaces (not necessarily null infinity) on which one could define ``localized charges'' \cite{Chandrasekaran:2018aop,Chandrasekaran:2021hxc,Chandrasekaran:2021vyu}. One of the most intriguing examples is the Killing horizon of black holes \cite{Donnay:2015abr,Donnay:2016ejv,Blau:2015nee,Haco:2018ske,Donnay:2019jiz}.
 From the intrinsic perspective of Carrollian manifold, there are no essential differences among various null hypersurfaces \cite{Hartong:2015xda,Ciambelli:2018xat, Ciambelli:2018ojf}. Therefore, one could define field theories and find quantum operators that represent Carrollian diffeomorphism for any null hypersurfaces in general dimensions. 

In this paper, we will study the quantum flux operators associated with Carrollian diffeomorphism in general dimensions. We will use two typical null hypersurfaces, the null infinity and the Rindler horizon, to present the results. More explicitly, we reduce the bulk massless scalar theory in Minkowski spacetime to future null infinity and construct the radiation fluxes. The flux operators form a faithful representation of the Carrollian diffeomorphism up to a central charge which may be regularized through the  
 spectral zeta function on the unit sphere. The result is also consistent with the  heat kernel method.  We will also study the scalar field theory in the Rindler wedge and reduce it to the null boundary, a Killing horizon associated with a Lorentz boost in Minkowski spacetime. Since the Killing horizon is not at infinity, we do not impose any artificial fall-off condition except that the field is finite at the null hypersurface. The field theory can be massive since any particles can cross through the null hypersurface. Similar to the case of null infinity, the quantum flux operators form a representation of Carrollian diffeomorphism at the Killing horizon. The supertranslation flux operators form a Virasoro algebra with a divergent central charge. We use the heat kernel method and Pauli-Villars method to regularize the central charge. They lead to the same conclusion that there is an effective central charge which is proportional to the area of the bifurcation surface of the Killing horizons. Using the conservation law of stress tensor and  Stokes' theorem, we identify the zero mode of supertranslation as the modular Hamiltonian of the Rindler wedge. The result indicates that Carrollian diffeomorphism may be related to quantum information theory. We try to calculate the expectation values of the quantum flux operators for the vacuum and excited states.  The relative fluxes, which are defined as the difference of the expectation values between two quantum states, may provide new observables in Unruh effect. 

The structure of the paper is as follows. 
In section \ref{reviewform}, we will introduce the basic ingredients on the  Carrollian manifold. In the following two sections, we will construct quantum flux operators on future null infinity and Rindler horizon and discuss their various properties.
We extend the results to general null hypersurfaces in section \ref{csfgeneralnull} and conclude in section \ref{conc}. Technical details are relegated to three appendices.

\textbf{Conventions and notations}

We will use the following conventions in this paper. 
Lorentz manifolds are denoted by $\mathbb{M},\mathbb{N}$, etc., with signature $(-,+,+,\cdots,+)$. Tensors on Lorentz manifolds are denoted by lowercase Greek indices $\mu,\nu,\cdots$. Timelike hypersurfaces are denoted by $\mathfrak{H},\mathfrak{I}$, and so on. A null hypersurface is a Carrollian manifold denoted  
as $\mathcal{H},\mathcal{I}$, etc..  Usually, they can be factorized as a  product such as 
    \be 
    \mathcal{H}=\textbf{R}\times S
    \ee where $\bf R$ is the retarded/advanced time direction and $S$ is a Riemannian manifold.
    Tensors on Carrollian manifolds are labeled by lowercase Roman indices $a,b,c$, etc. from the first half of the alphabet. 
Riemannian manifolds are denoted by  uppercase Roman letters. Tensors on Riemannian manifolds are indexed by uppercase Roman indices $A,B,C$, etc.. 

\section{Preliminaries}
 \label{reviewform}
In this section, we will review the minimal ingredients on Carrollian diffeomorphism which are relevant to later discussion. 
A Carrollian manifold $\mathcal{C}$ with topology ${\bf R}\times S$ has a degenerate metric 
\be 
ds^2\equiv \bm h=h_{AB}d\theta^A d\theta^B,\quad A,B=1,2,\cdots,d-2,
\ee as well as a null vector 
\bea 
\bm\chi=\partial_u.
\eea The metric $\bm{h}$ also defines the geometry of the $d-2$ dimensional manifold $S$, which is assumed to be Riemannian with  signature  $(+,+,\cdots,+)$. The  $d-2$ coordinates $\theta^{A},A=1,2,\cdots,d-2$ may also be collected as $\Omega=(\theta^1,\theta^2,\cdots,\theta^{d-2})$. The null vector $\bm\chi$ generates the time direction of the Carrollian manifold. We have used the coordinate $u$ to denote the time for  this manifold. The volume form for the manifold $S$  is denoted by
\be 
d\Omega=\sqrt{\det h} d\theta^1\wedge\cdots\wedge d\theta^{d-2}
\ee 
and  the unique volume $\bm\epsilon$ for the  Carrollian manifold $\mathcal{C}$ is given by
\be 
i_{\bm\chi}\bm\epsilon=d\Omega\quad\Rightarrow\quad \bm\epsilon=du\wedge d\Omega=\sqrt{\det h }du\wedge d\theta^1\wedge \cdots\wedge d\theta^{d-2}.
\ee For brevity, we will also write it as 
\be
\bm\epsilon=du d\Omega. 
\ee
Carrollian diffeomorphism is generated by the vectors $\bm\xi$  which satisfy
\be \mathcal{L}_{\bm\xi}\bm\chi=\mu\bm\chi.\label{Cdif}
\ee The general solution of the equation \eqref{Cdif} is 
\bea 
\bm\xi_{f,Y}=f(u,\Omega)\partial_u+Y^A(\Omega)\partial_A\label{cdifv}
\eea 
where $\bm\xi_f=f(u,\Omega)\partial_u$ generates general supertranslations (GSTs) and  $\bm\xi_Y=Y^A(\Omega)\partial_A$ generates special superrotations (SSRs) within the terminology of \cite{Liu:2022mne}. One may also consider general superrotations (GSRs) which are generated by $Y^A(u,\Omega)\partial_A$. However, this turns out to break the null structure of  the Carrollian manifold and we will not focus on it in this work. The supertranslations and superrotations are borrowed from BMS transformations. Nevertheless, in general situations, they are not related to the original BMS transformations which are only defined at future/past null infinity.
The vectors \eqref{cdifv} form an infinite dimensional Lie algebra 
\bea 
\ [\bm \xi_{f_1,Y_1},\bm \xi_{f_2,Y_2}]=\bm\xi_{f_1\dot{f}_2-f_2\dot{f}_1+Y_1^A\partial_A f_2-Y_2^A\partial_A f_1,[Y_1,Y_2]}\label{inflie}
\eea 
 where $\dot{f}_i\equiv \partial _uf_i,i=1,2$. In particular, for pure supertranslations, the corresponding  Lie algebra is 
\bea 
\ [\bm\xi_{f_1},\bm \xi_{f_2}]=\bm\xi_{f_1\dot{f}_2-f_2\dot{f}_1}
\eea 
which is called T-Witt algebra in \cite{Adami:2020amw}. For pure superrotations, the Lie algebra is 
\be 
\ [\bm\xi_{Y_1},\bm\xi_{Y_2}]=\bm \xi_{[Y_1,Y_2]}
\ee where $\bm\xi_Y$ generates diffeomorphism of the manifold $S$.
In \cite{Liu:2022mne, Liu:2023qtr, Liu:2023gwa}, the authors find that the Carrollian diffeomorphism generated by $\bm\xi$ may be realized by quantum flux operators $Q_{\bm\xi}$ whose algebra is exactly \eqref{inflie} up to anomalous quantum corrections 
\bea 
[Q_{\bm \xi_1},Q_{\bm \xi_2}]=iQ_{[\bm \xi_1,\bm \xi_2]}+K_{\bm\xi_1,\bm\xi_2}.
\eea 
We will use the conventions 
\be 
\mathcal{T}_f=Q_{\bm\xi_f} \quad\text{and}\quad\mathcal{M}_Y=Q_{\bm\xi_Y}
\ee to distinguish supertranslations and superrotations. The operators $Q_{\bm\xi}$ and the anomalous corrections $K_{\bm\xi_1,\bm\xi_2}$ may depend on the field $F$ defined on the Carrollian manifold in general 
\be 
Q_{\bm\xi}=Q_{\bm\xi}[F],\quad K_{\bm\xi_1,\bm\xi_2}=K_{\bm\xi_1,\bm\xi_2}[F].
\ee 
For real scalar theory, the unique anomalous term is a central charge of T-Witt algebra 
\be 
K_{\bm\xi_{f_1},\bm\xi_{f_2}}=C_T(f_1,f_2)=-\frac{i}{48\pi}\delta^{(d-2)}(0)\mathcal{I}_{f_1\dddot {f_2}-f_2\dddot {f_1}}\label{central}
\ee 
such that the algebra becomes a Virasoro algebra with additional transverse indices
\be 
[\mathcal{T}_{f_1},\mathcal{T}_{f_2}]=\mathcal{T}_{f_1\dot{f}_2-f_2\dot{f}_1}+C_T(f_1,f_2).
\ee 
 The identity operator in \eqref{central} is defined as 
\bea 
\mathcal{I}_f=\int du d\Omega f(u,\Omega),
\eea 
while the Dirac delta function with argument equalling zero
\be 
\delta^{(d-2)}(0)=\delta^{(d-2)}(\Omega-\Omega')|_{\Omega'=\Omega}
\ee is proportional to the density of states in the transverse direction, e.g. the density of states on the manifold $S$. Since the spectrum of the field in the transverse direction is continuous in general, the Dirac delta function $\delta^{(d-2)}(0)$ is divergent. Note that we have already written the results in $d$ dimensions. 
In the vector and gravitational theory, there are additional anomalous terms that may be interpreted as helicity flux operators whose explicit forms can be found in \cite{Liu:2023qtr, Liu:2023gwa}. The helicity flux operators generate super-duality transformations (angle-dependent duality transformations)  which are essential for the field theories with nonzero spins. 

To find the operators $Q_{\bm\xi}$, one may embed the Carrollian manifold into $d$ dimensional spacetime $\mathbb{M}$ where the bulk theory is well defined. The stress tensor $T_{\mu\nu}$ of the bulk theory is conserved such that one may integrate the current 
\be 
j_{\bm\xi}^\mu=T^{\mu}_{\ \nu}\xi^\nu
\ee over the Carrollian manifold $\mathcal{C}$ to obtain
\be 
Q_{\bm\xi}=\int_{\mathcal{C}} (d^{d-1}x)_\mu j_{\bm\xi}^\mu. \label{Carrollxi}
\ee 
The vector $\bm\xi$ may be chosen as the generator of Carrollian diffeomorphism \eqref{cdifv}. If $\bm\xi$ is a Killing vector  when extended into the bulk spacetime $\mathbb{M}$, the operator  $Q_{\bm\xi}$ is the corresponding charge which crosses the Carrollian manifold $\mathcal{C}$ during the whole time. Therefore, $Q_{\bm\xi}$ could be interpreted as a leaky flux from bulk to boundary.

The above results have been mainly found at null infinity in four-dimensional flat spacetime. In this paper, we will extend it to general null hypersurfaces  and general dimensions. 

\section{Scalar field theory at future null infinity}\label{csfnull}
In this section, we will construct various scalar field  theories at future null infinity ($\mathcal{I}^+$) from bulk reduction in general dimensions. The overall framework is to embed $\mathcal{I}^+$ into Minkowski spacetime $\mathbb{R}^{1,d-1}$  and  reduce the bulk theory from $\mathbb{R}^{1,d-1}$ to $\mathcal{I}^+$.
\subsection{Null infinity}
The future/past null infinity ($\mathcal{I}^+/\mathcal{I}^-$) is a Carrollian manifold in asymptotically flat spacetime. In this work, we will discuss $\mathcal{I}^+$ for  Minkowski spacetime $\mathbb{R}^{1,d-1}$ which  can be described in Cartesian coordinates $x^\mu=(t,x^i)$ 
 \bea 
 ds^2=-dt^2+dx^i dx^i=\eta_{\mu\nu}dx^\mu dx^\nu, 
 \eea where $\mu=0,1,\cdots,d-1$ denotes the components of spacetime coordinates and $i=1,2,\cdots,d-1$ labels the components of spatial coordinates. We also use the  retarded coordinate system $(u,r,\theta^A)$ 
 and write the metric of  $\mathbb{R}^{1,d-1}$ as 
 \bea 
 ds_{\mathbb{R}^{1,d-1}}^2=-du^2-2du dr+r^2\gamma_{AB}d\theta^Ad\theta^B,\quad A,B=1,2,\cdots, d-2.
 \eea 
  The future null infinity $\mathcal{I}^+$ is a $d-1$ dimensional Carrollian manifold 
  \bea 
  \mathcal{I}^+={\bf R}\times S^{d-2}
  \eea with a degenerate metric 
\bea 
ds_{\mathcal{I}^+}^2\equiv\bm{\gamma}=\gamma_{AB}d\theta^Ad\theta^B=ds^2_{S^{d-2}}.  \label{degemet}
\eea 
The spherical coordinates $\theta^A=(\theta_1,\cdots,\theta_{d-2})$ are used to describe the unit sphere $S^{d-2}$ whose metric $ds^2_{S^{d-2}}$ can be found in Appendix \ref{spc}.
We will also use the notation $\Omega=(\theta_1,\cdots,\theta_{d-2})$ to denote the spherical coordinates in the context. 
The covariant derivative $\nabla_A$ is  adapted to the metric $\gamma_{AB}$, while the covariant derivative $\nabla_\mu$ is adapted to the Minkowski metric in Cartesian frame. The integral measure on $\mathcal{I}^+$ is abbreviated as
\bea 
\int du d\Omega\equiv\int_{-\infty}^\infty du \int_{S^{d-2}}d\Omega,
\eea where the integral measure on $S^{d-2}$
is 
\bea 
\int d\Omega\equiv \int_{S^{d-2}}d\Omega=[\prod_{j=2}^{d-2}\int_0^\pi \sin^{j-1}\theta_j d\theta_j]\int_0^{2\pi}d\theta_1.
\eea Besides the metric \eqref{degemet}, there is also a distinguished null vector 
\be 
\bm\chi=\partial_u
\ee which is to generate the retarded time direction.

To obtain the metric of the Carrollian manifold \eqref{degemet} from  the bulk metric, one may choose a cutoff 
\bea 
r=R
\eea such that the induced metric on the hypersurface 
\bea 
\mathfrak{H}_R=\{p\in \mathbb{R}^{1,d-1}|\ p=(u,r,\theta_1,\cdots,\theta_{d-2}) \ \text{with}\ r=R\}
\eea is 
\bea 
ds^2=-du^2+R^2\gamma_{AB}d\theta^Ad\theta^B=R^2(-\frac{du^2}{R^2}+\gamma_{AB}d\theta^Ad\theta^B). \label{slice}
\eea  
 We use a Weyl scaling to remove the conformal factor and  take the limit 
\bea 
R\to\infty
\eea while keeping the retarded time $u$ finite such that \eqref{slice} becomes the metric of the Carrollian manifold $\mathcal{I}^+$. We define the limit 
\bea 
\lim{}\hspace{-0.8mm}_+=\lim_{r\to\infty,\ u \ \text{finite}}
\eea to send the quantities on $\mathfrak{H}_r$ to $\mathcal{I}^+$. Similarly, taking the limit below 
\bea 
\lim{}\hspace{-0.8mm}_-=\lim_{r\to\infty,\ v\ \text{finite}}
\eea  sends the quantities on $\mathfrak{H}_r$ to $\mathcal{I}^-$ where $v$ is the advanced time $v=t+r$.
\subsection{Bulk theory}
We may consider a general scalar theory with the action 
\bea 
S[\Phi]=\int d^dx \sqrt{-g}[-\frac{1}{2}\partial_\mu \Phi \partial^\mu \Phi-V(\Phi)].\label{actionPhi}
\eea
Only massless particles can arrive at $\mathcal{I}^+$, and therefore we consider the massless theory with the potential 
\be 
V(\Phi)=\sum_{k=3}^\infty \frac{\lambda_k}{k}\Phi^k.
\ee 
We impose the fall-off condition 
\be 
\Phi(t,\bm x)=\frac{\Sigma(u,\Omega)}{r^{\Delta}}+\frac{\Sigma^{(2)}(u,\Omega)}{r^{\Delta+1}}+\cdots\label{falloffscalar}
\ee for the bulk scalar field. The constant $\Delta$ is the scaling dimension of the bulk field and is related to the dimension $d$ through
\be \Delta=\frac{d-2}{2}.
\ee 

Now we consider the equation of motion (EOM)
\begin{align}
  \p_\mu\partial^\mu\Phi-V'(\Phi)=\p_\mu \p^\mu \Phi-\sum_{k=3}^{ \infty} {\lambda_k}\Phi^{k-1}=0.
\end{align}
For the first term, one can show that
\begin{align}
  \partial_\mu\partial^\mu=-2\partial_u\partial_r-\frac{2\Delta}{r}\partial_u+\partial_r^2+\frac{2\Delta}{r}\partial_r+\frac{1}{r^2}\nabla_A\nabla^A,
\end{align}
which leads to
\begin{align}
  \partial_\mu\partial^\mu \Phi=&\bigl(-2\partial_u\partial_r-\frac{2\Delta}{r}\partial_u+\partial_r^2+\frac{2\Delta}{r}\partial_r+\frac{1}{r^2}\nabla_A\nabla^A\bigr)\bigl[\frac{\Sigma(u,\Omega)}{r^{\Delta}}+\sum_{k\ge 2}^{ \infty}\frac{\Sigma^{(k)}(u,\Omega)}{r^{\Delta+k-1}}\bigr]\nn\\
  =&\ \sum_{k\ge2}^{ \infty}\frac{1}{r^{\Delta+k}}\Big[2(k-1)\dot\Sigma^{(k)} +(\Delta+k-2)(k-1-\Delta)\Sigma^{(k-1)}+\nabla_A\nabla^A\Sigma^{(k-1)}\Big].
\end{align}
For the potential term, we find
\begin{align}
  V'(\Phi)=\sum_{k\ge 2}^\infty\frac{1}{r^{\Delta+k}}\sum_{n=3}^{\infty}{\lambda_n}\sum_{k_1,\cdots,k_{n-1}\ge1}^{\sum_{j=1}^{n-1}k_j=\Delta+k-(n-1)(\Delta-1)}\Sigma^{(k_{1})}\cdots \Sigma^{(k_{n-1})}.
\end{align}
Note that it does not contribute to EOM for \(k=0\). Moreover, for $k=1$, the potential also has no contribution unless $\Delta=1$ (corresponding to dimension 4). If $\Delta=1$, the leading EOM is \(\lambda_3\Sigma^2=0\), which results in \(\lambda_3=0\). Otherwise, the leading EOM is trivial. Nevertheless, we write out the EOM at order  \(\mathcal{O}({1}/{r^{\Delta+k}})\)
\begin{align}
  0=&2(k-1)\dot\Sigma^{(k)} +(\Delta+k-2)(k-\Delta-1)\Sigma^{(k-1)}+\nabla_A\nabla^A\Sigma^{(k-1)}\nn\\
  &-\sum_{n=3}^{ \infty}{\lambda_n}\sum_{k_1,\cdots,k_{n-1}\ge1}^{k_1+\cdots+k_{n-1}=\Delta+k-(n-1)(\Delta-1)}\Sigma^{(k_{1})}\cdots \Sigma^{(k_{n-1})}.\label{constraintseqn}
\end{align}
In odd dimensions, $\Delta$ is a half integer. Therefore, only the terms with  $n=\text{even}$  have contributions in the second line. In general dimensions, the leading coefficients $\Sigma$ encode the dynamical propagating degree of freedom. All the subleading coefficients $\Sigma^{(k)},\ k\ge 2$ are determined by the EOM up to initial data $\Sigma^{(k)}(u_0,\Omega),\ k\ge 2$ where $u_0$ is the initial time on $\mathcal{C}$. 

\subsection{Boundary theory}
The pre-symplectic form of the theory is 
\bea 
\bm\Theta(\delta\Phi;\Phi)=-\int_{\mathfrak{H}_r} (d^{d-1}x)_\mu \partial^\mu\Phi \delta\Phi
\eea  where 
\bea 
(d^{d-1}x)_\mu=\frac{1}{(d-1)!}\epsilon_{\mu\nu_1\cdots\nu_{d-1}}dx^{\nu_1}\wedge \cdots\wedge dx^{\nu_{d-1}}=-r^{d-2}m_\mu du d\Omega.
\eea The normal co-vector $m_\mu$ of the hypersurface $\mathfrak{H}_r$ is 
\be 
dr=n_i dx^i\quad\Rightarrow\quad m_\mu=\frac{1}{2}(n_\mu+\bar{n}_\mu)
\ee where the null vectors $n_\mu$ and $\bar{n}_\mu$ are defined as 
\be 
n^\mu=(1,n^i),\quad \bar{n}^\mu=(-1,n^i)
\ee with $n^i={x^i}/{r}, i=1,2,\cdots,d-1$ being the normal vector of $S^{d-2}$ embedded into $\mathbb{R}^{d-1}$.
The symplectic form for the theory on $\mathfrak{H}_r$ is 
\be 
\bm\Omega^{\mathfrak{H}_r}(\delta\Phi;\delta\Phi;\Phi)=-\int_{\mathfrak{H}_r}(d^{d-1}x)_\mu \partial^\mu\delta \Phi \wedge\delta\Phi.
\ee With the fall-off condition \eqref{falloffscalar}, the symplectic form becomes 
\bea 
\bm\Omega(\delta\Sigma;\delta\Sigma;\Sigma)=\lim{}\hspace{-0.8mm}_{ +}\bm\Omega^{\mathfrak{H}_r}(\delta\Phi;\delta\Phi;\Phi)=\int du d\Omega \delta\Sigma\wedge\delta\dot\Sigma
\eea at $\mathcal{I}^+$. This is identified as the symplectic form of the Carrollian scalar field theory at $\mathcal{I}^+$. We may find the following commutators for the field $\Sigma$ 
\bs\begin{align}
    [\Sigma(u,\Omega),\Sigma(u',\Omega')]&=\frac{i}{2}\alpha(u-u')\delta^{(d-2)}(\Omega-\Omega'),\\
    [\Sigma(u,\Omega),\dot{\Sigma}(u',\Omega')]&=\frac{i}{2}\delta(u-u')\delta^{(d-2)}(\Omega-\Omega'),\\
    [\dot{\Sigma}(u,\Omega),\dot{\Sigma}(u',\Omega')]&=\frac{i}{2}\delta'(u-u')\delta^{(d-2)}(\Omega-\Omega'), \label{expSigma}
\end{align}\es where the function $\alpha(u-u')$ is \bea 
\alpha(u-u')=\frac{1}{2}[\theta(u'-u)-\theta(u-u')]
\eea 
with $\theta(x)$ being the Heaviside step function 
\be 
\theta(x)=\left\{\begin{array}{cc}1,&\ x>0\\
0,&\ x<0.\end{array}\right.
\ee 
The field $\Sigma$ may be expanded in a set of complete basis of $\mathcal{I}^+$ 
\bea 
\Sigma(u,\Omega)=\int_0^\infty \frac{d\omega}{\sqrt{4\pi\omega}}{ \sum_{\bm\ell}}[a_{\omega,\bm\ell}e^{-i\omega u}Y_{\bm \ell}(\Omega)+a_{\omega,\bm \ell}^\dagger e^{i\omega u}Y^*_{\bm \ell}(\Omega)]\label{expa}
\eea with the commutators 
\bs\begin{align}
    & [a_{\omega,\bm\ell},a_{\omega',\bm\ell'}]=[a_{\omega,\bm\ell}^\dagger,a_{\omega',\bm\ell'}^\dagger]=0,\\
& [a_{\omega,\bm\ell},a_{\omega',\bm\ell'}^\dagger]=\delta(\omega-\omega')\delta_{\bm\ell,\bm\ell'}.
\end{align}\es
In the expansion \eqref{expa}, the functions $Y_{\bm\ell}(\Omega)$ are spherical harmonic functions on $S^{d-2}$. Readers may find more details in Appendix \ref{sphe}. We have checked the mode expansion \eqref{expa} using canonical quantization method in Appendix \ref{modeexp}. With the canonical quantization, we may define the vacuum state $|0\rangle$ through 
\be 
a_{\omega,\bm\ell}|0\rangle=0.
\ee 
Therefore, we find the following correlators
\bs\begin{align}\langle 0|\Sigma(u,\Omega)\Sigma(u',\Omega')|0\rangle&=\beta(u-u')\delta^{(d-2)}(\Omega-\Omega'), \label{prop}\\
\langle 0|\Sigma(u,\Omega)\dot\Sigma(u',\Omega')|0\rangle&=\frac{1}{4\pi(u-u'-i\epsilon)}\delta^{(d-2)}(\Omega-\Omega'),\label{corsigdot}\\
\langle 0|\dot\Sigma(u,\Omega)\Sigma(u',\Omega')|0\rangle&=-\frac{1}{4\pi(u-u'-i\epsilon)}\delta^{(d-2)}(\Omega-\Omega'),\label{cordotsig}\\
\langle 0|\dot\Sigma(u,\Omega)\dot\Sigma(u',\Omega')|0\rangle&=-\frac{1}{4\pi(u-u'-i\epsilon)^2}\delta^{(d-2)}(\Omega-\Omega').\label{cordotsigdotsig}
\end{align}\es 

\subsection{Flux operators for Carrollian diffeomorphism}
In this section, we will define flux operators for Carrollian diffeomorphism in general dimensions. For any conserved current $j^\mu$ in the bulk, we may construct the corresponding flux radiated to $\mathcal{I}^+$ through the limit
\be
\mathcal{F}[j]=\lim{}\hspace{-0.8mm}_+\int_{\mathfrak{H}_r} (d^{d-1}x)_\mu j^\mu=-\lim{}\hspace{-0.8mm}_+\int_{\mathfrak{H}_r}du d\Omega r^{d-2}m_\mu j^\mu.
\ee 

The stress tensor for the theory \eqref{actionPhi} is 
\be 
T_{\mu\nu}=\partial_\mu\Phi \partial_\nu \Phi+ g_{\mu\nu}\mathcal{L}\label{stress}
\ee where $\mathcal{L}$ is the Lagrangian of the theory. Note that we have the expansion
\begin{align}
  \Phi=\sum_{k\ge 1}\frac{\Sigma^{(k)}(u,\Omega)}{r^{\Delta+k-1}},\quad \Sigma^{(1)}=\Sigma
\end{align}
for the field $\Phi$  which leads to
\begin{align}
  \partial_\mu\Phi=&\sum_{k\ge 1}\frac1{r^{\Delta+k-1}}\big[{-n_\mu\dot\Sigma^{(k)}-\frac{1}{2}(\Delta+k-2)(n_\mu+\bar{n}_\mu)\Sigma^{(k-1)}-Y_\mu^A\nabla_A\Sigma^{(k-1)}}\big],
\end{align} where we have assumed $\Sigma^{(0)}=0$.
The vector $Y_\mu^A$ is the  gradient of the null vector $n_\mu$
\be Y_\mu^A=-\nabla^An_\mu.
\ee 

Similarly, we have 
\begin{align}
  \partial_\mu\Phi\partial_\nu\Phi=\sum_{k_1,k_2\ge 1}&\frac{1}{r^{2\Delta+k_1+k_2-2}}\big[(n_\mu\dot\Sigma^{(k_1)}+(\Delta+k_1-2)m_\mu\Sigma^{(k_1-1)}+Y_\mu^A\nabla_A\Sigma^{(k_1-1)})\nn\\
  &\times(n_\nu\dot\Sigma^{(k_2)}+(\Delta+k_2-2)m_\mu\Sigma^{(k_2-1)}+Y_\nu^A\nabla_A\Sigma^{(k_2-1)})\big].\label{pppp}
\end{align}
The stress tensor has the following fall-off behaviour
\begin{align}
  T_{\mu\nu}=\frac{t_{\mu\nu}^{(2\Delta)}}{r^{2\Delta}}+\frac{t_{\mu\nu}^{(2\Delta+1)}}{r^{2\Delta+1}}+\cdots
\end{align}
where 
\begin{align}
  &t^{(2\Delta)}_{\mu\nu}=n_\mu n_\nu \dot\Sigma^2,\\ 
  &t^{(2\Delta+1)}_{\mu\nu}=(n_\mu\dot\Sigma^{(2)}+{\Delta}\Sigma m_\mu+Y_\mu^A\nabla_A\Sigma)n_\nu \dot\Sigma+(\mu\leftrightarrow\nu)-\eta_{\mu\nu}\Delta \dot\Sigma \Sigma.
\end{align}
For any Killing vector $\bm \xi$ in $\mathbb{R}^{1,d-1}$, we may find a conserved current 
\bea 
j^\mu_{\bm\xi}=T^{\mu\nu}\xi_\nu.
\eea Therefore, the flux corresponding to the Killing vector $\bm\xi$ is 
\bea 
\mathcal{F}_{\bm\xi}\equiv \mathcal{F}[j_{\bm\xi}]=-\lim{}\hspace{-0.8mm}_+\int_{\mathcal{H}_r}du d\Omega r^{d-2}m_\mu T^{\mu\nu}\xi_\nu.
\eea 
\begin{enumerate}
    \item For a translation generated by $\bm\xi_{\text{trans.}}=c^\mu\partial_\mu$, the energy-momentum flux is 
    \bea 
    \mathcal{F}_{\text{trans.}}=-\int dud\Omega c^\mu n_\mu \dot{\Sigma}^2.\label{fluxtrans}
    \eea 
    \item For a Lorentz rotation generated by $\bm\xi_{\text{LT}}=\omega^{\mu\nu}(x_\mu\partial_\nu-x_\nu\partial_\mu)$, the angular momentum or center of mass flux is 
    \bea 
    \mathcal{F}_{\text{LT}}=\int du d\Omega[ \frac{u}{d-2}\omega^{\mu\nu}\nabla_AY^A_{\mu\nu}\dot\Sigma^2+\omega^{\mu\nu}Y_{\mu\nu}^A\dot\Sigma\nabla_A\Sigma],\label{fluxrot}
    \eea 
   where $Y_{\mu\nu}^A$ is the conformal Killing vectors on $S^{d-2}$. Readers can find more details about the derivation in Appendix \ref{CKVSd2}.
\end{enumerate} 
From the fluxes \eqref{fluxtrans} and \eqref{fluxrot} we may define the energy density flux operator 
\bea 
T(u,\Omega)=:\dot\Sigma^2:
\eea and angular momentum density flux operator
\be 
M_A(u,\Omega)=\frac{1}{2}\left(:\dot\Sigma\nabla_A\Sigma-\Sigma\nabla_A\dot\Sigma:\right).
\ee We have used normal ordering $:\cdots:$ in the definition. The angular momentum density flux operator $M_A(u,\Omega)$ is only fixed up to a constant $\lambda$
\be 
M_A(u,\Omega;\lambda)=\frac{1}{2}\left(:\lambda\dot\Sigma\nabla_A\Sigma-(1-\lambda)\Sigma\nabla_A\dot\Sigma:\right)
\ee since they lead to the same fluxes in \eqref{fluxrot} after integration by parts. Following \cite{Liu:2022mne}, we fix $\lambda=\frac{1}{2}$ using the orthogonality condition 
\be 
\langle 0|T(u,\Omega)M_A(u',\Omega')|0\rangle=0.
\ee 

Note the Poincar\'e group is a subgroup of Carrollian diffeomorphism for $\mathcal{I}^+$. We may exploit \eqref{Carrollxi} and extend the fluxes to general Carrollian  diffeomorphism. 
After some algebra, we find the following extended flux operators
\bs\begin{align}
\mathcal{T}_f&=\int du d\Omega f(u,\Omega)T(u,\Omega),\\
\mathcal{M}_Y&=\int du d\Omega Y^A(\Omega)M_A(u,\Omega)
\end{align}\es for Carrollian diffeomorphism. The time and angle dependence of the density flux operators may  be transformed to the following smeared operators 
\bs\begin{align}
\bar{\mathcal{T}}_f&=\int du d\Omega f(u,\Omega)T(u,\Omega),\label{smeart}\\
\bar{\mathcal{M}}_Y&=\int du d\Omega Y^A(u,\Omega)M_A(u,\Omega).\label{smearr}
\end{align}\es
 We add a bar in the definition to distinguish them from the one about Carrollian diffeomorphism. The operator $\bar{\mathcal{T}}_f$ is equal to $\mathcal{T}_f$ exactly for any smooth function $f=f(u,\Omega)$. On the other hand, the operator $\bar{\mathcal{M}}_Y$ matches $\mathcal{M}_Y$ only for time-independent vectors $Y^A=Y^A(\Omega)$. As has been noticed, for general time-dependent vectors $Y^A=Y^A(u,\Omega)$, the corresponding operator generates non-local terms which may relate to the violation of the null structure under GSRs. 

There is a rather different way to obtain the same flux operators. From Hamiltonian theory, we learn that 
the Hamiltonian corresponding to a general vector $\bm\xi$ may be obtained from the Hamilton equation
\be 
\delta H_{\bm\xi}=i_{\bm\xi}\bm\Omega(\delta\Sigma;\delta\Sigma;\Sigma)\label{Hamcov}
\ee where $i_{\bm\xi}$ is the interior product  in phase space induced by $\bm\xi$. The variation of the boundary scalar field $\Sigma$ can be obtained from  bulk diffeomorphism 
\bea 
\delta_{\bm\xi}\Phi=\xi^\lambda\partial_\lambda\Phi\quad\Rightarrow\quad\delta_{\bm\xi}\Sigma=\lim{}\hspace{-0.8mm}_+r^{ \Delta}\delta_{\bm\xi}\Phi.
\eea 
In  Minkowski spacetime $\mathbb{R}^{1,d-1}$, the generator of a global translation along the direction $c^\mu$ is 
\be 
\bm\xi_{\text{trans.}}=c^\mu\partial_\mu=c^\mu[-n_\mu \partial_u+m_\mu\partial_r-\frac{1}{r}Y_\mu^A\partial_A].\label{partialxi}
\ee 
 We have inserted a constant vector $c^\mu$ into the expression to denote the direction of the translation, and transformed to retarded coordinates at the last step. Therefore, the  boundary field $\Sigma$ is transformed to 
\bea 
\delta_{\bm\xi_{\text{trans.}}}\Sigma=-c^\mu n_\mu \dot{\Sigma}.\label{sigmatrans}
\eea 
We observe that the generator of the translation $\bm\xi_{\text{trans.}}$ reduces to 
\bea 
-c^\mu n_\mu \partial_u
\eea at $\mathcal{I}^+$. Obviously, it is the Carrollian diffeomorphism generated by \be \bm\xi_f=f(u,\Omega)\partial_u
\ee with 
\bea 
f(u,\Omega)=-c^\mu n_\mu.
\eea 
It is natural to assume that 
the transformation of the boundary field $\Sigma$ under a GST is 
\be 
\delta_f\Sigma(u,\Omega)=f(u,\Omega)\dot{\Sigma}.
\ee 
Note this is also the transformation of a boundary scalar field under the Carrollian diffeomorphism generated by $\bm\xi_f$ 
\be 
\delta_f\Sigma(u,\Omega)=\bm\xi_f(\Sigma)=\mathcal{L}_{\bm\xi_f}\Sigma=f(u,\Omega)\dot{\Sigma}.
\ee In this derivation, there is no need to assume that $\bm\xi$ is a Killing vector or an asymptotic Killing vector near $\mathcal{I}^+$. 
There would be a corresponding Hamiltonian $H_f$ related to the GST through the Hamilton equation \eqref{Hamcov}. The infinitesimal variation of the Hamiltonian is
\bea \delta H_f=i_{\bm\xi_{f}}\bm\Omega(\delta\Sigma;\delta\Sigma;\Sigma)
\eea which is integrable. The integrable flux is 
\bea 
 H_f=\int du d\Omega f(u,\Omega)\dot\Sigma^2(u,\Omega).\label{Hamf}
\eea 
This is the classical version of the operator $\mathcal{T}_f$.
Similarly, the generator of a Lorentz transformation  parameterized by an antisymmetric constant tensor $\omega^{\mu\nu}$ is 
\bea 
\bm\xi_{\text{LT}}&=&\omega^{\mu\nu}(x_\mu\partial_\nu-x_\nu\partial_\mu)\nn\\&=&\omega^{\mu\nu}[-\frac{1}{2}u n_{\mu\nu}\partial_u+\frac{1}{2}(u+r)n_{\mu\nu}\partial_r+(Y_{\mu\nu}^A-\frac{u}{r}\bar{m}_{\mu\nu}^A)\partial_A]\label{ltgen}
\eea where the antisymmetric tensors $n_{\mu\nu}$ and $\bar{m}_{\mu\nu}$ are defined in Appendix \ref{CKVSd2}. The boundary field $\Sigma$ is transformed as
\bea 
\delta_{\bm\xi_{\text{LT}}}\Sigma=\omega^{\mu\nu}[\frac{u}{d-2}\nabla_AY_{\mu\nu}^A\dot{\Sigma}+Y_{\mu\nu}^A\nabla_A\Sigma+\frac{1}{2}\nabla_A Y_{\mu\nu}^A \Sigma]\label{sigmalt}
\eea under Lorentz transformation. We note that the generator \eqref{ltgen} reduces to 
\bea 
\omega^{\mu\nu}[\frac{u}{d-2}\nabla_AY^A_{\mu\nu} \partial_u+Y_{\mu\nu}^A\partial_A]
\eea at $\mathcal{I}^+$. This is a Carrollian diffeomorphism
\bea 
\bm\xi_{f,Y}=f(u,\Omega)\partial_u+Y^A(\Omega)\partial_A
\eea with 
\bea 
f=\frac{u}{d-2}\omega^{\mu\nu}\nabla\cdot Y_{\mu\nu}\quad\text{and}\quad Y^A=\omega^{\mu\nu}Y^A_{\mu\nu}.
\eea 
Therefore, it is natural to assume the variation of $\Sigma$ under SSR to be 
\bea 
\delta_{Y}\Sigma\equiv\delta_{\bm\xi_Y}\Sigma=\Delta(Y;\Sigma;u,\Omega)=Y^A\nabla_A\Sigma+\frac{1}{2}\nabla_AY^A \Sigma \label{deltaYsigmanull}
\eea which could be reduced to \eqref{sigmalt} for Lorentz transformations\footnote{We need to subtract a supertranslation which is generated by $\bm\xi_{f}$ with $f=\frac{u}{d-2}\nabla\cdot Y$.}. Note the transformation law \eqref{deltaYsigmanull} is independent of the dimension $d$. However, one should  notice that it is not equal to the following transformation
\be 
\bm\xi_Y(\Sigma)=Y^A\nabla_A\Sigma
\ee using boundary Lie derivative. 
To understand this transformation, we integrate the infinitesimal transformation \eqref{deltaYsigmanull} and find the finite transformation
\be 
\Sigma(u,\Omega)\to \Sigma'(u,\Omega')=\Big|\frac{\partial \theta'}{\partial\theta}\Big|^{-1/2}\left(\frac{\det\gamma(\theta)}{\det\gamma(\theta')}\right)^{1/4}\Sigma(u,\Omega)\,\,,\label{weight}
\ee where $|\frac{\partial \theta'}{\partial\theta}|$ is the Jacobian under finite superrotation
\be 
(u,\theta^A)\to(u,\theta'^A).\label{sr}
\ee 
Thus, the field $\Sigma$ is a scalar density under superrotations with weight $w=\frac{1}{2}$ due  to the  result \eqref{weight}. In contrast, $\Sigma$ is a scalar under general supertranslations. 
In general, we may define a scalar density $\Lambda(u,\Omega)$ of  weight $w$ under superrotation \eqref{sr} by the following transformation
\be 
\Lambda(u,\Omega)\to \Lambda'(u,\Omega')=\Big|\frac{\partial \theta'}{\partial\theta}\Big|^{-w}\left(\frac{\det\gamma(\theta)}{\det\gamma(\theta')}\right)^{w/2}\Lambda(u,\Omega).
\ee Infinitesimally, this is 
\be 
\delta_Y\Lambda(u,\Omega)=Y^A(\Omega)\nabla_A\Lambda(u,\Omega)+w\nabla_AY^A \Lambda(u,\Omega).\label{srweight}
\ee 

Interestingly, we could find the integrable flux for an SSR labeled by $Y^A(\Omega)$ 
\bea 
 H_Y=\int du d\Omega \dot\Sigma \Delta(Y;\Sigma;u,\Omega).\label{HamY}
\eea In quantum theory, 
the operators should be normal ordered and we find the same flux operator
\be 
\mathcal{M}_Y=\int du d\Omega : \dot\Sigma \Delta(Y;\Sigma;u,\Omega):\ =\int du d\Omega Y^A(\Omega)M_A(u,\Omega).
\ee 

Before we close this subsection,  we will shortly discuss the integrability problem for general scalar densities $\Lambda(u,\Omega)$. Assuming the boundary scalar theory is defined through the symplectic form
\be 
\bm\Omega(\delta\Lambda;\delta\Lambda;\Lambda)=\int du d\Omega \delta\Lambda\wedge\delta\dot\Lambda
\ee 
and the superrotation variation of the field $\Lambda$ is given by \eqref{srweight} where $w$ is free. The corresponding Hamiltonian is 
\bea 
\delta H_Y=\delta \int du d\Omega \dot\Lambda(Y^A\nabla_A\Lambda+w \nabla_AY^A\Lambda)+(1-2w)\int du d\Omega \nabla_AY^A\ \dot\Lambda \delta\Lambda.
\eea The first part of the above equation on the right-hand side is integrable while the second part is not except for $w=\frac{1}{2}$. 
\subsection{Commutators and regularization of central charge}
By computing the commutators 
\bs\begin{align}
    [\mathcal{T}_f,\Sigma(u,\Omega)]&=-if(u,\Omega)\dot\Sigma(u,\Omega),\\
    [\mathcal{M}_Y,\Sigma(u,\Omega)]&=-i\Delta(Y;\Sigma;u,\Omega),
\end{align}\es we confirm that $\mathcal{T}_f$ and $\mathcal{M}_Y$ are generators for the Carrollian diffeomorphism given that $\Sigma$ is a scalar density of weight $\frac{1}{2}$ under superrotations. More explicitly, $\mathcal{T}_f$ generates GSTs while $\mathcal{M}_Y$ generates SSRs. They should form Lie algebra \eqref{inflie} up to central extension terms due to quantum corrections. Indeed, we find 
\bs\label{comscalar}\begin{align}
\    [\mathcal{T}_{f_1},\mathcal{T}_{f_2}]&=C_T(f_1,f_2)+i\mathcal{T}_{f_1\dot f_2-f_2\dot f_1},\label{vir1}\\
\ [\mathcal{T}_f,\mathcal{M}_Y]&=-i\mathcal{T}_{Y^A\nabla_Af},\label{vir2}\\
\ [\mathcal{M}_Y,\mathcal{M}_Z]&=i\mathcal{M}_{[Y,Z]},\label{vir3}
\end{align}\es where the central charge is 
\be 
C_T(f_1,f_2)=-\frac{i}{48\pi}\delta^{(d-2)}(0)\mathcal{I}_{f_1\dddot {f_2}-f_2\dddot {f_1}}.\label{centralcharge}
\ee The Dirac delta function on $S^{d-2}$ with the argument equalling zero is divergent\footnote{Readers may find more details in \ref{generaladdthm}.}
\bea
\delta^{(d-2)}(0)&=&\delta^{(d-2)}(\Omega-\Omega')|_{\Omega'=\Omega}=\sum_{\bm\ell}Y_{\bm\ell}^*(\Omega)Y_{\bm\ell}(\Omega)=\frac{\sum\limits_{\bm \ell}1}{I(0)}
\eea where $\sum\limits_{\bm\ell}1$ counts the number of  all possible eigenstates on $S^{d-2}$ and $I(0)$ is the area of the unit sphere. Therefore, Dirac delta function $\delta^{(d-2)}(0)$ counts the number of states on $S^{d-2}$. Now we will try to regularize the Dirac delta function. Notice that the eigenstates on $S^{d-2}$ obey the Laplace equation 
\be 
\Delta_{S^{d-2}}Y_{\bm\ell}(\Omega)=-\ell_m(\ell_m+d-3))Y_{\bm\ell}(\Omega),
\ee where $Y_{\bm\ell}(\Omega)$ is the spherical harmonic function and $\Delta_{S^{d-2}}=\nabla^A\nabla_A$ is the  Laplace operator on $S^{d-2}$. The Laplace operator is a natural elliptic operator which may be used to define the spectral zeta function\footnote{Spectral zeta function has been used by Hawking \cite{1977CMaPh..55..133H} to regularize the partition function of quantum fields in curved spacetime.} on $S^{d-2}$
\be 
\zeta_{S^{d-2}}(s)\equiv \text{tr}'(-\Delta_{S^{d-2}})^{-s}
\ee where $\text{tr}'$ is to sum over all possible eigenstates except the one with $\ell_m=0$. The eigenstates with zero eigenvalue are excluded otherwise the summation would be divergent for any $\text{Re}(s)>0$. Using the degeneracy of the eigenstates  and the eigenvalue of the spherical harmonics, we find 
\be 
\zeta_{S^{d-2}}(s)=\sum_{\ell=1}^\infty g(\ell,d-2)\left(\ell(\ell+d-3)\right)^{-s}.
\ee The spectral zeta function $\zeta_{S^{d-2}}(s)$ is convergent for sufficient large $s$ and can be analytically continued to the complex plane except for a single pole. Since the degeneracy $g(\ell,d-2)$ is a polynomial of $\ell$ of degree $d-3$, we may always reduce $\zeta_{S^{d-2}}(s)$ to a finite summation of the functions 
\bea 
f(s;a,b,c)=\sum_{\ell=1}^\infty \ell^{-s+b}(\ell+a)^{-s+c}
\eea where $a,b,c$ are integers. In \cite{1994zrta.book.....E,elizalde2012ten}, this function has been treated carefully and expressed as a series of Riemann zeta functions (see Appendix \ref{zetaregu}). The number of eigenstates on $S^{d-2}$ may be regularized to 
\bea 
\#\, \text{of eigenstates}=\sum_{\bm \ell}1=1+\zeta_{S^{d-2}}(0).
\eea In odd dimensions, the regularization method leads to 0. In even dimensions, the results have been listed in table \ref{heatcoef}.
\begin{table}[h]
\renewcommand\arraystretch{1.3}
    \centering
    \begin{tabular}{|c|c|c|c|c|}  \hline\text{dimension $d$}&4&6&8&10\\\hline
$\#\,$ \text{of eigenstates}&$\frac{1}{3}$&$\frac{29}{90}$&$\frac{1139}{3780}$&$ \frac{32377}{113400}$\\\hline
    \end{tabular}
    \caption{\centering{Regularization of the number of eigenstates  for even dimensions}}
    \label{heatcoef}
\end{table}

We may also define a heat kernel\footnote{For a review on heat kernel method, we consult \cite{Vassilevich:2003xt}.} associated with the Laplacian 
\bea 
K_{S^{d-2}}(\sigma)=\text{tr}\ e^{\sigma\Delta_{S^{d-2}}}
\eea where $\text{tr}$ is the sum over all possible eigenstates, including the one with $\ell_m=0$. Therefore, 
\bea 
K_{S^{d-2}}(\sigma)=1+\sum_{\ell=1}^\infty g(\ell,d-2)e^{-\sigma \ell(\ell+d-3)}.
\eea The summation is convergent for any $\text{Re}(\sigma)>0$. It turns out that the heat kernel has a pole structure near $\sigma=0$
\bea 
K_{S^{d-2}}(\sigma)=\sum_{k=0}^\infty a_k(d) \sigma^{k-\frac{d-2}{2}}.
\eea There are a number of divergent terms in the expansion and we may extract the constant term by regularization. Interestingly, the constant term is always zero for odd dimensions. When $d$ is even, the asymptotic expansion of the heat kernel can be found in the  appendix of \cite{1987PhRvD..36.3037B}. The results match the ones from spectral zeta function regularization.
\section{Scalar field theory on Rindler horizon}\label{csfkilling}
\subsection{Rindler horizon} 
\begin{figure}
  \centering
  \begin{tikzpicture}
    \filldraw[fill=gray!20,draw,thick] (4,4) node[above right]{\footnotesize $\mathcal{H}^{++}$} -- (0,0) node[below]{\footnotesize $B$} -- (4,-4) node[below right]{\footnotesize $\mathcal{H}^{--}$};
    \filldraw[fill=gray!20,draw,thick] (-4,4) node[above left]{\footnotesize $\mathcal{H}^{-+}$} -- (0,0) node[below]{\footnotesize $B$} -- (-4,-4) node[below left]{\footnotesize $\mathcal{H}^{+-}$};
    \fill (0,0) circle (1pt);
    \node at (2,0) {\footnotesize right Rindler wedge};
    \node at (-2,0) {\footnotesize left Rindler wedge};
    \draw[<->] (0,4.2) node[above] {\footnotesize $t$} -- (0,3.6) -- (0.5,3.6) node[right] {\footnotesize $x^{d-1}$};
  \end{tikzpicture}
  \caption{Killing horizons and Rindler wedges. There are four Killing horizons associated with the Lorentz boost vector $\bm X$. The left and right Rindler wedges are separated by the null hypersurfaces $\mathcal{H}^{\pm}$. There is a bifurcation surface $B$ which is the entangling surface between left and right regions.}\label{LRRindler}
\end{figure}
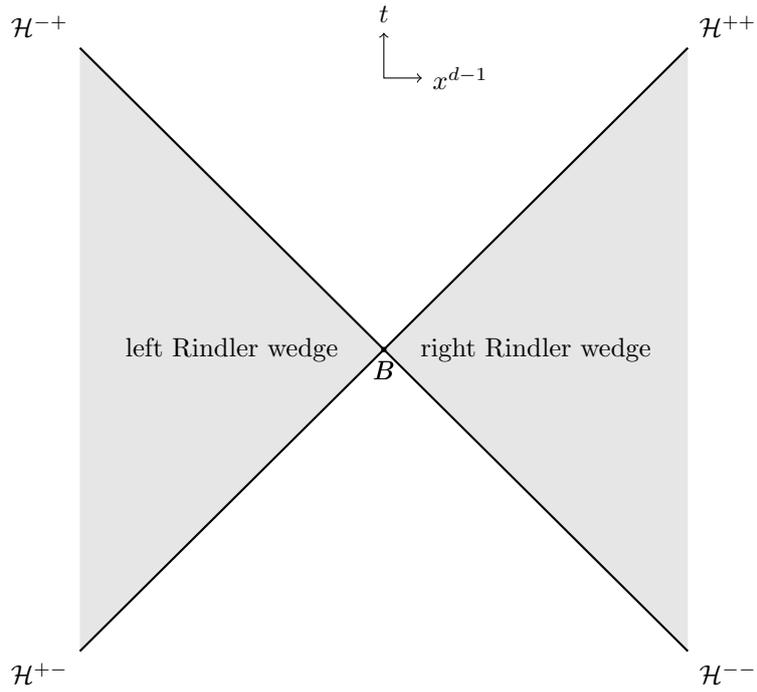

In Minkowski spacetime $\mathbb{R}^{1,d-1}$, the Killing horizon associated with the generator of Lorentz boost along the direction of $x^{d-1}$  is the set of points where 
\bea 
\bm X^2=0\quad\text{and}\quad \bm X\not=0\quad\text{with}\quad \bm X=x_{d-1}\partial_t+t\partial_{d-1}.
\eea 
The solution of null Killing vector condition $\bm X^2=0$ is the union of the following two hypersurfaces 
\bs\begin{align}
    \mathcal{H}^-(\bm X)&=\{t=-x^{d-1}\},\\
    \mathcal{H}^+(\bm X)&=\{t=x^{d-1}\},
\end{align}\es whose intersection is the bifurcation surface $B(\bm X)$ with $\bm X=0$ and 
\bea 
B(\bm X)=\{t=x^{d-1}=0\}.
\eea
This leads to the four Killing horizons associated with $\bm X$ 
\bs\begin{align}
\mathcal{H}^{--}(\bm X)&=\{t=-x^{d-1}, t<0\},\\
\mathcal{H}^{-+}(\bm X)&=\{t=-x^{d-1},t>0\},\\
\mathcal{H}^{+-}(\bm X)&=\{t=x^{d-1},t<0\},\\
\mathcal{H}^{++}(\bm X)&=\{t=x^{d-1},t>0\}.
\end{align}\es 
The above description is shown in figure \ref{LRRindler}.

Without loss of generality, we will study the Carrollian field theory on the  Killing horizon $\mathcal{H}^{--}(\bm X)$. This is one of the null boundaries of the Rindler spacetime which is also called Rindler horizon.  We will write it as $\mathcal{H}^{--}$ and omit $\bm X$ in the following. We note that it is also possible to define Carrollian field theories on the null hypersurface $\mathcal{H}^-=\mathcal{H}^{--}\cup \mathcal{H}^{-+}\cup B$. The interested reader may find more details in Appendix \ref{modeexpRindler}.

Rindler spacetime is a patch of Minkowski spacetime which may be obtained  from the coordinate transformation 
\be 
t=\rho \sinh\tau,\quad x^{d-1}=\rho\cosh\tau,\quad x_{\perp}^A=x^{A},\quad A=1,2,\cdots,d-2.\label{transrindler}
\ee 
The Rindler time has an imaginary period $2\pi$
\be 
\tau\sim \tau+2\pi i,
\ee whose inverse is the temperature of the thermal bath  in Unruh effect \cite{1976PhRvD..14..870U}.
The metric of Rindler spacetime is 
\be 
ds_{\text{Rindler}}^2=-\rho^2d\tau^2+d\rho^2+\delta_{AB}dx_{\perp}^Adx_{\perp}^B,\quad 0<\rho<\infty,\quad -\infty<\tau<\infty,\quad -\infty<x_{\perp}^A<\infty.
\ee We may define a new coordinate 
\be 
\lambda=\log\rho,\quad -\infty<\lambda<\infty,
\ee and the metric becomes 
\be 
ds_{\text{Rindler}}^2=e^{2\lambda}(-d\tau^2+d\lambda^2)+\delta_{AB}dx_{\perp}^Adx_{\perp}^B.
\ee 
With the retarded time in Rindler spacetime
\be 
u=\tau-\lambda,\quad -\infty<u<\infty, \label{retardedtimerindler}
\ee the metric of the Rindler spacetime may be written in retarded coordinates $(u,\lambda,x_{\perp}^A)$ as 
\be 
ds_{\text{Rindler}}^2=e^{2\lambda}(-du^2-2dud\lambda)+\delta_{AB}dx_{\perp}^Adx_{\perp}^B=-\rho^2du^2-2\rho du d\rho+\delta_{AB}dx_{\perp}^Adx_{\perp}^B.
\ee 
To obtain the metric of the Carrollian manifold $\mathcal{H}^{--}$ from  bulk metric, one may choose a cutoff 
\bea 
\rho=\epsilon
\eea such that the induced metric on the hypersurface 
\bea 
\mathfrak{H}_\epsilon=\{p\in \text{Rindler spacetime}|\ p=(u,\rho,x_{\perp}^1,\cdots,x_{\perp}^{d-2}) \ \text{with}\ \rho=\epsilon\}
\eea is 
\be 
ds^2=-\epsilon^2du^2+\delta_{AB}dx_{\perp}^Adx_{\perp}^B.\label{metricHep}
\ee 
The Killing horizon $\mathcal{H}^{--}$ can be parameterized by $d-1$ coordinates $(u,x_{\perp}^A)$ whose metric may be obtained by taking the limit $\epsilon\to 0$ for \eqref{metricHep}
\be 
ds^2_{\mathcal{H}^{--}}=\delta_{AB}dx_{\perp}^Adx_{\perp}^B.\label{dmRindler}
\ee Therefore, the Killing horizon $\mathcal{H}^{--}$ is a Carrollian manifold with a degenerate metric \eqref{dmRindler} as well as a null vector $\bm\chi=\partial_u$. The integration measure on $\mathcal{H}^{--}$ is 
\bea 
\int du d\bm x_{\perp}
\eea while in this case, we have 
\be 
\int d\bm x_{\perp}=\prod_{j=1}^{d-2}\int_{-\infty}^\infty dx_{\perp}^j.
\ee 

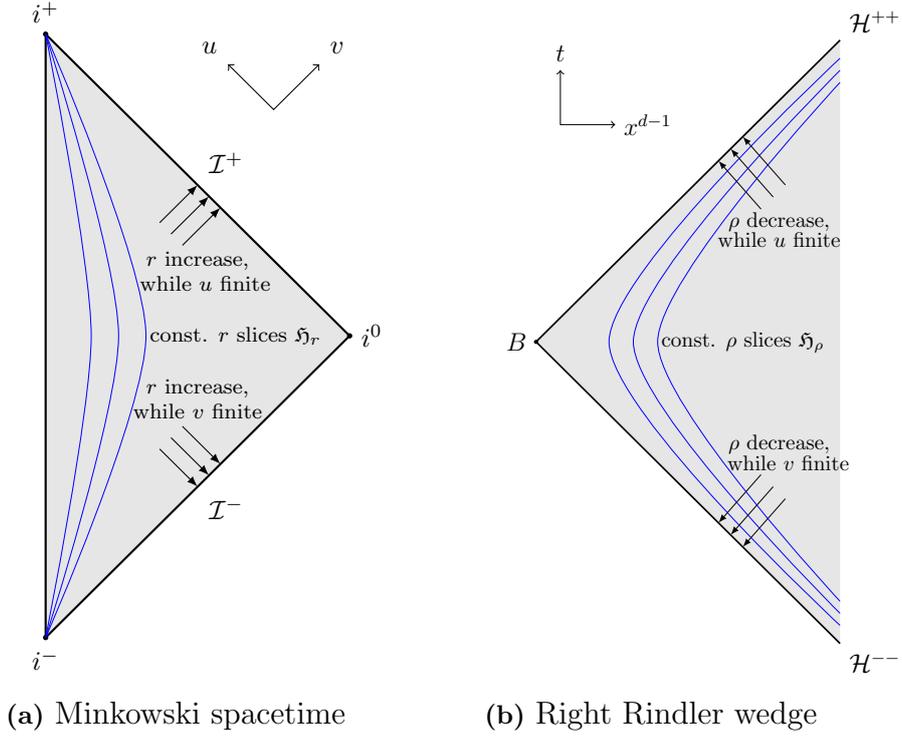
\begin{figure}
  \centering
  \subfloat[Minkowski spacetime]{
  \begin{tikzpicture}
    \filldraw[fill=gray!20,draw,thick] (0,4) node[above]{\footnotesize $i^+$} -- (2,2) node[above right]{\footnotesize $\mathcal{I}^+$} -- (4,0) node[right]{\footnotesize $i^0$}  -- (2,-2) node[below right]{\footnotesize $\mathcal{I}^-$} -- (0,-4) node[below]{\footnotesize $i^-$} -- cycle;
    \fill (4,0) circle (1pt);
    \fill (0,4) circle (1pt);
    \fill (0,-4) circle (1pt);
    \def\step{0.36};
    \foreach \i in {0,...,2}
    {\draw[very thin,blue] plot[smooth,tension=0.44] coordinates{(0,4) (0.6+\i*\step,0) (0,-4)};}
    \node at (2.5,0) {\scriptsize const. $r$ slices $\mathfrak{H}_r$};
    \draw[-latex] (1.8,1.2) -- (2.3,1.7);
    \draw[-latex] (1.65,1.35) -- (2.15,1.85);
    \draw[-latex] (1.5,1.5) -- (2,2);
    \draw[-latex] (1.8,-1.2) -- (2.3,-1.7);
    \draw[-latex] (1.65,-1.35) -- (2.15,-1.85);
    \draw[-latex] (1.5,-1.5) -- (2,-2);
    \node at (2,1) {\scriptsize $r$ increase,};
    \node at (2.1,0.7) {\scriptsize while $u$ finite};
    \node at (2,-1) {\scriptsize while $v$ finite};
    \node at (2,-0.7) {\scriptsize $r$ increase,};
    \draw[<->] (2.4,3.6) node[above left] {\footnotesize $u$} -- (3,3) -- (3.6,3.6) node[above right] {\footnotesize $v$};
  \end{tikzpicture}
  }\hspace{1cm}
  \subfloat[Right Rindler wedge]{\scalebox{0.8}{
    \begin{tikzpicture}
      \filldraw[fill=gray!20,draw,thick] (5,5) node[above right]{$\mathcal{H}^{++}$} -- (0,0) node[left]{$B$} -- (5,-5) node[below right]{$\mathcal{H}^{--}$};
      \fill (0,0) circle (1pt);
      \draw[very thin,blue] plot[smooth,tension=0.44] coordinates{(5,4.7) (1.2,0) (5,-4.7)};
      \draw[very thin,blue] plot[smooth,tension=0.44] coordinates{(5,4.5) (1.6,0) (5,-4.5)};
      \draw[very thin,blue] plot[smooth,tension=0.44] coordinates{(5,4.3) (2,0) (5,-4.3)};
      \node at (3.4,0){\footnotesize const. $\rho$ slices $\mathfrak{H}_\rho$};
      \draw[-latex] (3.7,2.2) -- (3,3);
      \draw[-latex] (3.9,2.4) -- (3.2,3.2);
      \draw[-latex] (4.1,2.6) -- (3.4,3.4);
      \draw[-latex] (3.7,-2.2) -- (3,-3);
      \draw[-latex] (3.9,-2.4) -- (3.2,-3.2);
      \draw[-latex] (4.1,-2.6) -- (3.4,-3.4);
      \node at (4,2){\footnotesize $\rho$ decrease,};
      \node at (4,1.7){\footnotesize while $u$ finite};
      \node at (4,-1.7){\footnotesize $\rho$ decrease,};
      \node at (4.15,-2){\footnotesize while $v$ finite};
      \draw[<->] (0.4,4.5) node[above]{$t$} -- (0.4,3.6) -- (1.3,3.6) node[right]{$x^{d-1}$};
    \end{tikzpicture}}
  }
  \caption{The figure (a) on the left-hand side is the Penrose diagram of Minkowski spacetime $\mathbb{R}^{1,d-1}$. We may reduce bulk quantities to future/past null infinity $\mathcal{I}^+/\mathcal{I}^-$ by choosing a series of constant $r$ slices $\mathfrak{H}_r$ and then taking corresponding limits. Figure (b) is the right Rindler wedge, which is a part of Minkowski spacetime. The null boundary of Rindler spacetime is denoted as $\mathcal{H}^{--}$ and $\mathcal{H}^{++}$. We may define Carrollian field theories  by reducing the bulk theory to the null hypersurfaces. In the figure, we have plotted constant $\rho$ slices which are hyperbolic curves in the spacetime diagram. }\label{Rindlerlimit}
\end{figure}

Similarly, we may define an advanced coordinate 
\be 
v=\tau+\lambda,
\ee and the metric of the Rindler spacetime in advanced coordinates $(v,\lambda,\theta^A)$ is 
\be 
ds^2_{\text{Rindler}}=e^{2\lambda}(-dv^2+2dv d\lambda)+\delta_{AB}dx_{\perp}^Adx_{\perp}^B=-\rho^2 dv^2+2\rho dvd\rho+\delta_{AB}dx_{\perp}^Adx_{\perp}^B.
\ee By taking the limit $\lambda\to-\infty$ while keeping the advanced time $v$ finite, we find the metric of the Carrollian manifold $\mathcal{H}^{++}$
\be 
ds^2_{\mathcal{H}^{++}}=\delta_{AB}dx_{\perp}^Adx_{\perp}^B.
\ee The null vector of the Carrollian manifold $\mathcal{H}^{++}$ is 
$\partial_v$.  We will define the limits $\lim{}\hspace{-0.8mm}_-$ and $\lim{}\hspace{-0.8mm}_+$ to reduce the quantities from bulk hypersurface $\mathfrak{H}_\epsilon$ to $\mathcal{H}^{--}$ and $\mathcal{H}^{++}$, respectively
\bs\begin{align}
\lim{}\hspace{-0.8mm}_-&=\lim_{\lambda\to-\infty,\ u\  \text{finite}}=\lim_{\rho\to0, \ u \ \text{finite}},\\
\lim{}\hspace{-0.8mm}_+&=\lim_{\lambda\to-\infty,\ v\  \text{finite}}=\lim_{\rho\to0,  \ v \ \text{finite}}.
\end{align}\es 
One can see these two limits in figure \ref{Rindlerlimit}.
An attentive reader may be confused about why we should reduce the theory from  Rindler spacetime to $\mathcal{H}^{--}$. It seems equivalent to reducing the theory straightforwardly from the theory in Minkowski spacetime. Indeed, one could do so, as has been shown in Appendix \ref{nullhm},  and what one actually gets is the theory  defined on $\mathcal{H}^-$. One should integrate out degrees of freedom on $\mathcal{H}^{-+}\cup B$ to find the final results.

\subsection{From bulk  to boundary}
In this section, we will construct  Carrollian scalar field theory at the Killing horizon $\mathcal{H}^{--}$ from bulk reduction in general dimensions. We still consider a general scalar theory with the action 
\eqref{actionPhi}.
The potential could include a mass term since massive particles can also arrive at $\mathcal{H}^{--}$. 
\be 
V(\Phi)=\sum_{k=2}^\infty \frac{\lambda_k}{k}\Phi^k,\quad \lambda_2=m^2.
\ee By embedding the theory in Minkowski spacetime, there is no need to impose any artificial boundary condition at $\mathcal{H}^{--}$ since they should be determined by the whole system. Instead, a scalar field should be finite at any position of Minkowski spacetime. Therefore, we just acquire the finiteness  of $\Phi$ at $\mathcal{H}^{--}$. Using Taylor expansion, we assume the series expansion of $\Phi$ near $\mathcal{H}^{--}$
\be 
\Phi(t,\bm x)=\Sigma(u,\bm x_{\perp})+\sum_{k=1}^\infty \Sigma^{(k)}(u,\bm x_{\perp})\rho^k\label{falloffscalarrindler}
\ee for the bulk scalar field. The pre-symplectic form of the theory is 
\bea 
\bm\Theta(\delta\Phi;\Phi)=-\int_{\mathfrak{H}_\rho} (d^{d-1}x)_\mu \partial^\mu\Phi \delta\Phi
\eea  where 
\bea 
(d^{d-1}x)_\mu=\frac{1}{(d-1)!}\epsilon_{\mu\nu_1\cdots\nu_{d-1}}dx^{\nu_1}\wedge \cdots\wedge dx^{\nu_{d-1}}=-\rho \hat{m}_\mu du d\bm x_{\perp}.
\eea The normal covector $\hat{m}_\mu$ of the hypersurface $\mathfrak{H}_\rho$ is $d\rho$ which implies
\be 
\hat{m}_\rho=1,\quad \hat m_u=\hat m_A=0,
\ee and we have normalized it to 1
\be 
\hat m\cdot \hat m=1.
\ee 
The symplectic form of the theory is 
\be 
\bm\Omega(\delta\Phi;\delta\Phi;\Phi)=-\int_{\mathfrak{H}_\rho}(d^{d-1}x)_\mu \partial^\mu\delta \Phi \wedge\delta\Phi.
\ee With the fall-off condition \eqref{falloffscalarrindler}, the symplectic form becomes\footnote{Similar to the constraint equations \eqref{constraintseqn} at null infinity, we may also use the differential operator 
\bea 
\partial_\mu \partial^\mu=\partial_\rho^2+\rho^{-1}\partial_\rho-2\rho^{-1}\partial_u\partial_\rho+\partial_A\partial^A, 
\eea and reduce the bulk EOM to 
\bs \begin{align}
&\Sigma^{(1)}-2\dot\Sigma^{(1)}=0,\\
&(k+2)^2\Sigma^{(k+2)}-2(k+2)\dot\Sigma^{(k+2)}+\nabla_A\nabla^A\Sigma^{(k)}-\sum_{n=2}^\infty\sum_{k_1,\cdots,k_{n-1}\ge 0}^{\sum_{j=1}^{n-1}k_j=k}\lambda_{n}\Sigma^{(k_1)}\cdots\Sigma^{(k_{n-1})}=0,\quad k\ge 0.
\end{align}\es Therefore, only the leading mode  $\Sigma$ is free from dynamical equations. All the subleading modes $\Sigma^{(k)},\ k\ge 1$ are determined by the equations up to initial data.}
\bea 
\bm\Omega(\delta\Sigma;\delta\Sigma;\Sigma)=\int du d\bm x_{\perp} \delta\Sigma\wedge\delta\dot\Sigma
\eea at $\mathcal{H}^{--}$. This is identified as the symplectic form of the Carrollian scalar field theory at $\mathcal{H}^{--}$. Interestingly, this is formally the same as the one at $\mathcal{I}^+$ though the Carrollian manifold and the boundary field $\Sigma$ are not identical.  We may find the following commutators for the field $\Sigma$ 
\bs\label{comusigmakilling}\begin{align}
    [\Sigma(u,\bm x_{\perp}),\Sigma(u',\bm x_{\perp}')]&=\frac{i}{2}\alpha(u-u')\delta^{(d-2)}(\bm x_{\perp}-\bm x_{\perp}'),\\
    [\Sigma(u,\bm x_{\perp}),\dot{\Sigma}(u',\bm x_{\perp}')]&=\frac{i}{2}\delta(u-u')\delta^{(d-2)}(\bm x_{\perp}-\bm x_{\perp}'),\\
    [\dot{\Sigma}(u,\bm x_{\perp}),\dot{\Sigma}(u',\bm x_{\perp}')]&=\frac{i}{2}\delta'(u-u')\delta^{(d-2)}(\bm x_{\perp}-\bm x_{\perp}').
\end{align}\es
The field $\Sigma$ may be expanded in a set of complete bases of $\mathcal{H}^{--}$ 
\bea 
\Sigma(u,\bm x_{\perp})=\int_0^\infty \frac{d\omega}{\sqrt{4\pi\omega}}\int \frac{d\bm k_{\perp}}{\sqrt{(2\pi)^{d-2}}}[a^{\rm R}_{\omega,\bm k_{\perp}}e^{-i\omega u+i\bm k_{\perp}\cdot\bm x_{\perp}}+a^{\rm R\dagger}_{\omega,\bm k_{\perp}}e^{i\omega u-i\bm k_{\perp}\cdot\bm x_{\perp}}]\label{expaHm}
\eea with the commutators 
\bs\begin{align}
    & [a^{\rm R}_{\omega,\bm k_{\perp}},a^{\rm R}_{\omega',\bm k_{\perp}'}]=[a_{\omega,\bm k_{\perp}}^{\rm R\dagger},a_{\omega',\bm k_{\perp}'}^{\rm R\dagger}]=0,\\
& [a^{\rm R}_{\omega,\bm k_{\perp}},a_{\omega',\bm k_{\perp}'}^{\rm R\dagger}]=\delta(\omega-\omega')\delta^{(d-2)}(\bm k_{\perp}-\bm k_{\perp}').
\end{align}\es
In the expansion \eqref{expaHm}, the functions $e^{-i\omega u+i\bm k\cdot\bm x_{\perp}}$  and their complex conjugates form a set of complete basis of $\mathcal{H}^{--}$. We have also  checked the mode expansion \eqref{expaHm} using canonical quantization method in Appendix \ref{modeexpRindler}. With the canonical quantization, we may define the vacuum state $|0\rangle_{\text{R}}$ through 
\be 
a^{\rm R}_{\omega,\bm\ell}|0\rangle_{\text{R}}=0.
\ee We have added a subscript $\text{R}$ to denote the vacuum state of Rindler spacetime. It should be distinguished from the vacuum state of Minkowski spacetime $|0\rangle$ defined in the previous section.
Therefore, we find the following correlators 
\bs\begin{align} \label{comKilling}{}_{\text{R}}\langle 0|\Sigma(u,\bm x_{\perp})\Sigma(u',\bm x_{\perp}')|0\rangle_{\text{R}}&=\beta(u-u')\delta^{(d-2)}(\bm x_{\perp}-\bm x_{\perp}'), \\
{}_{\text{R}}\langle 0|\Sigma(u,\bm x_{\perp})\dot\Sigma(u',\bm x_{\perp}')|0\rangle_{\text{R}}&=\frac{1}{4\pi(u-u'-i\epsilon)}\delta^{(d-2)}(\bm x_{\perp}-\bm x_{\perp}'),\\
{}_{\text{R}}\langle 0|\dot\Sigma(u,\bm x_{\perp})\Sigma(u',\bm x_{\perp}')|0\rangle_{\text{R}}&=-\frac{1}{4\pi(u-u'-i\epsilon)}\delta^{(d-2)}(\bm x_{\perp}-\bm x_{\perp}'),\\
{}_{\text{R}}\langle 0|\dot\Sigma(u,\bm x_{\perp})\dot\Sigma(u',\bm x_{\perp}')|0\rangle_{\text{R}}&=-\frac{1}{4\pi(u-u'-i\epsilon)^2}\delta^{(d-2)}(\bm x_{\perp}-\bm x_{\perp}').
\end{align}\es  
\subsection{Flux operators for Carrollian diffeomorphism}
In this section, we will apply the formula \eqref{Carrollxi} to the Killing horizon and calculate the flux operators for Carrollian diffeomorphism. To understand the physical meaning of the flux operators, we may study the flux operators associated with Killing vectors at first. 
Rindler spacetime is locally flat which has $d(d+1)/2$ Killing vectors  inherited from  Minkowski spacetime. Therefore, we may use these generators to discuss the symmetries of Rindler spacetime. The Killing vectors are listed below which are expressed in terms of the quantities in retarded coordinates $(u,\rho,\bm x_{\perp})$
\bs\begin{align}
    &\partial_t=e^u\partial_u-\frac{1}{2}(\rho e^u-\rho^{-1}e^{-u})\partial_\rho,\\
    &\partial_{d-1}=-e^u\partial_u+\cosh\tau\partial_\rho=-e^u \partial_u+\frac{1}{2}(\rho e^u+\rho^{-1}e^{-u})\partial_\rho,\\
    &\partial_{A}=\frac{\partial}{\partial x_{\perp}^A},\\
    &t\partial_{d-1}+x_{d-1}\partial_t=\partial_u,\\
    &t\partial_A+x_A\partial_t=e^u (x_{\perp})_A\partial_u-\frac{1}{2}(x_{\perp})_A (e^u \rho-e^{-u}\rho^{-1})\partial_\rho+\frac{1}{2}(e^u\rho^2-e^{-u})\partial_A,\\
    &x_{d-1}\partial_A-x_A\partial_{d-1}=-e^u (x_{\perp})_A\partial_u+\frac{1}{2}(x_{\perp})_A(e^u \rho+e^{-u}\rho^{-1})\partial_\rho+\frac{1}{2}(e^u\rho^2+e^{-u})\partial_A,\\
   & x_A\partial_B-x_B\partial_A=(x_{\perp})_A\partial_B-(x_{\perp})_B\partial_A.
\end{align}\es
In the light cone coordinates 
\bea 
x^\pm=t\pm x^{d-1}\quad\Leftrightarrow\quad x_{\pm}=-\frac{1}{2}x^{\mp}, 
\eea  we find the following generators
\bs\label{geneRind}\begin{align} 
\bm\xi_-&=\partial_-= \frac{1}{2}(\partial_t-\partial_{d-1})=e^u\partial_u-\frac{1}{2}\rho e^u\partial_\rho\sim e^u \partial_u,\\ 
\bm\xi_+&=\partial_+=\frac{1}{2}( \partial_t+\partial_{d-1})=\frac{e^{-u}}{2\rho}\partial_\rho,\\
\bm\xi_A&=\partial_A,\\
\bm\xi_{+-}&=2(x_+\partial_--x_-\partial_+)=\partial_u,\\
\bm\xi_{+A}&=(x_+\partial_A-x_A\partial_+)=\frac{1}{2}e^{-u}\partial_A-x_A\frac{e^{-u}}{2\rho}\partial_\rho,\\
\bm\xi_{-A}&=(x_-\partial_A-x_A\partial_-)=-x_A e^u\partial_u+\frac{1}{2}\rho x_A e^u \partial_\rho-\frac{1}{2}\rho^2 e^{u}\partial_A\sim -x_A e^u\partial_u,\\
\bm\xi_{AB}&=x_A\partial_B-x_B\partial_A.
\end{align}\es 
The vectors $\bm\xi_+$ and $\bm\xi_{+A}$ blow up near $\mathcal{H}^{--}$ and they do not generate Carrollian diffeomorphism. To be more precise, we note the
field $\Phi$ transforms along the direction of the Killing vectors as 
\be 
\delta_{\bm\xi}\Phi=\xi^\mu\partial_\mu\Phi.
\ee Note that  
the variation of the field $\Phi$ along the direction of $\bm\xi_+$ or $\bm\xi_{+A}$ blows up and violates the fall-off condition \eqref{falloffscalarrindler}. Therefore, we should exclude the generators $\bm\xi_+$ and $\bm\xi_{+A}$. On the other hand, the variations of the field $\Phi$ along the directions $\bm\xi_-,\bm\xi_A,\bm\xi_{+-},\bm\xi_{-A},\bm\xi_{AB}$ preserve the fall-off conditions and we find the following variations for the boundary field $\Sigma$
\bs\label{vasigma}\begin{align}
    \delta_{\bm\xi_-}\Sigma&=e^u \dot\Sigma,\\
    \delta_{b^A\bm\xi_A}\Sigma&=b^A\partial_A\Sigma,\\
    \delta_{\bm\xi_{+-}}\Sigma&=\dot\Sigma,\\
    \delta_{c^A\bm\xi_{-A}}\Sigma&=-c^Ax_A e^u \dot\Sigma,\\
    \delta_{\omega^{AB}\bm\xi_{AB}}\Sigma&=\omega^{AB}(x_A\partial_B-x_B\partial_A)\Sigma.
\end{align}\es We have inserted constant vectors $b^A, c^A$
or antisymmetric tensor $\omega^{AB}$ to balance the indices. From the explicit form of the generators, we conclude that $\bm\xi_-,\bm\xi_A,\bm\xi_{+-},\bm\xi_{-A},\bm\xi_{AB}$ could reduce to special choices of Carrollian diffeomorphism at the boundary $\mathcal{H}^{--}$ 
\bs\label{corresfY}\begin{align}
    \bm\xi_-&=\bm\xi_{f,Y}\quad\text{with}\quad  f=e^u,\ Y^A=0,\label{xi-}\\
    b^A\bm\xi_A&=\bm\xi_{f,Y}\quad\text{with}\quad f=0,\ Y^A=b^A,\\
    \bm\xi_{+-}&=\bm\xi_{f,Y}\quad \text{with}\quad f=1,\ Y^A=0,\\
    c^A\bm\xi_{-A}&=\bm\xi_{f,Y}\quad \text{with}\quad f=-c^Ax_A e^u,\ Y^A=0,\label{xi-A}\\
    \omega^{AB}\bm\xi_{AB}&=\bm\xi_{f,Y}\quad \text{with}\quad f=0,\ Y^A=2\omega^{CA}x_C.
\end{align}\es All the transformations \eqref{vasigma} could be unified  in the form
\bea 
\delta_{\bm\xi_{f,Y}}\Sigma=f(u,\bm x_{\perp})\dot{\Sigma}+Y^A(\bm x_{\perp})\partial_A\Sigma+w \nabla_AY^A\ \Sigma \label{carrollscalar}
\eea with the special choices  of the $f$ and $Y^A$
listed in \eqref{corresfY}. Note that the last term is always zero due to the vanishing of the term $\nabla_AY^A$ for all the choices of $Y^A$ in \eqref{corresfY}. Therefore, The constant $w$ is free at this moment. 
Interestingly, the function $f(u,\bm x_{\perp})$ is automatically time-dependent from \eqref{xi-} and \eqref{xi-A}. This fact strongly indicates that one should consider time-dependent supertranslations. Another noticeable fact is that the translation along Rindler time is a Lorentz boost in Minkowski spacetime. Therefore, the corresponding flux should be interpreted as center of mass flux in Minkowski spacetime. Similarly, a superrotation flux with $Y^A=b^A$ should be interpreted as a momentum flux in Minkowski spacetime. 
Nevertheless, we can construct the leaky fluxes across $\mathcal{H}^{--}$ for the Killing vectors \eqref{corresfY} by 
\bea 
\mathcal{F}_{\bm\xi}\equiv\mathcal{F}[j_{\bm\xi}]=-\lim{}_- \int_{\mathfrak{H}_\rho} (d^{d-1}x)_\mu j^\mu_{\bm\xi} =-\lim{}_- \int_{\mathfrak{H}_\rho} du d\bm x_{\perp}\ \rho  T^\rho_{\ \mu}\xi^\mu, \label{fluxformularindler}
\eea where 
\bea 
j^\mu_{\bm\xi}=T^{\mu\nu}\xi_\nu
\eea  and $T^{\mu\nu}$ is the stress tensor given by \eqref{stress} and $\bm\xi$ is a Killing vector of Rindler spacetime. 
Given the fall-off conditions \eqref{falloffscalarrindler}, we find 
\bs\begin{align}
    \partial_u\Phi&=\sum_{k=0}^\infty \dot{\Sigma}^{(k)} \rho^k,\\
    \partial_\rho\Phi&=\sum_{k=0}^\infty (k+1) \Sigma^{(k+1)}\rho^{k},\\
    \partial_A\Phi&=\sum_{k=0}^\infty \partial_A\Sigma^{(k)}\rho^k,\\
    \partial^u\Phi&=-\rho^{-1}\partial_\rho\Phi=-\sum_{k=0}^\infty (k+1) \Sigma^{(k+1)}\rho^{k-1},\\
    \partial^\rho\Phi&=-\rho^{-1}\partial_u\Phi+\partial_\rho\Phi=-\sum_{k=0}^\infty \dot{\Sigma}^{(k)}\rho^{k-1}+\sum_{k=0}^\infty (k+1)\Sigma^{(k+1)}\rho^{k},\\
    \partial^A\Phi&=\sum_{k=0}^\infty \partial^A\Sigma^{(k)}\rho^k
\end{align}\es 
and 
\bs\begin{align}
\partial^\rho\Phi \partial_u\Phi&=-\rho^{-1}\dot\Sigma^2+\cdots,\\
\partial^\rho\Phi\partial_\rho\Phi&=-\rho^{-1}\dot\Sigma\Sigma^{(1)}+\cdots,\\
\partial^\rho\Phi\partial_A\Phi&=-\rho^{-1}\dot\Sigma\nabla_A\Sigma+\cdots,\\
\partial_\mu\Phi \partial^\mu\Phi&=-2\rho^{-1}\Sigma^{(1)}\dot\Sigma+\cdots.
\end{align}\es To compute \eqref{fluxformularindler}, we need the 
 following fall-off behaviour of stress tensor 
\bs\begin{align}
    T^\rho_{\ u}&=-\rho^{-1}\dot\Sigma^2+\mathcal{O}(\rho^0),\\
    T^\rho_{\ \rho}&=t^\rho_\rho+\mathcal{O}(\rho),\\
    T^\rho_{\ A}&=-\rho^{-1}\dot\Sigma\nabla_A\Sigma+\mathcal{O}(\rho^0).
\end{align}\es where 
\bea 
t^\rho_{\rho}=\frac{1}{2}(\Sigma^{(1)})^2-\frac{1}{2}\partial_A\Sigma\partial^A\Sigma-\sum_{k=2}^\infty\frac{ \lambda_k}{k}\Sigma^k.
\eea 
Therefore, we find the fluxes corresponding to the Killing vectors \eqref{geneRind}
\bs\begin{align}
    \mathcal{F}_{\bm\xi_-}&=\int du d\bm x_{\perp} e^u \dot\Sigma^2,\\
    \mathcal{F}_{\bm \xi_+}&=-\frac{1}{2}\int du d\bm x_{\perp}e^{-u}t^\rho_{\rho},\\
    \mathcal{F}_{b^A\bm\xi_A}&=\int du d\bm x_{\perp}b^A\dot\Sigma\nabla^A\Sigma,\\
    \mathcal{F}_{\bm\xi_{+-}}&=\int du d\bm x_{\perp} \dot\Sigma^2,\\
    \mathcal{F}_{\tilde{c}^A\bm\xi_{+A}}&=\frac{1}{2}\int du d\bm x_{\perp}e^{-u}\tilde{c}^A[\dot\Sigma\nabla_A\Sigma+x_A t^\rho_\rho],\\
    \mathcal{F}_{c^A\bm\xi_{-A}}&=\int du d\bm x_{\perp} c^Ax_A e^u \dot\Sigma^2,\\
    \mathcal{F}_{\omega^{AB}\bm\xi_{AB}}&=\int du d\bm x_{\perp}2\omega^{CA}x_C\dot\Sigma\nabla_A\Sigma.
\end{align}\es 

The fluxes $\mathcal{F}_{\bm \xi_-},\mathcal{F}_{\bm\xi_{+-}},\mathcal{F}_{c^A\bm\xi_{-A}}$ are exactly the previous $\mathcal{T}_f$ with $f=e^u,\ 1$ and $c^Ax_A e^u$, respectively. Similarly, the fluxes $\mathcal{F}_{b^A\bm\xi_A},\mathcal{F}_{\omega^{AB}\bm\xi_{AB}}$ are exactly the  previous $\mathcal{M}_Y$ with $Y^A=b^A$ and $2\omega^{CA}x_C$ respectively. On the other hand, the fluxes $\mathcal{F}_{\bm\xi_+}$ and $\mathcal{F}_{\tilde{c}^A\bm\xi_{+A}}$ are finite but not related to Carrollian diffeomorphism. 

Now we can construct the flux operators associated with Carrollian diffeomorphism. For any GST, the flux  is 
\bea 
\mathcal{F}_{\bm\xi_f}=\int du d\bm x_{\perp} f(u,\bm x_{\perp})\dot\Sigma^2(u,\bm x_{\perp}).
\eea We may uplift  it to a quantum operator  using normal ordering 
\be 
\mathcal{T}_f=\int du  d\bm x_{\perp} f(u,\bm x_{\perp}):\dot\Sigma^2(u,\bm x_{\perp}):.\label{fluxst}
\ee Similarly, we can find the flux associated with any SSR
\be 
\mathcal{F}_{\bm\xi_Y}=\int du d\bm x_{\perp} Y^A(\bm x_{\perp})\dot\Sigma \nabla_A\Sigma.
\ee The corresponding flux operator is 
\bea 
\mathcal{M}_Y=\int du d\bm x_{\perp} Y^A(\bm x_{\perp})M_A(u,\bm x_{\perp}),\label{fluxsr}
\eea where we have chosen 
\be 
M_A(u,\bm x_{\perp})=\frac{1}{2}:(\dot\Sigma\nabla_A\Sigma-\Sigma\nabla_A\dot\Sigma):.
\ee 

We should mention that  one could also use Hamilton equation \eqref{Hamcov} to obtain the same results for Carrollian diffeomorphism. 
For a GST generated by 
\be 
\bm\xi_f=f(u,\bm x_{\perp})\partial_u\quad\text{with}\quad \delta_f\Sigma(u,\bm x_{\perp})=f(u,\bm x_{\perp})\dot{\Sigma}(u,\bm x_{\perp}),
\ee the corresponding Hamiltonian is exactly the flux operator \eqref{fluxst} after normal ordering. For an SSR generated by \be 
\bm\xi_Y=Y^A(\bm x_{\perp})\partial_A\quad\text{with}\quad \delta_Y\Sigma(u,\bm x_{\perp})=Y^A(\bm x_{\perp})\nabla_A\Sigma(u,\bm x_{\perp})+w \nabla_AY^A(\bm x_{\perp})\ \Sigma(u,\bm x_{\perp}),\label{deltaYsigma}\ee 
we find an integrable flux only for $w=\frac{1}{2}$ which is again exactly the flux operator $\eqref{fluxsr}$.
Note that the constant $w$ is absent in the variation 
\be 
\bm\xi_Y(\Sigma)=Y^A\nabla_A\Sigma,
\ee which we  have already met with this problem at future null infinity. In that case, we can  fix the value $w=\frac{1}{2}$ by considering Lorentz  transformations in the bulk. However, one cannot fix it in the Killing horizon $\mathcal{H}^{--}$ as has been shown in \eqref{carrollscalar}. Nevertheless, the value $w$ should be $\frac{1}{2}$ due to the following reasons.
\begin{enumerate}
    \item The Hamiltonian for superrotation is integrable only for $w=\frac{1}{2}$.
    \item The Dirac delta function $\delta^{(d-2)}(\bm x_{\perp}-\bm x_{\perp}')$ is a scalar density of weight $1$. Therefore, the consistency for the commutator \eqref{comusigmakilling}  has already  fixed  $w=\frac{1}{2}$.
\end{enumerate}
It turns out that the commutators between $\mathcal{T}_f$ and $\mathcal{M}_Y$ are exactly the same as \eqref{comscalar}. The central charge \eqref{centralcharge} is also divergent. However, in this case, we find 
\bea
\delta^{(d-2)}(\bm 0)&=&\delta^{(d-2)}(\bm x_{\perp}-\bm x_{\perp}')|_{\bm x_{\perp}'=\bm x_{\perp}}=\frac{1}{(2\pi)^{d-2}}\int d\bm k_{\perp} e^{i\bm k_{\perp}(\bm x_{\perp}-\bm x_{\perp}')}|_{\bm x_{\perp}'=\bm x_{\perp}}=\frac{1}{(2\pi)^{d-2}}\int d\bm k_{\perp}.\nn 
\eea It shows that the central charge also counts the number of eigenstates in the transverse direction. To regularize it, we introduce a UV cutoff $2\pi/a$  where $a$ has the dimension of a microscopic length scale. In lattice theory, $a$ is  the  distance of adjacent lattice points. Therefore, the Dirac function becomes
\bea 
\delta^{(d-2)}(\bm 0)=\frac{1}{(2\pi)^{d-2}}\int_0^{2\pi/a} d\bm k_{\perp}=\frac{1}{a^{d-2}}=\frac{1}{V_a}.\label{regdelta}
\eea At the last step, we have used the fact that $a^{d-2}$ is the volume of the smallest cell of the lattice, 
\be 
V_a=a^{d-2}.
\ee Therefore, the central charge should be proportional to the density of states in the transverse direction, which has exactly the same physical meaning as the one at $\mathcal{I}^+$. 
We may also use the heat kernel method to define
\bea 
K(\sigma)=\text{tr}\ e^{\sigma \Delta}
\eea with  $\Delta$ being the Laplacian in the transverse direction.  The spectrum of the Laplacian operator is $-\bm k_{\perp}^2$ which is continuous. Then the heat kernel becomes 
\bea 
K(\sigma)=\frac{1}{(2\pi)^{d-2}}\int d^{d-2}\bm k_{\perp}e^{-\sigma \bm k_{\perp}^2}=\frac{1}{(4\pi\sigma)^{(d-2)/2}}.\label{Ksigma}
\eea 
As $\sigma\to 0$, the heat kernel approaches the Dirac delta function $\delta^{(d-2)}(\bm 0)$.  Since the dimension of $\sigma$ is length square, we may introduce a UV cutoff $a$ with dimension length and identify 
\be 
\sigma=\frac{a^2}{4\pi}.\label{identify}
\ee Combining \eqref{Ksigma} and \eqref{identify}, we find the same result as \eqref{regdelta}. 

Now we choose the bases for the supertranslation function
\be 
f_{\omega,\bm k_{\perp}}(u,\bm x_{\perp})=e^{-i\omega u}e^{i\bm k_{\perp}\cdot\bm x_{\perp}}, 
\ee 
and define the Fourier transformation of the flux operator 
\be 
\mathcal{T}_{\omega,\bm k_{\perp}}=\int du d\bm x_{\perp} f_{\omega,\bm k_{\perp}}(u,\bm x_{\perp}):\dot\Sigma^2:.
\ee 
Then we find the Virasoro algebra
\bea 
\hspace{-3pt} [\mathcal{T}_{\omega,\bm k_{\perp}},\mathcal{T}_{\omega',\bm k_{\perp}'}]=(\omega'-\omega)\mathcal{T}_{\omega+\omega',\bm k_{\perp}+\bm k_{\perp}'}-\frac{\omega^3}{24\pi}\times (2\pi)^{d-1}\delta^{(d-2)}(\bm 0)\delta(\omega+\omega')\delta^{(d-2)}(\bm k_{\perp}+\bm k_{\perp}').
\eea 
The Dirac delta function $\delta^{(d-2)}(\bm k_{\perp}+\bm k_{\perp}')$ is zero for $\bm k_{\perp}'\not=-\bm k_{\perp}$. The divergence for the central charge only appears for $\bm k_{\perp}'=-\bm k_{\perp}$. In this case, we find 
\bea 
(2\pi)^{d-2}\delta^{(d-2)}(\bm k_{\perp}+\bm k_{\perp}')|_{\bm k_{\perp}'=-\bm k_{\perp}}=\int d\bm x_{\perp}=L^{d-2}=\text{Area}(B)\label{regdeltak}
\eea where we have introduced an IR cutoff $L$ which is the large length scale in the transverse direction. At the last step, we have switched the $L^{d-2}$ to the area of the bifurcation surface $B$\footnote{This regularization is similar to restricting the particles in a  box \cite{weinberg_1995}.}. Combining with the regularization \eqref{regdelta}, we find the Virasoro algebra
\bea 
\ [\mathcal{T}_{\omega,\bm k_{\perp}},\mathcal{T}_{\omega',-\bm k_{\perp}}]=(\omega'-\omega)\mathcal{T}_{\omega+\omega',0}-\frac{\omega^3}{12}\times\frac{\text{Area}(B)}{a^{d-2}}\delta(\omega+\omega').
\eea The quantity 
\be \bar{c}=\frac{\text{Area}(B)}{a^{d-2}}
\ee is exactly the number of states in the transverse directions. We add a bar to distinguish from the central charge $c=\delta^{(d-2)}(\bm 0)$ which is the density of states in the transverse directions. We will call it an effective central charge. 

\subsection{Relation to Modular Hamiltonian}
We have defined flux operators associated with supertranslations and superrotations for the Rindler wedge. In Rindler wedge, the fundamental quantity is the modular Hamiltonian  \cite{Haag:1992} 
\be 
H_{\text{modular}}=2\pi\int_{V} d^{d-1}\bm x x^{d-1}T_{tt}(t=0,\bm x)\label{modular}
\ee 
where $T_{tt}(t=0,\bm x)$ is the $tt$ component of the stress tensor on the surface $V$ which is defined by 
\be V=\{p=(t,x^1,\cdots,x^{d-2})|t=0,\ x^{d-1}>0\}.
\ee The modular Hamiltonian \eqref{modular} is defined for general relativistic field theories and plays a central role in the context of geometric entanglement entropy. As a subregion of Minkowski spacetime,  $V$ defines the Rindler wedge whose 
null boundaries are exactly $\mathcal{H}^{--}$ and $\mathcal{H}^{++}$. There could be relations between the flux operators associated with Carrollian diffeomorphism and modular Hamiltonian. We will explore this problem in the following.

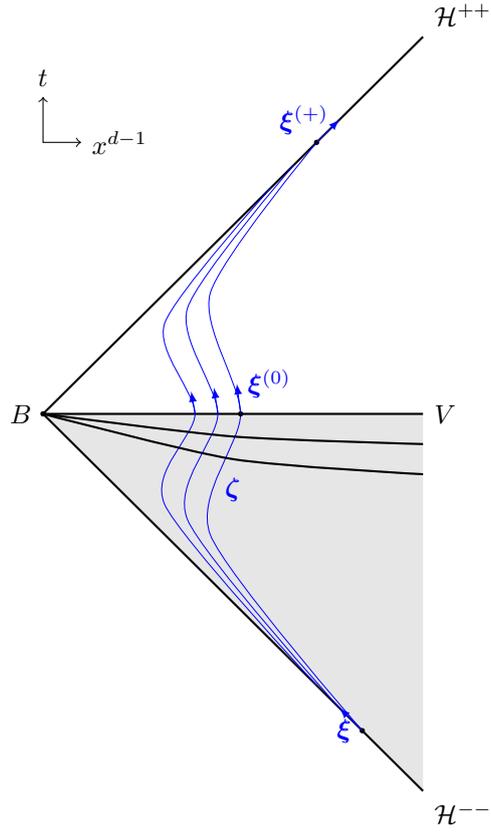
\begin{figure}
  \centering
  \begin{tikzpicture}
    \filldraw[fill=gray!20,draw,thick] (5,0) node[right]{\footnotesize $V$} -- (0,0) node[left]{\footnotesize $B$} -- (5,-5) node[below right]{\footnotesize $\mathcal{H}^{--}$};
    \draw[thick] (0,0) --  (5,5) node[above right]{\footnotesize $\mathcal{H}^{++}$};
    \draw[very thin,blue] plot[smooth,tension=0.44] coordinates{(3,3) (1.6,1.2) (2,0) (1.6,-1.2) (3.6,-3.6)};
    \draw[very thin,blue] plot[smooth,tension=0.44] coordinates{(3.3,3.3) (1.9,1.4) (2.3,0) (1.9,-1.4) (3.9,-3.9)};
    \draw[very thin,blue] plot[smooth,tension=0.44] coordinates{(3.6,3.6) (2.2,1.6) (2.6,0) (2.2,-1.6) (4.2,-4.2)};
    \draw[thick] plot[smooth,tension=0.44] coordinates{(0,0) (2.5,-0.6) (5,-0.8)};
    \draw[thick] plot[smooth,tension=0.44] coordinates{(0,0) (2.5,-0.3) (5,-0.4)};
    \node[right,blue] at (2.25,-1) {\footnotesize $\bm\zeta$};
    \fill (0,0) circle (1pt);
    \fill (3.6,3.6) circle (1pt);
    \fill (4.2,-4.2) circle (1pt);
    \fill (2.6,0) circle (1pt);
    \draw[-latex,thin,blue] (2.6,0) -- (2.55,0.4) node[above,right]{\footnotesize $\bm\xi^{(0)}$};
    \draw[-latex,thin,blue] (2.3,0) -- (2.25,0.35);
    \draw[-latex,thin,blue] (2,0) -- (1.95,0.3);
    \draw[-latex,thin,blue] (4.2,-4.2) node[left]{\footnotesize $\bm\xi$} -- (3.9,-3.9);
    \draw[-latex,thin,blue] (3.6,3.6) -- (3.9,3.9) node[below,left]{\footnotesize $\bm\xi^{(+)}$};
    \draw[<->] (0,4.2) node[above]{\footnotesize $t$} -- (0,3.6) -- (0.5,3.6) node[right]{\footnotesize $x^{d-1}$};
  \end{tikzpicture}
  \caption{The flow generated by the vector field $\bm\zeta$ in Rindler wedge. The vector field reduces to the generator of Carrollian diffeomorphism  $\bm\xi$ on $\mathcal{H}^{--}$ and  $\bm\xi^{(+)}$ on $\mathcal{H}^{++}$. It crosses the Cauchy surfaces and reduces to the vector field $\bm\xi^{(0)}$ on $V$. }\label{VRindler}
\end{figure}

We will derive a general formula for the flux operators at first. In figure \ref{VRindler}, we draw a vector field $\bm\zeta$ in the Rindler wedge.  The vector $\bm\zeta$ reduces to the generator of Carrollian diffeomorphism $\bm\xi/\bm\xi^{(+)}$ on $\mathcal{H}^{--}/\mathcal{H}^{++}$ 
\bea 
\bm\zeta=\bm\xi+\mathcal{O}(\rho),\quad \bm\zeta=\bm\xi^{(+)}+\mathcal{O}(\rho).
\eea It will cross $V$ and define a vector $\bm\xi^{(0)}$ on this spacelike hypersurface
\bea 
\bm\xi^{(0)}=\bm\zeta|_{V}.
\eea From the conservation of stress tensor, we can find the following identity from Stokes' theorem 
\bea 
\int_{\text{bulk}} d^dx T^{\mu\nu}\nabla_\mu\zeta_\nu=\int_{\mathcal{H}^{++}} (d^{d-1}x)_\mu T^{\mu\nu}\zeta_\nu-\int_{\mathcal{H}^{--}} (d^{d-1}x)_\mu T^{\mu\nu}\zeta_\nu,\label{stokes}
\eea where the domain of integration on the left hand side is the Rindler wedge which is denoted as ``bulk''. We have dropped the terms at infinity since the modes are suppressed exponentially in Rindler spacetime which is checked in Appendix \ref{modeexpRindler}. Since the stress tensor is symmetric 
\be 
T_{\mu\nu}=T_{\nu\mu}
\ee and the vector $\bm\zeta$ generates a diffeomorphism in the bulk with 
\be 
\delta_{\bm\zeta}g_{\mu\nu}=\nabla_\mu\zeta_\nu+ \nabla_\nu\zeta_\mu,
\ee we may recap \eqref{stokes} as 
\bea 
\frac{1}{2}\int_{\text{bulk}}d^d x T^{\mu\nu}
\delta_{\bm\zeta}g_{\mu\nu}=Q_{\bm\xi}-Q_{\bm\xi^{(+)}}
\eea where we have used the definition of flux operators associated with a Carrollian diffeomorphism generated by $\bm\xi$ or $\bm\xi^{(+)}$. Interestingly, for the Killing vectors $\bm\zeta$, the flux operators should be equal to each other 
\be 
Q_{\bm\xi}=Q_{\bm\xi^{(+)}},\quad \bm\zeta\ \text{is a Killing vector}.
\ee The derivation could be extended to another  bulk region. For example, when we choose the shaded region which is bounded by $\mathcal{H}^{--}$ and $V$ in figure \ref{VRindler}, the identity becomes 
\bea 
\frac{1}{2}\int_{\text{shaded region}} d^dx T^{\mu\nu}\delta_{\bm\zeta}g_{\mu\nu}=Q_{\bm\xi}-Q_{\bm\xi^{(0)}}
\eea where we have defined a ``charge'' associated with $ \bm\xi^{(0)}$  on $V$
\bea 
Q_{ \bm\xi^{(0)}}=\int (d^{d-1}x)_\mu T^{\mu\nu}\xi^{(0)}_\nu.
\eea 
Again, for a Killing vector $\bm\zeta$, we can identify the flux operators on $\mathcal{H}^{--}$ with the charges on $V$
\bea 
Q_{\bm\xi}=Q_{\bm\xi^{(0)}},\quad \bm\zeta\ \text{is a Killing vector}.
\eea 
The modular Hamiltonian is generated by the modular flow $\partial_\tau$, which is  exactly the Killing vector $\bm\xi_-=x_{d-1}\partial_t+t\partial_{d-1}$. In retarded coordinates, we have found 
\be 
\bm\xi_-=\partial_u
\ee which  generates the supertranslation with 
\be 
f(u,\bm x_{\perp})=1.
\ee On $V$, the Killing vector $\partial_\tau$ becomes 
\be 
\bm\xi^{(0)}=x_{d-1}\partial_t,\quad t=0.
\ee Therefore, we can identify the modular Hamiltonian as the flux operator $\mathcal{T}_{f=1}$ in Rindler wedge 
\bea
\mathcal{T}_{f=1}=Q_{\bm\xi^{(0)}}=\int_{V}(d^{d-1}x)_\mu T^{\mu}_{\ \nu}\xi^{(0)\nu}=\int_{V}d^{d-1}\bm x x^{d-1}T_{tt}(t=0,\bm x)=\frac{1}{2\pi}H_{\text{modular}}.\label{moduidentify}
\eea This identity strongly indicates that the flux operators $\mathcal{T}_f$ and $\mathcal{M}_Y$ are physically important. We may regard them as an extension of modular Hamiltonian for subsystems. The reduced density matrix for a Rindler wedge becomes 
\be 
\rho_{\text{Rindler}}=e^{-H_{\text{modular}}}=e^{-2\pi\mathcal{T}_{f=1}}.\label{modulardensity}
\ee 
As a by-product, we list the one-to-one correspondence between the  operators on $\mathcal{H}^{--}$ and  $V$ for the Killing vectors \eqref{corresfY}.
\begin{table}
\begin{center}
\renewcommand\arraystretch{1.5}
    \begin{tabular}{|c||c|c|}\hline
Killing vectors $\bm\zeta$&Flux operators $(f,Y^A)$& Operators in the subregion\\\hline\hline
$\frac{1}{2}(\partial_t-\partial_{d-1})$&$(e^u,0)$&$\int_V d^{d-1}\bm x (T_{tt}-T_{t,d-1})$\\\hline
$b^A\partial_A$&$(0,b^A)$&$b^A\int_V d^{d-1}\bm x T_{tA}$\\\hline
$2(x_+\partial_--\partial_-\partial_+)$&$(1,0)$&$ \frac{1}{2\pi}H_{\text{modular}}$\\\hline
$c^A(x_-\partial_A-x_A\partial_-)$&$(-c^Ax_A e^u,0)$&$\frac{c^A}{2}\int_V d^{d-1}\bm x (x_A T_{t,d-1}-x_{d-1}T_{tA})$\\\hline
$\omega^{AB}(x_A\partial_B-x_B\partial_A)$&$(0,\omega^{CA}x_C)$&$\omega^{CA}\int_V d^{d-1}\bm x T_{tA}x_C$\\\hline
\end{tabular}
\caption{\centering{Correspondence between operators on $\mathcal{H}^{--}$ and $V$.}}\label{physicalconst}
\end{center}
\end{table}

\subsection{Observables in  Unruh effect}\label{Unruh}
The Rindler wedge is the spacetime observed by an accelerated observer with constant acceleration $a=1$. The Unruh effect says that the accelerated observer could detect a thermal bath in the Minkowski vacuum $|0\rangle$. From mode expansion which is reviewed in Appendix \ref{modeexpRindler}, the vacuum defined in $\mathcal{H}^{--}$ is the Rindler vacuum $|0\rangle_{\rm R}$ rather than  $|0\rangle$. In this section, we will compute the expectation values of the flux operators $\mathcal{T}_f$ and $\mathcal{M}_Y$ in the Minkowski vacuum $|0\rangle$ and excited states to propose that they are physical observables for the accelerated observer. 

The starting point is the identification of the modular Hamiltonian and the zero mode of supertranslation flux operator \eqref{moduidentify}. In quantum information theory, considering a subsystem $R$ which is obtained by tracing out the degrees of freedom of its complement $\bar{R}$, we may find the reduced density matrix $\mathfrak{\rho}_R$ for the subsystem 
\be 
\mathfrak{\rho}_{R}=\text{tr}_{\bar{R}}\mathfrak{\rho}\label{trace}
\ee where $\mathfrak{\rho}$ is the density matrix of the whole system $R\cup\bar{R}$. The reduced density matrix may be rewritten as an exponential of the modular Hamiltonian
\be 
\mathfrak{\rho}_R=e^{-H_{\text{modular}}}
\ee which is given by \eqref{moduidentify} for the  Rindler wedge. The entanglement entropy for the subsystem is 
\be 
S_{R}=-\text{tr}_R \mathfrak{\rho}_R\log\mathfrak{\rho}_R=\text{tr}_R \mathfrak{\rho}_R H_{\text{modular}}.
\ee Switching to the whole system with the relation \eqref{trace}, we conclude that the entanglement entropy is the expectation  value of modular Hamiltonian. Now for the Rindler wedge, we may replace the modular Hamiltonian with the supertranslation flux operator with $f=1$ 
\be 
S_R=2\pi\text{tr}_R (\mathfrak{\rho}_R \mathcal{T}_{f=1})= 2\pi\text{tr}(\mathfrak{\rho}\mathcal{T}_{f=1}).
\ee 
Now we may extend the modular Hamiltonian to general flux operators $\mathcal{T}_f$ and $\mathcal{M}_Y$, and then it is natural to define the following expectation values 
 \bea 
 \mathfrak{T}_f(\mathfrak{\rho})\equiv\text{tr}_R(\mathfrak{\rho}_R\mathcal{T}_f)= \text{tr}(\mathfrak{\rho}\mathcal{T}_f),\qquad \mathfrak{M}_Y(\mathfrak{\rho})\equiv\text{tr}_R(\mathfrak{\rho}_R\mathcal{M}_Y)= \text{tr}(\mathfrak{\rho}\mathcal{M}_Y).\label{entanglecarroll}
 \eea They are direct extensions of entanglement entropy for Rindler wedge. 
 These are also the fluxes across the Killing horizon observed by the accelerated observer. When the system is in the vacuum state $|0\rangle$, the density matrix is 
 \be 
 \rho_0=|0\rangle\langle 0|.
 \ee 
 
 Sometimes, we compare the fluxes for different states. Given two states characterized by density matrices $\mathfrak{\rho}_1$ and $\mathfrak{\rho}_2$, we may define the following relative fluxes 
 \bs\label{relativeflux}\begin{align}
 \Delta\mathfrak{T}_f(\mathfrak{\rho}_1|\mathfrak{\rho}_2)&=\mathfrak{T}_f(\mathfrak{\rho}_1)-\mathfrak{T}_f(\mathfrak{\rho}_2)=\text{tr}[(\mathfrak{\rho}_1-\mathfrak{\rho}_2)\mathcal{T}_f],\\
  \Delta\mathfrak{M}_Y(\mathfrak{\rho}_1|\mathfrak{\rho}_2)&=\mathfrak{M}_Y(\mathfrak{\rho}_1)-\mathfrak{M}_Y(\mathfrak{\rho}_2)=\text{tr}[(\mathfrak{\rho}_1-\mathfrak{\rho}_2)\mathcal{M}_Y].
 \end{align}\es
In the following, we will calculate the quantities \eqref{entanglecarroll}
in vacuum and excited states and find the relative fluxes according to \eqref{relativeflux}.

\subsubsection{Minkowski vacuum}
We use the mode expansion to find the operators in  momentum space
\bs\begin{align}
    \mathcal{T}_f=&-\frac{1}{2}\int_0^\infty d\omega \int_0^\infty d\omega' \sqrt{\omega\omega'}\int d\bm k_{\perp} \int d\bm k_{\perp}'[\mathfrak{f}(\omega+\omega',\bm k_{\perp}+\bm k_{\perp}') a^{\rm R}_{\omega,\bm k_{\perp}}a^{\rm R}_{\omega',\bm k_{\perp}'}\nn\\&-2\mathfrak{f}^*(\omega-\omega',\bm k_{\perp}-\bm k '_{\perp})a^{\rm R\dagger}_{\omega,\bm k_{\perp}}a^{\rm R}_{\omega',\bm k_{\perp}'}+\mathfrak{f}^*(\omega+\omega',\bm k_{\perp}+\bm k_{\perp}')a^{\rm R\dagger}_{\omega,\bm k_{\perp}}a^{\rm R\dagger}_{\omega',\bm k_{\perp}'}],\label{rindlerTf}\\
    \mathcal{M}_Y=&-\int_0^\infty d\omega \int_0^\infty d\omega'\int d\bm k_{\perp}\int d\bm k_{\perp}' a^{\rm R\dagger}_{\omega,\bm k_{\perp}} a^{\rm R}_{\omega',\bm k_{\perp}'} k_{\perp}^A \mathfrak{Y}^*_A(\bm k_{\perp}-\bm k_{\perp}')\nn\\
    =&
     -\frac{1}{2}\int_0^\infty d\omega \int d\bm k_{\perp}\int d\bm k_{\perp}'a^{\rm R\dagger}_{\omega,\bm k_{\perp}} a^{\rm R}_{\omega,\bm k_{\perp}'} (k_{\perp}^A+k_{\perp}^{'A}) \mathfrak{Y}^*_A(\bm k_{\perp}-\bm k_{\perp}') \label{rindlerMY}
\end{align}\es where $\mathfrak{f}(\omega,\bm k_{\perp})$ is the Fourier transform  of the function $f(u,\bm x_{\perp})$
\be 
\mathfrak{f}(\omega,\bm k_{\perp})=\frac{1}{(2\pi)^{d-1}}\int du d\bm x_{\perp} f(u,\bm x_{\perp})e^{-i\omega u+i\bm k_{\perp}\cdot\bm x_{\perp}}
\ee 
and $\mathfrak{Y}_A(\bm k_{\perp})$ is the Fourier transform  of $Y_A(\bm x_{\perp})$ with
\be 
\mathfrak{Y}_A(\bm k_{\perp})=\frac{1}{(2\pi)^{d-2}}\int d\bm x_{\perp}Y_A(\bm x_{\perp})e^{i\bm k_{\perp}\cdot\bm x_{\perp}}.
\ee The operators $a^{\rm R}_{\omega,\bm k_{\perp}},a^{\rm R\dagger}_{\omega,\bm k_{\perp}} $ are related to $b_{\bm p},b_{\bm p}^\dagger$  by Bogoliubov coefficients
\bs\begin{align}
    a^{\rm R}_{\omega,\bm k_{\perp}}&=\int d\bm p (\alpha_{\omega,\bm k_{\perp};\bm p}^* b_{\bm p}-\beta_{\omega,\bm k_{\perp};\bm p}^* b_{\bm p}^\dagger),\\
    a^{\rm R\dagger}_{\omega,\bm k_{\perp}}&=\int d\bm p (\alpha_{\omega,\bm k_{\perp};\bm p} b^\dagger_{\bm p}-\beta_{\omega,\bm k_{\perp};\bm p} b_{\bm p}).
\end{align}\es The Bogoliubov coefficients are 
\bs \label{Bog}
\begin{align}
    \alpha_{\omega,\bm k_{\perp};\bm p}&=\frac{1}{2\pi}\sqrt{\frac{\omega}{E}}e^{\pi\omega/2} \Gamma(i\omega)(\frac{E+p^{d-1}}{2})^{-i\omega}\delta^{(d-2)}(\bm k_{\perp}-\bm p_{\perp}).\\
\beta_{\omega,\bm k_{\perp};\bm p}&=-\frac{1}{2\pi}\sqrt{\frac{\omega}{E}}e^{-\pi\omega/2} \Gamma(i\omega)(\frac{E+p^{d-1}}{2})^{-i\omega}\delta^{(d-2)}(\bm k_{\perp}+\bm p_{\perp}).
\end{align}\es 
Readers can find the derivation in Appendix \ref{modeexpRindler}.
Therefore, we find the following expectation values in Minkowski vacuum
\bs\begin{align}
\langle 0|a^{\rm R}_{\omega,\bm k_{\perp}}a^{\rm R}_{\omega',\bm k_{\perp}'}|0\rangle&=-\int d\bm p\  \alpha^*_{\omega,\bm k_{\perp};\bm p}\beta^*_{\omega',\bm k_{\perp}';\bm p},\\
\langle 0|a^{\rm R\dagger}_{\omega,\bm k_{\perp}}a^{\rm R}_{\omega',\bm k_{\perp}'}|0\rangle&=\int d\bm p \ \beta_{\omega,\bm k_{\perp};\bm p}\beta^*_{\omega',\bm k_{\perp}';\bm p},\\
   \langle 0|a^{\rm R\dagger}_{\omega,\bm k_{\perp}}a^{\rm R\dagger}_{\omega',\bm k_{\perp}'}|0\rangle&=-\int d\bm p \ \beta_{\omega,\bm k_{\perp};\bm p}\alpha_{\omega',\bm k_{\perp}';\bm p}. 
\end{align}\es Utilizing the integral identity 
\be 
\int_{-\infty}^\infty \frac{dx}{\sqrt{b^2+x^2}}(\sqrt{b^2+x^2}+x)^{iz}=2\pi\delta(z),
\ee 
we find 
\bs\begin{align}
    \langle 0|a^{\rm R}_{\omega,\bm k_{\perp}}a^{\rm R}_{\omega',\bm k_{\perp}'}|0\rangle&=\frac{1}{2\pi}e^{\pi\omega}\sqrt{\omega\omega'}\Gamma(-i\omega)\Gamma(i\omega)\delta(\omega+\omega')\delta^{(d-2)}(\bm k_{\perp}+\bm k_{\perp}'),\\
    \langle 0|a^{\rm R\dagger}_{\omega,\bm k_{\perp}}a^{\rm R}_{\omega',\bm k_{\perp}'}|0\rangle&=\frac{1}{2\pi}e^{-\pi\omega}\omega \Gamma(-i\omega)\Gamma(i\omega)\delta(\omega-\omega')\delta^{(d-2)}(\bm k_{\perp}-\bm k_{\perp}'),\\
   \langle 0|a^{\rm R\dagger}_{\omega,\bm k_{\perp}}a^{\rm R\dagger}_{\omega',\bm k_{\perp}'}|0\rangle&=\frac{1}{2\pi}e^{-\pi\omega}\sqrt{\omega\omega'}\Gamma(-i\omega)\Gamma(i\omega) \delta(\omega+\omega')\delta^{(d-2)}(\bm k_{\perp}+\bm k_{\perp}')
   \end{align}\es
 In \eqref{rindlerTf}, the integration variables $\omega,\omega'$ are non-negative. Therefore, the contributions  from $ \langle 0|a^{\rm R}_{\omega,\bm k_{\perp}}a^{\rm R}_{\omega',\bm k_{\perp}'}|0\rangle$ and $ \langle 0|a^{\rm R\dagger}_{\omega,\bm k_{\perp}}a^{\rm R\dagger}_{\omega',\bm k_{\perp}'}|0\rangle$ are zeros and 
 \bea 
 \mathfrak{T}_f(\mathfrak{\rho}_0) &=&\int_0^\infty d\omega \int_0^\infty d\omega'\int d\bm k_{\perp} \int d\bm k_{\perp}'\sqrt{\omega\omega'}\mathfrak{f}^*(\omega-\omega',\bm k-\bm k_{\perp}')  \langle 0|a^{\rm R\dagger}_{\omega,\bm k_{\perp}}a^{\rm R}_{\omega',\bm k_{\perp}'}|0\rangle\nn\\&=&\frac{1}{2\pi}\int_0^\infty d\omega \int d\bm k_{\perp} \mathfrak{f}^*(0,\bm 0)\omega^2 e^{-\pi\omega} \Gamma(-i\omega)\Gamma(i\omega)\nn\\&=&\int_0^\infty d\omega \frac{\omega}{e^{2\pi\omega}-1}\int d\bm k_{\perp} \mathfrak{f}^*(0,\bm 0).\label{rindlerTfth}
 \eea At the last step, we used the identity for Gamma function
 \be 
 \Gamma(ix)\Gamma(-ix)=\frac{\pi}{x\sinh\pi x}.
 \ee 
 Similarly, one can obtain
 \bea 
\mathfrak{M}_Y(\mathfrak{\rho}_0) &=&- \delta(0)\int_0^\infty d\omega \frac{1}{e^{2\pi\omega}-1}\int d\bm k_{\perp} k_{\perp}^A \mathfrak{Y}_A^*(\bm 0).\label{rindlerMYth}
 \eea
 Since the integrand is odd, we conclude that the vacuum  expectation value of $\mathfrak{M}_Y(\mathfrak{\rho}_0)$ vanishes.
 This shows that the observer will detect a thermal bath without any time and angular independent information. We notice that the expectation value $\mathfrak{T}_f(\mathfrak{\rho}_0)$  is divergent in Minkowski vacuum.
 The integral in the transverse  direction can be  regularized by choosing a cutoff in the UV region, similar to \eqref{regdelta}. To be precise, we notice $\mathfrak{f}(0,\bm 0)$ is a constant and the divergent integral is exactly 
 \be 
 \int d\bm k_{\perp}=\frac{(2\pi)^{d-2}}{a^{d-2}}.
 \ee The length dimension of $\bm x_{\perp}$ is 1 while the retarded time $u$ is dimensionless, following from the transformations \eqref{transrindler} and \eqref{retardedtimerindler}. Therefore the supertranslation function $f(u,\bm x_{\perp})$ is dimensionless. From the definition of $\mathfrak{f}$, we conclude that $\mathfrak{f}^*(0,\bm 0)$ has the dimension of area in the transverse direction. We parameterize \bea 
 \mathfrak{f}^*(0,\bm 0)=\gamma \frac{\text{Area}(B)}{(2\pi)^{d-2}}\label{paragamma}
 \eea by a dimensionless constant $\gamma$ which may be expressed as 
 \bea 
 \gamma=(2\pi)^{d-2}\frac{\int du d\bm x_{\perp} f(u,\bm x_{\perp})}{\int d\bm x_{\perp}}.
 \eea Therefore, the expectation value $\mathfrak{T}_f(\mathfrak{\rho}_0)$ becomes
 \be 
 \mathfrak{T}_f(\mathfrak{\rho}_0)=\gamma\frac{\text{Area}(B)}{a^{d-2}}\int_0^\infty d\omega \frac{\omega}{e^{2\pi\omega}-1}.
 \ee To check the consistency of the regularization, we choose $f=1$ and its Fourier transform is 
 \bea 
 \mathfrak{f}^*(0,\bm 0)=\frac{1}{(2\pi)^{d-1}}\int du d\bm x_{\perp}=\frac{T L^{d-2}}{(2\pi)^{d-1}},
 \eea where $T$ and $L$ are the IR cutoffs for the retarded time and transverse directions, respectively. Compared with \eqref{paragamma} we read 
 \be 
 \gamma=\frac{T}{2\pi}.
 \ee Note that $T$ is dimensionless by definition. As we have mentioned, the expectation value $\mathfrak{T}_{f=1}(\rho_0)$ is the entanglement entropy for the Rindler wedge. It is well known that the entanglement entropy for a continuous subsystem is proportional to the area of the entanglement surface \cite{1986PhRvD..34..373B,Srednicki:1993im,Callan:1994py}. Our regularization method also leads to this conclusion.
 \subsubsection{Excited states in Minkowski spacetime}
 
 Considering the one particle state with definite momentum $\bm p$ 
 \be 
|\bm p\rangle= b_{\bm p}^\dagger |0\rangle,
 \ee we could find the following expectation values 
 \bs\begin{align}
     \langle \bm p|a_{\omega,\bm k_{\perp}}^{\rm R} a^{\rm R}_{\omega',\bm k_{\perp}'}|\bm p\rangle&=-(\alpha_{\omega,\bm k_{\perp};\bm p}^* \beta_{\omega',\bm k_{\perp}';\bm p}^*+\beta_{\omega,\bm k_{\perp};\bm p}^* \alpha^*_{\omega',\bm k_{\perp}';\bm p})+\langle 0|a_{\omega,\bm k_{\perp}}^{\rm R} a^{\rm R}_{\omega',\bm k_{\perp}'}|0\rangle \langle \bm p|\bm p\rangle,\\
     \langle \bm p|a_{\omega,\bm k_{\perp}}^{\rm R\dagger} a^{\rm R}_{\omega',\bm k_{\perp}'}|\bm p\rangle&=\alpha_{\omega,\bm k_{\perp};\bm p}\alpha^*_{\omega',\bm k'_{\perp};\bm p}+\beta_{\omega,\bm k_{\perp};\bm p}\beta^*_{\omega',\bm k'_{\perp};\bm p}+\langle 0|a_{\omega,\bm k_{\perp}}^{\rm R\dagger} a^{\rm R}_{\omega',\bm k_{\perp}'}|0\rangle \langle \bm p|\bm p\rangle,\\
     \langle \bm p|a_{\omega,\bm k_{\perp}}^{\rm R\dagger} a^{\rm R\dagger}_{\omega',\bm k_{\perp}'}|\bm p\rangle&=-(\alpha_{\omega,\bm k_{\perp};\bm p}\beta_{\omega',\bm k_{\perp}';\bm p}+\beta_{\omega,\bm k_{\perp};\bm p}\alpha_{\omega',\bm k_{\perp}';\bm p})+\langle 0|a_{\omega,\bm k_{\perp}}^{\rm R\dagger} a^{\rm R\dagger}_{\omega',\bm k_{\perp}'}|0\rangle \langle \bm p|\bm p\rangle.
 \end{align}\es  We can easily obtain the following relative fluxes
 \bs\begin{align}
     \Delta\mathfrak{T}_f(\mathfrak{\rho}|\mathfrak{\rho}_0)&\equiv \frac{\langle \bm p|\mathcal{T}_f|\bm p\rangle}{\langle \bm p|\bm p\rangle}{-\bra0\mathcal{T}_f\ket0},\\
     \Delta \mathfrak{M}_Y(\mathfrak{\rho}|\mathfrak{\rho}_0)&\equiv \frac{\langle \bm p|\mathcal{M}_Y|\bm p\rangle}{\langle \bm p|\bm p\rangle},
 \end{align}\es where 
 \bs\begin{align}
     \langle\bm p|\mathcal{T}_f|\bm p\rangle=&-\frac{1}{4\pi^2}\int_0^\infty d\omega \int_0^\infty d\omega' \frac{\omega\omega'}{E}\cosh\frac{\pi(\omega-\omega')}{2}\Gamma(-i\omega)\Gamma(-i\omega')(\frac{E+p^{d-1}}{2})^{i(\omega+\omega')}\mathfrak{f}(\omega+\omega',\bm 0)\nn\\&-\frac{1}{4\pi^2}\int_0^\infty d\omega\int_0^\infty d\omega'\frac{\omega\omega'}{E}\cosh\frac{\pi(\omega-\omega')}{2}\Gamma(i\omega)\Gamma(i\omega')(\frac{E+p^{d-1}}{2})^{-i(\omega+\omega')}\mathfrak{f}^*(\omega+\omega',\bm 0)\nn\\&+\frac{1}{2\pi^2}\int_0^\infty d\omega\int_0^\infty d\omega'\frac{\omega\omega'}{E}\cosh\frac{\pi(\omega+\omega')}{2}\Gamma(i\omega)\Gamma(-i\omega')(\frac{E+p^{d-1}}{2})^{-i(\omega-\omega')}\mathfrak{f}^*(\omega-\omega',\bm 0),\\
     \langle \bm p|\mathcal{M}_Y|\bm p\rangle=& -\frac{1}{2\pi}\frac{p_{\perp}^A\mathfrak{Y}_A^*(0)}{E}\int_0^\infty d\omega. 
 \end{align}\es 
Note that the expectation value of $\mathcal{M}_Y$ is divergent. However, 
 the normalization of the state $|\bm p\rangle$ 
 \bea 
 \langle \bm p|\bm p\rangle=\delta^{(d-1)}(\bm 0)
 \eea 
 is also divergent.  
 We may consider the following wave packet 
 \bea 
 |\sigma\rangle=\int d\bm p \sigma_{\bm p}|\bm p\rangle
 \eea which is the superposition of the state with definite momentum. The coefficient $\sigma_{\bm p}$ is chosen to normalize the state $|\sigma\rangle$
 \be 
 \langle\sigma|\sigma\rangle=1\quad\Rightarrow\quad \int d\bm p \sigma_{\bm p}^* \sigma_{\bm p}=1.
 \ee The density matrix becomes 
 \bea 
 \mathfrak{\rho}_{\sigma}=|\sigma\rangle \langle\sigma|.
 \eea 
 Using the transition matrices 
 \bs\begin{align}
     \langle \bm p|a^{\rm R}_{\omega,\bm k_{\perp}}a^{\rm R}_{\omega',\bm k_{\perp}'}|\bm p'\rangle&=-(\alpha^*_{\omega,\bm k_{\perp};\bm p'}\beta^*_{\omega',\bm k_{\perp}';\bm p}+\alpha^*_{\omega',\bm k_{\perp}';\bm p'}\beta^*_{\omega,\bm k_{\perp};\bm p})+\langle 0|a^{\rm R}_{\omega,\bm k_{\perp}}a^{\rm R}_{\omega',\bm k_{\perp}'}|0\rangle\langle \bm p|\bm p'\rangle,\\
     \langle \bm p|a^{\rm R\dagger}_{\omega,\bm k_{\perp}}a^{\rm R}_{\omega',\bm k_{\perp}'}|\bm p'\rangle&=\alpha_{\omega,\bm k_{\perp};\bm p}\alpha^*_{\omega',\bm k_{\perp}';\bm p'}+\beta_{\omega,\bm k_{\perp};\bm p'}\beta^*_{\omega',\bm k_{\perp}';\bm p}+\langle 0|a^{\rm R\dagger}_{\omega,\bm k_{\perp}}a^{\rm R}_{\omega',\bm k_{\perp}'}|0\rangle\langle \bm p|\bm p'\rangle,\\
     \langle\bm p|a^{\rm R\dagger}_{\omega,\bm k_{\perp}}a^{\rm R\dagger}_{\omega',\bm k_{\perp}'}|\bm p'\rangle&=-(\alpha_{\omega,\bm k_{\perp};\bm p}\beta_{\omega',\bm k_{\perp}';\bm p'}+\alpha_{\omega',\bm k_{\perp}';\bm p}\beta_{\omega,\bm k_{\perp};\bm p'})+\langle 0|a^{\rm R\dagger}_{\omega,\bm k_{\perp}}a^{\rm R\dagger}_{\omega',\bm k_{\perp}'}|0\rangle\langle \bm p|\bm p'\rangle,
 \end{align}\es the relative fluxes are  
 \bs\begin{align}
     \Delta\mathfrak{T}_f(\mathfrak{\rho}_{\sigma}|\mathfrak{\rho}_0)=&-\frac{1}{8\pi^2}\int d\bm p\int d\bm p' \sigma^*_{\bm p}\sigma_{\bm p'}\int_0^\infty d\omega \int_0^\infty d\omega'\frac{\omega\omega'}{\sqrt{EE'}}\Gamma(-i\omega)\Gamma(-i\omega')\mathfrak{f}(\omega+\omega',\bm p_{\perp}'-\bm p_{\perp})\nn\\&\times   \ [(\frac{E+p^{d-1}}{2})^{i\omega'}(\frac{E'+p^{'d-1}}{2})^{i\omega}e^{\pi(\omega-\omega')/2}+(\frac{E+p^{d-1}}{2})^{i\omega}(\frac{E'+p^{'d-1}}{2})^{i\omega'}e^{-\pi(\omega-\omega')/2}]\nn\\&-\frac{1}{8\pi^2}\int d\bm p\int d\bm p'\sigma^*_{\bm p}\sigma_{\bm p'}\int_0^\infty d\omega \int_0^\infty d\omega' \frac{\omega\omega'}{\sqrt{EE'}}\Gamma(i\omega)\Gamma(i\omega')\mathfrak{f}^*(\omega+\omega',\bm p_{\perp}-\bm p_{\perp}')\nn\\&\times[(\frac{E+p^{d-1}}{2})^{-i\omega}(\frac{E'+p^{'d-1}}{2})^{-i\omega'}e^{\pi(\omega-\omega')/2}+(\frac{E+p^{d-1}}{2})^{-i\omega'}(\frac{E'+p^{'d-1}}{2})^{-i\omega}e^{-\pi(\omega-\omega')/2}]\nn\\&+\frac{1}{4\pi^2}\int d\bm p\int d\bm p'\sigma_{\bm p}^* \sigma_{\bm p'}\int_0^\infty d\omega \int_0^\infty d\omega' \frac{\omega\omega'}{\sqrt{EE'}}\Gamma(i\omega)\Gamma(-i\omega')\nn\\&\times[\mathfrak{f}^*(\omega-\omega',\bm p_{\perp}-\bm p_{\perp}')(\frac{E+p^{d-1}}{2})^{-i\omega}(\frac{E'+p^{'d-1}}{2})^{i\omega'}e^{\pi(\omega+\omega')/2}\nn\\&+\mathfrak{f}^*(\omega-\omega',\bm p_{\perp}'-\bm p_{\perp})(\frac{E+p^{d-1}}{2})^{i\omega'}(\frac{E'+p^{'d-1}}{2})^{-i\omega}e^{-\pi(\omega+\omega')/2}],\\
     \Delta\mathfrak{M}_Y(\mathfrak{\rho}_{\sigma}|\mathfrak{\rho}_0)=
     &-\frac{1}{4\pi}\int d\bm p\int d\bm p'\sigma_{\bm p}^*\sigma_{\bm p'} \int_0^\infty d\omega \frac{(p_{\perp}+p_{\perp}')^A}{\sqrt{EE'}}\nn\\
  & \times \Big[\frac{\mathfrak{Y}_A^*(\bm p_{\perp}-\bm p'_{\perp})}{1-e^{-2\pi\omega}}\left(\frac{E+p^{d-1}}{E'+p^{'d-1}}\right)^{-i\omega}-\frac{\mathfrak{Y}_A^*(\bm p'_{\perp}-\bm p_{\perp})}{e^{2\pi\omega}-1}\left(\frac{E+p^{d-1}}{E'+p^{'d-1}}\right)^{i\omega}\Big].
 \end{align}
 \es The relative fluxes are finite and encode the time and  angle-dependent information of the excited states.

\section{Null hypersurfaces}\label{csfgeneralnull}
We consider a general null hypersurface $\mathcal{N}$ with topology ${\bf{R}} \times N$ whose metric is degenerate 
\be 
ds^2_{\mathcal{N}}=h_{AB}d\theta^Ad\theta^B.
\ee The metric $h_{AB}$ is also the metric of the $d-2$ dimensional manifold $N$. We need a null vector $
\bm \chi=\partial_u$ to generate the time direction of the Carrollian manifold $\mathcal{N}$. The null hypersurface may be  embedded into $d$ dimensional spacetime whose metric may be expanded near $\mathcal{N}$ \cite{Moncrief:1983xua,chrusciel2020geometry,2019CQGra..36p5002D}
\bea 
ds^2=K du^2-2du d\rho+H_{AB}(d\theta^A+\Lambda^A du)(d\theta^B+\Lambda^B du)\label{embmetric}
\eea where $K,\Lambda^A,H_{AB}$ depend on the coordinates $u,\rho,\theta^A,\ A=1,2,\cdots,d-1$ and satisfy the conditions
\bea 
K|_{\rho=0}=0,\quad \Lambda^A|_{\rho=0}=0,\quad H_{AB}|_{\rho=0}=h_{AB}.
\eea We may expand $K,\Lambda^A,H_{AB}$ by Taylor series 
\bs\begin{align} 
K(u,\rho,\theta)&=-2\kappa(u,\theta)\rho-2\sum_{k=2}^\infty \kappa^{(k)}(u,\theta)\rho^k,\\
\Lambda^A(u,\rho,\theta)&=\lambda^A(u,\theta)\rho+\sum_{k=2}^\infty \lambda^{A(k)}(u,\theta)\rho^k,\\
H_{AB}(u,\rho,\theta)&=h_{AB}+\sum_{k=1}^\infty h_{AB}^{(k)}(u,\theta)\rho^k
\end{align}\es 
such that the near $\mathcal{N}$ metric is 
\bea 
ds^2=-2\kappa \rho du^2-2dud\rho+2\lambda_A\rho du d\theta^A+(h_{AB}+\rho h^{(1)}_{AB})d
\theta^A d\theta^B+\mathcal{O}(\rho^2),
\eea
 where $\lambda_A=\lambda^Bh_{AB}$.
The coordinate system is usually called null Gaussian coordinates. The hypersurface $\mathcal{N}$ sits at $\rho=0$. The components of the metric could be written out explicitly as
\bs\begin{align}
    g_{uu}&=K+H_{AB}\Lambda^A\Lambda^B=-2\kappa\rho+\mathcal{O}(\rho^2),\\
    g_{u\rho}&=-1,\quad g_{\rho\rho}=0,\quad g_{\rho A}=0,\\
    g_{uA}&=H_{AB}\Lambda^B=\lambda_{A}\rho+\mathcal{O}(\rho^2),\\
    g_{AB}&=H_{AB}=h_{AB}+\mathcal{O}(\rho).
\end{align}\es  The determinant of the metric \eqref{embmetric} is 
\bea 
-\det g=\det H=\det h (1+\mathcal{O}(\rho))
\eea where $\det H$  and $\det h$ are the determinants of the metric $H_{AB}$ and $h_{AB}$, respectively.
The inverse metric of $g_{\mu\nu}$ is 
\bs\begin{align}
    g^{uu}&=0,\quad g^{u\rho}=-1,\quad g^{uA}=0,\\
    g^{\rho\rho}&=-K=2\kappa\rho+\mathcal{O}(\rho^2),\\ g^{\rho A}&=-H^{AB}\Lambda_B=-\lambda^A \rho+\mathcal{O}(\rho^2),\\
    g^{AB}&=H^{AB}=h^{AB}+\mathcal{O}(\rho).
\end{align}\es The integral measure on $\mathcal{N}$ is abbreviated as 
\be 
\int du d\bm\theta\equiv \int_{-\infty}^\infty du \int_{N}d\bm\theta
\ee with 
\bea 
\int_{N} d\bm\theta=\int_N \sqrt{\det h}d\theta^1\cdots d\theta^{d-2}.
\eea The normal covector of the hypersurface $\mathcal{N}$ is 
\bea 
\mathfrak{m}=(-K)^{-1/2}d\rho\quad\Rightarrow\quad \mathfrak{m}_\rho=(-K)^{-1/2}=(2\kappa\rho)^{-1/2},\quad \mathfrak{m}_u=\mathfrak{m}_A=0.
\eea 
We may consider a general scalar theory with the action \eqref{actionPhi} whose field $\Phi$ may be expanded as  
\be 
\Phi(u,\rho,\bm \theta)=\Sigma(u,\bm \theta)+\rho \Sigma^{(1)}(u,\bm \theta)+\cdots
\ee near $\mathcal{N}$. However, we may allow a mass term since the null hypersurface $\mathcal{N}$ is placed at a finite position in spacetime which could be approached by massive particles similar to the Rindler spacetime. Therefore, we may expand the potential perturbatively as 
\be 
V(\Phi)=\sum_{k=2}^\infty \frac{\lambda_k}{k}\Phi^k,
\ee where $\lambda_2$ is the mass square
\be 
\lambda_2=m^2.
\ee The pre-symplectic form of the scalar theory on the null hypersurface is 
\bea 
\bm\Theta(\delta\Phi;\Phi)=-\int_{\mathcal{N}} (d^{d-1}x)_\mu \partial^\mu\Phi \delta\Phi,
\eea  where $(d^{d-1}x)_\mu$ is defined as
\bea 
(d^{d-1}x)_\mu=\frac{1}{(d-1)!}\epsilon_{\mu\nu_1\cdots\nu_{d-1}}dx^{\nu_1}\wedge \cdots\wedge dx^{\nu_{d-1}}=-\sqrt{2\kappa\rho}\mathfrak{m}_\mu du d\bm\theta=-\delta_\mu^\rho dud\bm\theta.
\eea From the expansion of the field $\Phi$, we find 
\bs\begin{align}
    \partial_u\Phi&=\sum_{k=0}^\infty \dot{\Sigma}^{(k)} \rho^k=\dot\Sigma+\mathcal{O}(\rho),\\
    \partial_\rho\Phi&=\sum_{k=1}^\infty k \Sigma^{(k)}\rho^{k-1}=\Sigma^{(1)}+\mathcal{O}(\rho),\\ 
    \partial_A\Phi&=\sum_{k=0}^\infty \partial_A\Sigma^{(k)}\rho^k=\partial_A\Sigma+\mathcal{O}(\rho).
\end{align}
\es Therefore, we obtain
\bs\begin{align}
    \partial^u\Phi&=g^{u\rho}\partial_\rho\Phi=-\Sigma^{(1)}+\mathcal{O}(\rho),\\ 
    \partial^\rho\Phi&=g^{\rho u}\partial_u\Phi+g^{\rho\rho}\partial_\rho\Phi+g^{\rho A}\partial_A\Phi=-\dot\Sigma+\mathcal{O}(\rho),\\ 
    \partial^A\Phi&=g^{A\rho}\partial_\rho\Phi+g^{AB}\partial_B\Phi=h^{AB}\partial_B\Sigma+\mathcal{O}(\rho).
\end{align}\es
It follows that the symplectic form is 
\bea 
\bm\Omega(\delta\Sigma;\delta\Sigma;\Sigma)&=&-\int_{\mathcal{N}} (d^{d-1}x)_\mu \partial^\mu\delta\Phi \wedge \delta\Phi\nn\\&=&\int du d\bm\theta \delta\Sigma\wedge\delta\dot\Sigma,
\eea 
which implies that the commutators between $\Sigma$ and $\dot\Sigma$ are
\bs\label{comSigmautheta}\begin{align}
    & [ \Sigma(u,\bm \theta),\Sigma(u',\bm\theta')]=\frac{i}{2}\alpha(u-u')\delta^{(d-2)}(\bm \theta-\bm \theta'),\\ 
  & [\Sigma(u,\bm \theta),\dot\Sigma(u',\bm \theta')]=\frac{i}{2}\delta(u-u')\delta^{(d-2)}(\bm \theta-\bm \theta'),\\ 
   & [\dot\Sigma(u,\bm \theta),\dot\Sigma(u',\bm \theta')]=\frac{i}{2}\delta'(u-u')\delta^{(d-2)}(\bm \theta-\bm \theta').
\end{align}\es The field $\Sigma$ may be expanded by the superposition of positive frequency modes and negative frequency modes 
\bea 
\Sigma(u,\bm\theta)=\int_0^\infty \frac{d\omega}{\sqrt{4\pi\omega}}[a_{\omega}(\bm\theta)e^{-i\omega u}+a^\dagger_{\omega}(\bm\theta)e^{i\omega u}]
\eea with the following commutators between $a_{\omega}(\bm\theta)$ and $a^\dagger_{\omega}(\bm\theta)$
\be
    [a_{\omega}(\bm\theta),a_{\omega'}(\bm\theta')]= [a^\dagger_{\omega}(\bm\theta),a^\dagger_{\omega'}(\bm\theta')]=0,\quad  [a_{\omega}(\bm\theta),a^\dagger_{\omega'}(\bm\theta')]=\delta(\omega-\omega')\delta^{(d-2)}(\bm\theta-\bm\theta').\label{comaadagger}
\ee
Indeed, one can check the commutation relations \eqref{comSigmautheta} are satisfied by virtue of  the commutators \eqref{comaadagger}. Therefore, we may interpret $a_{\omega}(\bm\theta)$ and  $a^\dagger_{\omega}(\bm\theta)$ as annihilation and creation operators. More explicitly, we may define the vacuum state as 
\bea 
a_{\omega}(\bm\theta)|0\rangle_{\mathcal{N}}=0
\eea and the creation operator $a_{\omega}^\dagger(\bm\theta)$ acts on the vacuum $|0\rangle_{\mathcal{N}}$ should create a particle with frequency $\omega$ at the position $\bm\theta$. We add a subscript $\mathcal{N}$ to emphasize that the vacuum state may depend on the null hypersurface $\mathcal{N}$. 
The correlators are \bs\begin{align}
{}_{\mathcal{N}}\langle 0|\Sigma(u,\bm \theta)\Sigma(u',\bm \theta')|0\rangle_{\mathcal{N}}&=\beta(u-u')\delta^{(d-2)}(\bm \theta-\bm \theta'),\\
{}_{\mathcal{N}}\langle 0|\Sigma(u,\bm \theta)\dot\Sigma(u',\bm \theta')|0\rangle_{\mathcal{N}}&=\frac{1}{4\pi(u-u'-i\epsilon)}\delta^{(d-2)}(\bm \theta-\bm \theta'),\\
{}_{\mathcal{N}}\langle 0|\dot\Sigma(u,\bm \theta)\Sigma(u',\bm \theta')|0\rangle_{\mathcal{N}}&=-\frac{1}{4\pi(u-u'-i\epsilon)}\delta^{(d-2)}(\bm \theta-\bm \theta'),\\
{}_{\mathcal{N}}\langle 0|\dot\Sigma(u,\bm \theta)\dot\Sigma(u',\bm \theta')|0\rangle_{\mathcal{N}}&=-\frac{1}{4\pi(u-u'-i\epsilon)^2}\delta^{(d-2)}(\bm \theta-\bm \theta').
\end{align}\es
Now it is straightforward to use the formula \eqref{Carrollxi} to find the flux operators associated with Carrollian diffeomorphisms.  The results are 
\bs\begin{align}
    \mathcal{T}_f&=\int du d\bm\theta f(u,\bm\theta):\dot\Sigma^2:,\\
    \mathcal{M}_Y&=\frac{1}{2}\int du d\bm\theta Y^A(\bm\theta)(:\dot\Sigma\nabla_A\Sigma-\Sigma\nabla_A\dot\Sigma:).
\end{align}
\es We can also obtain similar algebra \eqref{comscalar} with a divergent central charge 
\be 
c=\delta^{(d-2)}(\bm 0)=\delta^{(d-2)}(\bm\theta-\bm\theta')|_{\bm\theta'=\bm\theta}.
\ee There is a set of eigenfunctions $f_{\bm q}(\bm\theta)$ of the Laplace operator on the manifold $N$
\be 
\Delta_{N}f_{\bm q}(\bm\theta)=-\lambda_{\bm q}f_{\bm q}(\bm\theta)
\ee where $\lambda_{\bm q}$ is the corresponding eigenvalue. The eigenfunctions are normalized to 1 and orthogonal to each other\footnote{We have assumed the manifold is compact and the eigenvalues are discrete in the following.}
\be 
\int_N d\bm\theta f_{\bm q}(\bm\theta)f^*_{\bm q'}(\bm\theta)=\delta_{\bm q,\bm q'}.\label{eigenfunctions}
\ee 
The completeness relation for the eigenfunctions is 
\be 
\sum_{\bm q}f_{\bm q}(\bm\theta)f^*_{\bm q}(\bm\theta')=\delta^{(d-2)}(\bm\theta-\bm\theta').
\ee  Summing over all possible eigenstates in \eqref{eigenfunctions}, we obtain 
\bea 
\text{tr}\ 1=\sum_{\bm q}1=\sum_{\bm q}\sum_{\bm q'}\delta_{\bm q,\bm q'}=\sum_{\bm q}\int_N d\bm\theta f_{\bm q}(\bm\theta)f^*_{\bm q}(\bm\theta)=\int_N d\bm\theta \delta^{(d-2)}(\bm 0).
\eea Therefore, the central charge is 
\bea 
c=\delta^{(d-2)}(\bm 0)=\frac{\sum\limits_{\bm q} 1}{\int_N d\bm\theta}=\frac{\text{tr}\ 1}{\text{Area of $N$}}=\text{density of states on $N$}.
\eea We may still use the heat kernel method \cite{Iosif:2000} to regularize the central charge for general Riemannian manifold without boundary.
\section{Conclusion and discussion}\label{conc}
In this work, we have constructed the quantum flux operators  for scalar field theory which form a faithful representation of Carrollian diffeomorphism for null hypersurfaces up to a non-zero  central charge.
We have checked the results mainly for two interesting null hypersurfaces and then extended them to general null hypersurfaces. In the first case, the flux operators   associated with Carrollian diffeomorphism at future null infinity are finite 
in general dimensions. Therefore, our results overcome various problems in the context of higher dimensional asymptotic symmetry analysis and are also consistent with soft graviton theorems. Secondly, we extend our method to the Rindler horizon and find similar flux operators  associated with the accelerated observer. They are natural operators for the Rindler spacetime, a subsystem in the context of quantum information. We conclude that the zero mode of the supertranslation flux is actually the modular Hamiltonian of the subsystem.
These results build a connection between Carrollian diffeomorphism and quantum information theory. We could define
the expectation values of the supertranslation and superrotation flux operators associated with the subsystem, which extend the geometric entanglement entropy. By computing the expectation value of $\mathcal{T}_f$ for the vacuum state explicitly, we could check the regularization method of the central charge using the area law of entanglement entropy. We also define the relative fluxes between different states which may characterize the difference between any two states, and may be  new observables in Unruh effect. Finally, we also propose to regularize the central charge using spectral zeta function or heat kernel method. We could find a finite result for the theory at null infinity. For the Rindler wedge, we  conclude that there is an effective central charge  proportional to the area of the bifurcation surface 
    \be 
    \bar{c}=\frac{\text{Area}(B)}{a^{d-2}}.\label{centralarea}
    \ee This result is consistent with entanglement entropy by identifying the  modular Hamiltonian and the zero mode of supertranslation flux. The result shows that the effective central charge $\bar{c}$ counts the number of microstates in the transverse directions which is similar to the area law of black hole entropy. 
There are various open questions in this direction.
\begin{itemize}
    \item It should be able to extend our results to the theory with nonzero spins. In \cite{Liu:2023qtr,Liu:2023gwa}, it has been shown that more anomalous terms should be included for these theories. It would be nice to work out the details for general null hypersurfaces in general dimensions.
    \item Spectral geometry of Carrollian manifold. To study the regularization of the central charge, we compute the spectral zeta function associated with the Carrollian manifold $\mathcal{C}=\textbf{R}\times N$
    \be 
    \zeta_{\mathcal{C}}(s)=\text{tr}'(-\Delta_N)^{-s}
    \ee where $\Delta_N$  is the Laplace operator of the Riemannian manifold $N$. We read out the central charge from $1+\zeta_{\mathcal{C}}(0)$. It would be nice to extract more physical information from the spectral zeta function for general Carrollian manifolds. A related question is to study the heat kernel associated with a general Carrollian manifold 
    \be 
    K_{\mathcal{C}}(\sigma)=\text{tr}\ e^{\sigma\Delta_N}.
    \ee 
    \item Periodicity of imaginary time. For a general subsystem or a black hole, the imaginary time should be periodic and the inverse period is the temperature of the system.  We still do not discuss the consequences of this condition. Taking Rindler spacetime as an example,  the retarded time $u$ is periodic in imaginary direction
\be u\sim u+2\pi i.
\ee To be consistent with the periodic condition, the basis function should obey the equation 
\be 
f_{\omega,\bm k_{\perp}}(u,\bm x_{\perp})=f_{\omega,\bm k_{\perp}}(u+2\pi i,\bm x_{\perp})
\ee which implies 
\be 
\omega=i n,\quad n=0,\pm 1,\pm 2,\cdots.
\ee We may redefine 
\be 
L_{n,\bm k_{\perp}}=i\mathcal{T}_{in,\bm k_{\perp}},
\ee then the Virasoro algebra becomes 
\bs\begin{align}
    \ [L_{n,\bm k_{\perp}},L_{n',\bm k_{\perp}'}]&=(n-n')L_{n+n',\bm 0}+\frac{\bar{c}}{12}n^3\delta_{n,-n'}\quad \text{for}\quad \bm k_{\perp}'=-\bm k_{\perp},\\
    \ [L_{n,\bm k_{\perp}},L_{n',\bm k_{\perp}'}]&=(n-n')L_{n+n',\bm k_{\perp}+\bm k_{\perp}'}\quad\text{for}\quad  \bm k_{\perp}'\not=-\bm k_{\perp}.
\end{align}
\es We find an infinite tower of two-dimensional Virasoro algebras whose label $n$ is discrete, which is exactly the same as two-dimensional conformal field theory. It would be nice to explore this topic in the future. 
\item Event horizon and Hawking radiation. Our method can be extended to the event horizon of black holes. It would be very interesting to study the Carrollian diffeomorphism and its associated quantum flux operators at the event horizon. A better understanding of the representation theory of the algebra  may provide the key insight on Hawking radiation.
\item Relation to quantum information theory. We have noticed that the quantum flux operators defined on the null hypersurface are direct generalizations of the null modular Hamiltonians defined in \cite{Casini:2017roe}. For each spatial subregion, there is a null boundary which is exactly described by a Carrollian manifold. It would be interesting to explore the physical interpretation of the quantum flux operators in the context of quantum information theory. 
\end{itemize}
\vspace{10pt}
{\noindent \bf Acknowledgments.} 
The work of J.L. is supported by NSFC Grant No. 12005069.
\appendix
\section{Unit sphere \texorpdfstring{$S^m$}{}}\label{spc}
In this Appendix, we will review various aspects of the unit sphere $S^m$ with $m=d-2$.
\subsection{Spherical coordinates} \label{scoord}
The isometry group is $SO(m+1)$ and the conformal group is $SO(m+1,1)$ for $S^m$. We note that the conformal group is isomorphic to $SO(d-1,1)$ which is the isometry group of Minkowski spacetime $\mathbb{R}^{1,d-1}$. 
 The unit sphere $S^m$ embedded into Minkowski spacetime $\mathbb{R}^{1,d-1}$ introduces the coordinate transformation from Cartesian coordinates of $d$ dimensional Minkowski spacetime to spherical coordinates $\left( t,r,\theta ^A \right) ,A=1,2,\cdots ,m
$ as follows
\be 
x^1=r\prod_{j=1}^{m}\sin\theta_{j},\quad x^i=r(\prod_{j=i}^{m}\sin\theta_{j})\cos\theta_{i-1},\quad 2\le i\le m+1.
\ee 
 More explicitly, we have
\begin{equation}
    \begin{aligned}\label{G.1.20}
x^1&=r\sin\theta_{m}\sin\theta_{m-1}\cdots\sin\theta_3\sin\theta_2\sin\theta_1,
\\
x^2&=r\sin \theta_{m}\sin\theta_{m-1}\cdots\sin\theta_3\sin\theta_2\cos \theta _1,\\
x^3&=r \sin\theta_{m}\sin\theta_{m-1}\cdots\sin\theta_3\cos\theta_2,
\\
&\cdots 
\\
x^{m}&=r\sin \theta _{m}\cos\theta_{m-1},
\\
x^{m+1}&=r\cos \theta _{m},
    \end{aligned}
\end{equation}
where \be 
r\ge 0;\quad 0\le \theta _{1}\le 2\pi; \quad 0\le \theta _i\le \pi ,\quad 2\le i\le m.\ee 

The normal vector of the sphere $S^{m}$ is
\begin{align}
  n^i=\frac{x^i}{r}.
\end{align}
In spherical coordinates, the metric of Minkowski spacetime is 
\be
  ds^2=-dt^2+(dx^i)^2=-dt^2+dr^2+r^{2}ds^2_{S^{m}},
\ee
where the metric of the unit sphere $S^{m}$ is
\begin{align}
  ds^2_{S^{m}}=\gamma_{AB}d\theta^Ad\theta^B,\quad A,B=1,2,\cdots,m.
\end{align}
The components of the metric $\gamma_{AB} $ are
\be 
\gamma_{AB}=r^{-2}\delta_{ij}\frac{\p x^i}{\p \theta^A}\frac{\p x^j}{\p \theta^B}=\delta_{ij}Y_A^iY_B^j,
\ee 
where $Y_A^i$ is the gradient of the normal vector $n^i$
\begin{align}
  Y^i_A=-\frac{\p n^i}{\p 
\theta^A}=-\nabla_An^i.
\end{align}
The explicit metric of the sphere may be found by the following recursion formula
\bea 
ds_{S^1}^2=d\theta_1^2,\quad ds_{S^{n}}^2=d\theta_{n}^2+\sin\theta_{n}^2ds_{S^{n-1}}^2,\quad n\ge 2.
\eea The determinant of the metric is 
\bea 
\sqrt{\gamma}=\prod_{j=2}^{m}\sin^{j-1}\theta_j.
\eea 
\subsection{Spherical harmonics}\label{sphe}
The scalar spherical harmonic functions on $S^m$ are given by \cite{Higuchi:1986wu}
\bea 
Y_{\bm \ell}(\Omega)=\frac{e^{i\ell_1\theta_1}}{\sqrt{2\pi}}[\prod_{j=2}^{m}c(j,\ell_j,\ell_{m-1})(\sin\theta_j)^{-(j-2)/2} P_{\ell_j+\frac{j-2}{2}}^{-(\ell_{m-1}+\frac{j-2}{2})}(\cos\theta_j)],
\eea where the vector $\bm \ell$ is 
\be 
\bm\ell=(\ell_1,\ell_2,\cdots,\ell_m)
\ee with $\ell_j,\ j=1,2,\cdots,m$ being integers that satisfy 
\be 
\ell_m\ge \ell_{m-1}\ge\cdots\ge \ell_2\ge |\ell_1|.
\ee The associated Legendre function of the first kind is defined through the hypergeometric function:
\bea 
P_\nu^{\mu}(x)=\frac{1}{\Gamma(1-\mu)}\left(\frac{1-x}{1+x}\right)^{-\mu/2}{}_2F_1(-\nu,\nu+1;1-\mu;\frac{1-x}{2}).
\eea The coefficient $c(a_1,a_2,a_3)$ is chosen as 
\bea 
c\left( a_1,a_2,a_3 \right) =\sqrt{\frac{2a_2+a_1-1}{2}\frac{\left( a_2+a_3+a_1-2 \right) !}{\left( a_2-a_3 \right) !}}
\eea such that the spherical harmonic functions are normalized and orthogonal to each other
\be \label{orthonormalityofhar}
\int_{S^m} d\Omega Y_{\bm\ell}^*(\Omega)Y_{\bm \ell'}(\Omega)=\delta_{\bm\ell,\bm\ell'},
\ee 
where  $Y_{\boldsymbol{\ell }}^{\ast}\left( \Omega \right) $ is the complex conjugate of $Y_{\boldsymbol{\ell }}\left( \Omega \right) $. The completeness of spherical harmonic functions is 
\be 
\sum_{\bm \ell}Y_{\bm \ell}^*(\Omega)Y_{\bm \ell}(\Omega')=\delta(\Omega-\Omega').
\ee 
The harmonic functions obey the following Laplace equation on $S^m$
\be 
\nabla^A\nabla_A Y_{\bm\ell}(\Omega)=-\ell_m(\ell_m+m-1)Y_{\bm\ell}(\Omega).
\ee Therefore, for any fixed quantum number $\ell_m=L$, the degeneracy of the spherical harmonic functions is \cite{avery:2017hyperspherical}
\bea \label{degeneracyofsh}
g(L,m)= \sum_{\ell_{m-1}=0}^{\ell_{m-1}=L}\cdots\sum_{\ell_2=0}^{\ell_3}\sum_{\ell_1=-\ell_2}^{\ell_2}1=\frac{(2L+m-1)(L+m-2)!}{L!(m-1)!}.
\eea The generating function of the degeneracy in general dimension is 
\bea 
G(x,m)\equiv\sum_{L=0}^\infty g(L,m)x^L=\frac{1+x}{(1-x)^m}.
\eea

\subsection{Addition theorem}\label{generaladdthm}
As is well known, the addition theorem in four dimensions states that 
\be
 C_{\ell}^{1/2}\left(\cos\gamma\right) =\frac{4\pi}{2\ell +1}\sum_{m=-\ell}^{\ell}Y_{\ell ,m}^{\ast}\left( \Omega \right) Y_{\ell ,m}\left( \Omega' \right),\quad \boldsymbol{\ell }=\left( m,\ell \right),
\ee where $\gamma$ is the intersection angle between the two directions labeled by $\Omega$ and $\Omega'$. The notation $C_{\ell}^k(x)$ is the so-called Gegenbauer polynomial which may be defined by hypergeometric function
\be\label{definitionofgegenbauer}
C_{\ell}^{k}\left( x \right) =\frac{\Gamma \left( 2k+\ell \right)}{\ell !\Gamma \left( 2k \right)}\,\,_2F_1\left( -\ell ,2k+\ell ;\frac{1}{2}+k;\frac{1-x}{2} \right).
\ee
In $4$ dimensions,  $k$ takes the value $1/2$ and the intersection angle $\gamma$ becomes $0$ when $\Omega'=\Omega$, thus the addition theorem reduces to 
\be
C_{\ell}^{1/2}\left( 1 \right) =\frac{4\pi}{2\ell +1}\sum_{m=-\ell}^{\ell}{Y_{\ell ,m}^{\ast}\left( \Omega \right) Y_{\ell ,m}\left( \Omega \right)},\quad \boldsymbol{\ell }=\left( m,\ell \right) ,
\ee
The left hand side could be reduced to the familiar Legendre polynomial of order $\ell$. In general $d$ dimensions, we have a  similar addition theorem, 
\be\label{generalizedaddthm1}
C_{{\ell_m }}^{(m-1)/2}\left( \cos\gamma \right) =K_{{\ell_m }}{\sum_{\boldsymbol{\ell }}}'{Y_{\boldsymbol{\ell }}^{\ast}\left( \Omega \right) Y_{\boldsymbol{\ell }}\left( \Omega' \right)},
\ee where $k$ is related to the dimension $d$ by \be k=\frac{d-3}{2}=\frac{m-1}{2}\ee  and $K_{{\ell_m}}$ is a constant  which is not important in this work.  The summation ${\sum\limits_{\boldsymbol{\ell }}}'$ is to sum over all possible quantum numbers while fixing the value of  $\ell_m$
\be 
{\sum_{\boldsymbol{\ell }}}'=\sum_{\ell_{m-1}=0 }^{\ell_m}\cdots\sum_{\ell_{2}=0}^{\ell_3}\sum_{\ell_1=-\ell_2}^{\ell_2}.
\ee 
When $\Omega'=\Omega$, the addition theorem becomes
\be\label{generalizedaddthm}
C_{{\ell_m }}^{(m-1)/2}\left( 1 \right) =K_{{\ell_m }}{\sum_{\boldsymbol{\ell }}}'{Y_{\boldsymbol{\ell }}^{\ast}\left( \Omega \right) Y_{\boldsymbol{\ell }}\left( \Omega \right)},
\ee
The special value $C_{{\ell_m }}^{(m-1)/2}\left( 1 \right)$ can be found from the definition \eqref{definitionofgegenbauer} 
\be 
C_{{\ell_m }}^{(m-1)/2}\left( 1 \right)=\frac{\Gamma(m+\ell_m-1)}{\ell_m!(m-1)!}.
\ee  Integrating both sides of \eqref{generalizedaddthm} over the $S^m$  and considering the orthonormality \eqref{orthonormalityofhar} yields
\begin{equation}
\begin{split}
&\int_{S^m}{d\Omega}C_{{\ell_m }}^{(m-1)/2}\left( 1 \right) =K_{{\ell_m }}{\sum_{\boldsymbol{\ell }}}'{\int_{S^m}{d\Omega}Y_{\boldsymbol{\ell }}^{\ast}\left( \Omega \right) Y_{\boldsymbol{\ell }}\left( \Omega \right)},
\\
\Rightarrow \quad &C_{\ell_m }^{(m-1)/2}\left( 1 \right) I(0)=K_{\ell _m}{\sum_{\boldsymbol{\ell }}}'{1}=K_{\ell_m} g(\ell_m,m).
\end{split}    
\end{equation}
where $I(0)$ is the area of the unit sphere $S^m$:
\be
I\left( 0 \right) \equiv \int_{S^m}{d\Omega}=\frac{2\pi ^{\left( m+1 \right) /2}}{\Gamma \left[ \left( m+1 \right) /2 \right]},
\ee
 Finally substituting the result above into \eqref{generalizedaddthm} yields 
\be
{ \delta^{(d-2)}(0)}=\sum_{\boldsymbol{\ell }}{Y_{\boldsymbol{\ell }}^{\ast}\left( \Omega \right) Y_{\boldsymbol{\ell }}\left( \Omega \right)}=\sum_{\ell_m=0}^\infty\frac{ g(\ell_m,m)}{I(0)}=\frac{\sum\limits_{\boldsymbol{\ell}} 1}{I(0)}.
\ee
This is the  density of states on $S^m$. 

As a matter of fact, this conclusion is immediately derived if we take the sum for the orthogonality relation of the spherical harmonics 
\begin{align}
  \int_{S^m} d\Omega Y_{\bm\ell}^*(\Omega)Y_{\bm \ell'}(\Omega)=\delta_{\bm\ell,\bm\ell'},\label{yoth}
\end{align}
and notice the completeness relation
\begin{align}
  \sum_{\bm\ell}Y_{\bm\ell}(\Omega)Y^*_{\bm \ell}(\Omega')=\delta(\Omega-\Omega').\label{yoth2}
\end{align}
It follows that
\begin{align}
  \sum_{\bm\ell,\bm\ell'}\delta_{\bm\ell,\bm\ell'}\int_{S^m} d\Omega Y_{\bm\ell}^*(\Omega)Y_{\bm \ell'}(\Omega)=\sum_{\bm\ell,\bm\ell'}\delta_{\bm\ell,\bm\ell'}\quad\Rightarrow\quad \delta^{(d-2)}(0)\int_{S^m} d\Omega=\sum_{\bm\ell}1.
\end{align}
This method may be generalized to any compact manifold where the eigenfunctions have the similar orthogonality and completeness relations.

\subsection{Conformal Killing vectors}\label{CKVSd2}
The conformal group of $S^m$ is $SO(m+1,1)$ which is isomorphic to the Lorentz group of $\mathbb{R}^{1,d-1}$. The conformal algebra $so(m+1,1)$ is generated by \be \frac{(m+2)(m+1)}{2}=\frac{d(d-1)}{2}\ee  conformal Killing vectors (CKVs) of $S^m$. The CKVs could be found by embedding $S^{m}$ into $\mathbb{R}^{1,d-1}$ and making use of the Lorentz algebra of $\mathbb{R}^{1,d-1}$. We may define two null vectors { $n^{\mu},\bar{n}^{\mu}$} in Cartesian coordinates 
\be 
n^\mu=(1,n^i),\qquad \bar{n}^\mu=(-1,n^i).
\ee They satisfy the relations 
\be 
n^2=\bar{n}^2=0,\qquad n\cdot\bar{n}=2.
\ee The spacelike vector     $m^\mu$ and timelike vector $\bar{m}^\mu$ are related to $n^\mu$ and $\bar{n}^\mu$ by the relations
\be 
m^\mu=\frac{1}{2}(n^\mu+\bar{n}^\mu)=(0,n^i),\qquad \bar{m}^\mu=\frac{1}{2}(n^\mu-\bar{n}^\mu)=(1,0).
\ee They are normalized and orthogonal to each other 
\be 
m^2=-\bar{m}^2=1,\qquad m\cdot\bar{m}=0.
\ee
 Hence, the Cartesian coordinates are related to the retarded coordinates by 
\be x^\mu=r n^\mu+u \bar{m}^\mu ,\label{xur}
\ee and the partial derivatives in Cartesian coordinates can be rewritten as 
\be 
\partial_\mu=-n_\mu \partial_u+m_\mu\partial_r-\frac{1}{r}Y_\mu^A\partial_A,\label{partialmu}
\ee where 
\be 
Y_\mu^A=-\nabla^A n_\mu=-\nabla^A \bar{n}_\mu=-\nabla^A m_\mu.
\ee 
The partial derivatives $\partial_\mu$ can also be regarded as the translation operator in $\mathbb{R}^{1,d-1}$. Therefore, we may also write it as
\bea 
P_\mu=\partial_\mu=-n_\mu \partial_u+m_\mu\partial_r-\frac{1}{r}Y_\mu^A\partial_A.
\eea 
We can construct the following antisymmetric tensors  
\bea 
n_{\mu\nu}=n_\mu\bar{n}_\nu-n_\nu\bar{n}_\mu,\quad Y_{\mu\nu}^A=Y_\mu^A n_\nu-Y^A_\nu n_\mu,\quad \bar{m}^A_{\mu\nu}=Y_\mu^A \bar{m}_\nu-Y_\nu^A\bar{m}_\mu,
\eea 
in which the tensors $n_{\mu\nu}$ and $\bar{m}_{\mu\nu}^A$ are related to $Y_{\mu\nu}^A$ through 
\be 
n_{\mu\nu}=-\frac{2}{d-2}\nabla_AY^A_{\mu\nu},\qquad \bar{m}^A_{\mu\nu}= -\frac{1}{d-2}\nabla^A\nabla_C Y^C_{\mu\nu}.
\ee 
The Lorentz algebra is generated by 
\bea 
L_{\mu\nu}=x_\mu\partial_\nu-x_\nu\partial_\mu.
\eea Using retarded coordinates and the relations \eqref{xur}-\eqref{partialmu}, we find 
\bea 
L_{\mu\nu}=-\frac{1}{2}u n_{\mu\nu}\partial_u+\frac{1}{2}(u+r)n_{\mu\nu}\partial_r+(Y_{\mu\nu}^A{ +}\frac{u}{r}\bar{m}_{\mu\nu}^A)\partial_A.\label{lorentztrans}
\eea To find the CKVs of $S^m$, we may set $u=0$ and $r=1$. Therefore, the CKVs of $S^m$ are $Y_{\mu\nu}^A$. More explicitly, they are denoted as $Y_i^A$ and $Y_{ij}^A $ in the context
\bea 
Y_i^A\equiv Y_{0i}^A,\quad Y_{ij}^A\equiv Y_i^A n_j-Y_j^A n_i.
\eea The vectors $Y_{ij}^A$ are Killing vectors of $S^m$ and they form the algebra $so(m+1)$. The vectors $Y_i^A$ are strictly CKVs of $S^m$ and they obey the equation 
\bea\nabla_AY^i_{B}=\nabla_BY^i_A=\frac{\gamma_{AB}}{d-2}\nabla_CY^{iC}. 
\eea
In the following, we collect several important identities about the CKVs.
\begin{enumerate}
\item Products among $n^\mu,\bar{n}^\mu,m^\mu,\bar{m}^\mu$
\bs\begin{align} 
&n^2=\bar{n}^2=0,\quad m^2=-\bar{m}^2=1,\\
& n\cdot \bar{n}=2,\quad n\cdot m=1,\quad n\cdot\bar{m}=-1,\quad \bar n\cdot m=1,\quad \bar{n}\cdot \bar{m}=1,\quad m\cdot \bar m=0.
\end{align}\es
    \item Orthogonality
    \be 
    Y_\mu^A n^\mu=Y_\mu^A\bar{n}^\mu=Y_\mu^A m^\mu=Y_\mu^A\bar{m}^\mu=0.
    \ee 
    \item Products between $Y_{\mu\nu}^A$ and  $n^\mu,\bar{n}^\mu,m^\mu,\bar{m}^\mu$
    \bea 
    Y_{\mu\nu}^A n^\nu=0,\quad Y_{\mu\nu}^A\bar{n}^\nu=2Y_\mu^A,\quad Y^A_{\mu\nu}m^\nu=Y_\mu^A,\quad Y_{\mu\nu}^A \bar{m}^\nu=-Y_{\mu}^A.
    \eea 
    \item Derivatives of  $n_\mu,\bar{n}_\mu,m_\mu,\bar{m}_\mu$
    \begin{align}
       &\nabla_A\nabla_B {n}_\mu=\nabla_A\nabla_B \bar{n}_\mu=\nabla_A\nabla_B {m}_\mu=-\gamma_{AB}m_\mu ,\\
       &\nabla^A\nabla_A n_\mu=\nabla^A\nabla_A \bar n_\mu=\nabla^A\nabla_A m_\mu=-(d-2)m_\mu,\\
       &\nabla_A\nabla_B\bar{m}_\mu=0,\qquad \nabla^A\nabla_A\bar{m}_\mu=0.
    \end{align}
    \item Products  between $Y_\mu^A$ and $Y_{\mu\nu}^A$
    \bs\begin{align}
   & Y_\mu^A Y^{\mu B}=\gamma^{AB},\quad Y_\mu^AY_{\nu A}=\eta_{\mu\nu}-\frac{1}{2}(n_\mu\bar{n}_\nu+n_\nu\bar{n}_\mu),\\
   & Y_{\mu\nu}^A Y^{\mu B}=\gamma^{AB}n_\nu.
    \end{align}\es
    \item Derivatives of $Y_\mu^A$
    \bs\begin{align}
    &\nabla_A Y_B^\mu=\gamma_{AB}m^\mu,\quad \nabla_AY^{A\mu}=(d-2)m^\mu,\\
    & Y_\mu^A\nabla_AY^{\mu B}=0.
    \end{align}\es
    \item Derivatives of $Y_{\mu\nu}^A$
    \begin{align}
        \nabla^AY^B_{\mu\nu}=-\frac{1}{2}\gamma^{AB}n_{\mu\nu}+2Y^A_{[\mu} Y^B_{\nu]},\qquad \nabla_AY_{\mu\nu}^{ A}=-\frac{(d-2)}{2}n_{\mu\nu}.
    \end{align}
\end{enumerate}

\section{Mode expansion}
\subsection{Null infinity}\label{modeexp}
In Minkowski spacetime $\mathbb{R}^{1,d-1}$, the scalar field may be expanded in plane waves 
\bea
\Phi(t,\bm x)=\int \frac{d^{d-1}\bm k}{\sqrt{(2\pi)^{d-1}}}\frac{1}{\sqrt{2\omega_{\bm{k}}}}(e^{-i\omega t+i\bm{k}\cdot\bm{x}}b_{\bm{k}}+e^{i\omega t-i\bm{k}\cdot\bm{x}}b_{\bm{k}}^\dagger)\label{modeexpMink}
\eea where $b_{\bm k}$ and $b_{\bm k}^\dagger$ are annihilation and creation operators, respectively. They obey the standard commutators 
\bs\begin{align}
    [b_{\bm k},b_{\bm k'}]&=[b_{\bm k}^\dagger,b_{\bm k'}^\dagger]=0,\\
    [b_{\bm k},b_{\bm k'}^\dagger]&=\delta^{(d-1)}(\bm k-\bm k').
\end{align}\es 
The plane wave can be expanded in spherical waves by \cite{avery:2017hyperspherical}
\begin{equation}
   \begin{aligned}
    e^{i \bm{k}\cdot \bm x}=\frac{2(d-3)!!\pi^{(d-1)/2}}{\Gamma((d-1)/2)}\sum_{\bm\ell}i^{\ell_{m}} j^{d-1}_{\ell_{m}}(\omega r)Y^*_{\bm\ell}(\Omega_k)Y_{\bm\ell}(\Omega),
   \end{aligned} 
\end{equation} where the summation $\sum\limits_{\bm \ell}^{}$ is over all possible combinations labeled by $\bm\ell$.
The $d-1$ dimensional vectors $\bm k$ and $\bm x$ are written in spherical coordinates 
\be 
\bm k=(\omega,\Omega_k),\qquad \bm x=(r,\Omega).
\ee 
The function $j_{\ell_m}^{d-1}(kr)$ is the $d-1$ dimensional generalization of spherical Bessel function which is related to the ordinary Bessel function of the first kind by 
\bea 
j_{\ell_m}^{d-1}(x)=\frac{\Gamma(\frac{m-1}{2})2^{(m-1)/2}J_{(m-1)/2+\ell_m}(x)}{(m-3)!!x^{(m-1)/2}}.
\eea 
 For example, when $d=4$, the function $j_{\ell_m}^{d-1}(x)$ becomes 
the ordinary spherical Bessel function.
The Bessel function of the first kind $J_n(x)$ has the following asymptotic behaviour for large $x$
\bea 
J_n(x)\sim \sqrt{\frac{{2}}{\pi x}}\cos(x-\frac{n\pi}{2}-\frac{\pi}{4})+\cdots.
\eea Therefore, the mode expansion $\Phi(t,\bm x)$ has the expected fall-off condition 
\bea 
\Phi(t,\bm x)\sim \frac{\Sigma(u,\Omega)}{r^{(d-2)/2}}
\eea 
with the expansion \eqref{expa} 
\bea \Sigma(u,\Omega)=\sum_{\bm\ell}\int_0^\infty \frac{d\omega}{\sqrt{4\pi\omega}}[a_{\omega,\bm\ell}e^{-i\omega u}Y_{\bm \ell}(\Omega)+a_{\omega,\bm \ell}^\dagger e^{i\omega u}Y^*_{\bm \ell}(\Omega)]
\eea where
\bs\begin{align} 
a_{\omega,\bm\ell}&=\omega^{m/2}e^{-i\pi m/4}\int d\Omega_k b_{\bm k}Y^*_{\bm \ell}(\Omega_k),\\
a_{\omega,\bm\ell}^\dagger&=\omega^{m/2}e^{i\pi m/4}\int d\Omega_k b^\dagger_{\bm k}Y_{\bm \ell}(\Omega_k).
\end{align}\es

We could also use mode expansion to calculate the commutator between boundary scalar fields. The first step is to notice that
\begin{align}
    [a_{\omega,\bm\ell},a^\dagger_{\omega',\bm\ell'}]=&\delta(\omega-\omega')\delta_{\bm \ell,\bm \ell'}.
\end{align}
Then one can compute the commutator straightforwardly
\begin{align}
  [\Sigma(u,\Omega),\Sigma(u',\Omega')] =&\sum_{\bm\ell}\int_0^\infty \frac{d\omega}{{4\pi\omega}}\left(e^{-i\omega (u-u')}Y_{\bm \ell}(\Omega)Y^*_{\bm \ell}(\Omega')+e^{i\omega (u-u')}Y^*_{\bm \ell}(\Omega)Y_{\bm \ell}(\Omega')\right)\nn\\
  =&\delta(\Omega-\Omega')\int_{-\infty}^\infty \frac{d\omega}{{4\pi\omega}}e^{-i\omega (u-u')}\nn\\
  =&\frac{i}{2}\alpha(u-u')\delta(\Omega-\Omega').
\end{align}

\subsection{Null hypersurface \texorpdfstring{$\mathcal{H}^-$}{}}\label{nullhm}
The null hypersurface $\mathcal{H}^-$ corresponds to the surface $t+x^{d-1}=0$ in Cartesian coordinates. On this hypersurface, we may define the retarded time $u=t-x^{d-1}=-2x^{d-1}$ and expand the bulk field $\Phi(t,\bm x)$ as 
\be 
\Phi(t,\bm x)=\Sigma(u,\bm x_{\perp})+\sum_{k=1}^\infty \Sigma^{(k)}(u,\bm x_{\perp})v^k
\ee 
with $v=t+x^{d-1}$. Therefore, we can reduce the expansion \eqref{modeexpMink} to $\mathcal{H}^-$ by setting $t+x^{d-1}=0$. This leads to 
\bea 
\Sigma(u,\bm x_{\perp})=\int \frac{d^{d-1}\bm p}{\sqrt{(2\pi)^{d-1}}}\frac{1}{\sqrt{2E}}[b_{\bm p}e^{-i\frac{E+p^{d-1}}{2}u +i\bm p_{\perp}\cdot\bm x_{\perp}}+b_{\bm p}^\dagger e^{i\frac{E+p^{d-1}}{2}u-i\bm p_{\perp}\cdot\bm x_{\perp}}],\label{expsigma2}
\eea 
where 
\be 
E=\sqrt{|\bm p|^2+m^2}=\sqrt{\bm p_{\perp}^2+(p^{d-1})^2+m^2}.
\ee 
It is much more convenient to define the momentum in light cone coordinates \cite{Compere:2019rof}
\be 
p_\pm=\frac{E\pm p^{d-1}}{2}
\ee such that 
\bea 
dp_+=\frac{p_+}{E}dp^{d-1}\quad \text{for fixed}\ \bm p_{\perp}.
\eea We may write $\Sigma$ as 
\bea 
\Sigma(u,\bm x_\perp)=\int_0^\infty \frac{dp_+}{\sqrt{4\pi p_+}}\frac{d\bm p_{\perp}}{\sqrt{(2\pi)^{d-2}}}[d_{p_+,\bm p_{\perp}}e^{-ip_+u+i\bm p_{\perp}\cdot\bm x_{\perp}}+d^\dagger_{p_+,\bm p_{\perp}}e^{ip_+u-i\bm p_{\perp}\cdot\bm x_{\perp}}]
\eea with 
\bea 
d_{p_+,\bm p_\perp}=\sqrt{\frac{E}{p_+}}b_{\bm p},\quad d^\dagger_{p_+,\bm p_\perp}=\sqrt{\frac{E}{p_+}}b^\dagger_{\bm p},\quad E=p_++\frac{\bm p_{\perp}^2+m^2}{4p_+}.
\eea 
The commutators between $d_{p_+,\bm p_{\perp}}$ and $d^\dagger_{p_+,\bm p_{\perp}}$ are
\bea 
[d_{p_+,\bm p_{\perp}},d_{p'_+,\bm p'_{\perp}}]=[d^\dagger_{p_+,\bm p_{\perp}},d^\dagger_{p'_+,\bm p'_{\perp}}]=0,\quad [d_{p_+,\bm p_{\perp}},d^\dagger_{p'_+,\bm p'_{\perp}}]=\delta(p_+-p_+')\delta^{(d-2)}(\bm p_\perp-\bm p_{\perp}').
\eea 
Therefore, we find the following commutators  
\bs\begin{align}
    [\Sigma(u,\bm x_{\perp}),\Sigma(u',\bm x_{\perp}')]&=\frac{i}{2}\alpha(u-u')\delta^{(d-2)}(\bm x_{\perp}-\bm x_{\perp}'),\\
     [\Sigma(u,\bm x_{\perp}),\dot\Sigma(u',\bm x_{\perp}')]&=\frac{i}{2}\delta(u-u')\delta^{(d-2)}(\bm x_{\perp}-\bm x_{\perp}'),\\
      [\dot\Sigma(u,\bm x_{\perp}),\dot\Sigma(u',\bm x_{\perp}')]&=\frac{i}{2}\delta'(u-u')\delta^{(d-2)}(\bm x_{\perp}-\bm x_{\perp}').
\end{align} 
\es  

\subsection{Rindler horizon \texorpdfstring{$\mathcal{H}^{--}$}{}}\label{modeexpRindler}
In this subsection, we will review the mode expansion of scalar field in Rindler wedge \cite{Crispino:2007eb, Compere:2019rof}. 
In Rindler wedge, the massless/massive scalar field can be written in the coordinate system $(\tau,\lambda,\bm x_{\perp})$ as 
\be 
-e^{-2\lambda}\partial_\tau^2\Phi^{\rm R}+e^{-2\lambda}\partial_\lambda^2\Phi^{\rm R}+\partial_{\perp}^2\Phi^{\rm R}-m^2\Phi^{\rm R}=0.
\ee We add an upper index R  to label the right Rinder wedge modes.
The solution in the Rindler wedge may be expanded as 
\bea 
\Phi^{\rm R}(t,\bm x)=\int d\omega d\bm k_{\perp}\chi_{\omega,\bm k_\perp}(\kappa)(c_{\omega,\bm k_\perp} e^{-i\omega\tau+i\bm k_{\perp}\cdot\bm x_{\perp}}+c^{\dagger}_{\omega,\bm k_\perp} e^{i\omega\tau-i\bm k_{\perp}\cdot\bm x_{\perp}}),
\eea 
where the function $\chi_{\omega,k_{\perp}}$ satisfies the equation 
\bea 
e^{-2\kappa}\frac{d^2}{d\kappa^2}\chi_{\omega,\bm k_\perp}+\omega^2 e^{-2\kappa}\chi_{\omega,\bm k_\perp}-(\bm k_{\perp}^2+m^2)\chi_{\omega,\bm k_\perp}=0.
\eea This equation may be transformed to Bessel equation of order $\omega$ using the coordinate $\rho=e^\kappa$
\be 
\frac{d^2}{d\rho^2}\chi_{\omega,\bm k_\perp}+\rho^{-1}\frac{d}{d\rho}\chi_{\omega,\bm k_\perp}-(k_{\perp}^2+m^2)\chi_{\omega,\bm k_{\perp}}+\omega^2\rho^{-2}\chi_{\omega,\bm k_{\perp}}=0,\quad k_{\perp}=|\bm k_{\perp}|,
\ee whose solution is 
\bea 
\chi_{\omega,\bm k_\perp}(\rho)=\sqrt{\frac{4\sinh\pi \omega }{(2\pi)^d}}K_{i\omega}(\bar{k}_{\perp}\rho),\quad \bar{k}_{\perp}=\sqrt{k_{\perp}^2+m^2},
\eea where the function $K_{i\omega}(\bar{k}_{\perp}\rho)$ is the Modified Bessel function of the second kind. $\Phi^{\rm R}$ is bounded near $\rho\to\infty$ since $K_{i\omega}(\bar{k}_{\perp}\rho)$ decays in this region. More explicitly, 
\bea 
K_{i\omega}(\bar{k}_{\perp}\rho)\sim \sqrt{\frac{\pi}{2\bar{k}_{\perp}\rho}}e^{-\bar{k}_{\perp}\rho},\quad \rho\to\infty.
\eea 
Therefore, the stress tensor $T_{\mu\nu}$ decays exponentially when $\rho\to\infty$.
The normalization of the solution is chosen such that the operators $c_{\omega,\bm k_{\perp}}$ and $c_{\omega,\bm k_{\perp}}^\dagger$ obey the commutator 
\bs\begin{align}
\ [c_{\omega,\bm k_{\perp}},c^\dagger_{\omega',\bm k_{\perp}'}]&=\delta(\omega-\omega')\delta(\bm k_{\perp}-\bm k'_{\perp}).
\end{align}
\es The asymptotic behaviour of the function $K_{i\omega}(k_{\perp}\rho)$ is 
\bea 
K_{i\omega}(\bar{k}_{\perp}\rho)\sim 2^{-1-i\omega}\Gamma(-i\omega)(\bar{k}_{\perp}\rho)^{i\omega}+2^{-1+i\omega}\Gamma(i\omega)(\bar{k}_{\perp}\rho)^{-i\omega},\quad \rho\to 0.
\eea 
The $\Phi$ is finite near $\mathcal{H}^{--}$ and 
we find the following mode expansion of the boundary field $\Sigma$
\be 
\Sigma(u,\bm x_{\perp})=\int_0^\infty \frac{d\omega}{\sqrt{4\pi\omega}}\frac{d\bm k_{\perp}}{\sqrt{(2\pi)^{d-2}}}[a_{\omega,\bm k_{\perp}}^{\rm R} e^{-i\omega u+i\bm k_{\perp}\cdot\bm x_{\perp}}+a_{\omega,\bm k_{\perp}}^{\rm R\dagger} e^{i\omega u-i\bm k_{\perp}\cdot\bm x_{\perp}}]\label{expsig1}
\ee with 
\bs\begin{align}
   a_{\omega,\bm k_{\perp}}^{\rm R}&=\sqrt{\frac{\omega\sinh\pi\omega}{\pi}}(\frac{\bar{k}_{\perp}}{2})^{i\omega}\Gamma(-i\omega)c_{\omega,\bm k_{\perp}},\\
   a_{\omega,\bm k_{\perp}}^{\rm R\dagger}&=\sqrt{\frac{\omega\sinh\pi\omega}{\pi}}(\frac{\bar{k}_{\perp}}{2})^{-i\omega}\Gamma(i\omega)c^{\dagger}_{\omega,\bm k_{\perp}}.
\end{align}
\es
It is easy to find that
\begin{align}
  [a_{\omega,\bm k_{\perp}}^{\rm R},a_{\omega,\bm k_{\perp}}^{\rm R\dagger}]=\delta(\omega-\omega')\delta^{(d-2)}(\bm k_{\perp}-\bm k'_{\perp}),
\end{align}
which implies
\begin{align}
\left[ \Sigma \left( u,\boldsymbol{x}_{\bot} \right) ,\Sigma \left( u',\boldsymbol{x}_{\bot}' \right) \right] =&\delta ^{\left( d-2 \right)}\left( \boldsymbol{x}_{\bot}-\boldsymbol{x}_{\bot}' \right) \int_{-\infty}^{\infty}{\frac{d\omega}{4\pi \omega}e^{-i\omega \left( u-u' \right)}}\nn\\
=&\frac{i}{2}\alpha \left( u-u' \right) \delta ^{\left( d-2 \right)}\left( \boldsymbol{x}_{\bot}-\boldsymbol{x}_{\bot}' \right) .
\end{align}

To match the expansion \eqref{expsig1} with \eqref{expsigma2}, we find the following Bogoliubov transformation 
\bs\begin{align}
    a^{\rm R}_{\omega,\bm k_{\perp}}&=\int d\bm p (\alpha_{\omega,\bm k_{\perp};\bm p}^* b_{\bm p}-\beta_{\omega,\bm k_{\perp};\bm p}^* b_{\bm p}^\dagger),\\
    a^{\rm R\dagger}_{\omega,\bm k_{\perp}}&=\int d\bm p (\alpha_{\omega,\bm k_{\perp};\bm p} b^\dagger_{\bm p}-\beta_{\omega,\bm k_{\perp};\bm p} b_{\bm p}),
\end{align}\es with 
\bs \label{Bog1}
\begin{align}
    \alpha_{\omega,\bm k_{\perp};\bm p}&=\frac{1}{2\pi}\sqrt{\frac{\omega}{E}}e^{\pi\omega/2} \Gamma(i\omega)(\frac{E+p^{d-1}}{2})^{-i\omega}\delta^{(d-2)}(\bm k_{\perp}-\bm p_{\perp}),\\
\beta_{\omega,\bm k_{\perp};\bm p}&=-\frac{1}{2\pi}\sqrt{\frac{\omega}{E}}e^{-\pi\omega/2} \Gamma(i\omega)(\frac{E+p^{d-1}}{2})^{-i\omega}\delta^{(d-2)}(\bm k_{\perp}+\bm p_{\perp}).
\end{align}\es 
The Bogoliubov coefficients can also be expressed as the commutators 
 \bs\begin{align}
     [b_{\bm p},a^{\rm R}_{\omega,\bm k_{\perp}}]&=-\beta^*_{\omega,\bm k_{\perp};\bm p},\\
     [b_{\bm p},a^{\rm R\dagger}_{\omega,\bm k_{\perp}}]&=\alpha_{\omega,\bm k_{\perp};\bm p},\\
     [b^\dagger_{\bm p},a^{\rm R}_{\omega,\bm k_{\perp}}]&=-\alpha_{\omega,\bm k_{\perp};\bm p}^*,\\
     [b^\dagger_{\bm p},a^{\rm R\dagger}_{\omega,\bm k_{\perp}}]&=\beta_{\omega,\bm k_{\perp};\bm p}.
 \end{align}\es
We may also expand the field $\Phi^{\rm R}$ near $\mathcal{H}^{++}$
\be 
\Sigma^{+}(v,\bm x_{\perp})=\int_0^\infty \frac{d\omega}{\sqrt{4\pi\omega}}\frac{d\bm k_{\perp}}{\sqrt{(2\pi)^{d-2}}}[\widetilde{a}_{\omega,\bm k_{\perp}}^{\rm R} e^{-i\omega v+i\bm k_{\perp}\cdot\bm x_{\perp}}+\widetilde{a}_{\omega,\bm k_{\perp}}^{\rm R\dagger} e^{i\omega v-i\bm k_{\perp}\cdot\bm x_{\perp}}]
\ee with 
\bs\begin{align}
   \widetilde{a}_{\omega,\bm k_{\perp}}^{\rm R}&=\sqrt{\frac{\omega\sinh\pi\omega}{\pi}}(\frac{\bar{k}_{\perp}}{2})^{-i\omega}\Gamma(i\omega)c_{\omega,\bm k_{\perp}},\\
   \widetilde{a}_{\omega,\bm k_{\perp}}^{\rm R\dagger}&=\sqrt{\frac{\omega\sinh\pi\omega}{\pi}}(\frac{\bar{k}_{\perp}}{2})^{i\omega}\Gamma(-i\omega)c^{\dagger}_{\omega,\bm k_{\perp}}.
\end{align}
\es
The relations between the  operators ${a}_{\omega,\bm k_{\perp}}^{\rm R},{a}_{\omega,\bm k_{\perp}}^{\rm R\dagger}$ on $\mathcal{H}^{--}$ and $\widetilde{a}_{\omega,\bm k_{\perp}}^{\rm R},\widetilde{a}_{\omega,\bm k_{\perp}}^{\rm R\dagger}$ on $\mathcal{H}^{++}$ are 
\bea 
{a}^{\rm R}_{\omega,\bm k_{\perp}}=\Gamma(-i\omega)\Gamma(i\omega)^{-1}(\frac{\bar{k}_{\perp}}{2})^{2i\omega}\widetilde{a}^{\rm R}_{\omega,\bm k_{\perp}},\quad {a}^{\rm R\dagger}_{\omega,\bm k_{\perp}}=\Gamma(i\omega)\Gamma(-i\omega)^{-1}(\frac{\bar{k}_{\perp}}{2})^{-2i\omega}\widetilde{a}^{\rm R\dagger}_{\omega,\bm k_{\perp}}.
\eea 
The relation can be transformed to 
\be 
\Sigma^+(\omega,\bm k_{\perp})=\Gamma(-i\omega)\Gamma(i\omega)^{-1}(\frac{\bar{k}_{\perp}}{2})^{2i\omega}\Sigma(\omega,\bm k_{\perp}),
\ee where $\Sigma(\omega,\bm k_{\perp})$ and $\Sigma^+(\omega,\bm k_{\perp})$ are Fourier transform of $\Sigma(u,\bm x_{\perp})$ and $\Sigma^+(v,\bm x_{\perp})$, respectively 
\bs\begin{align}
\Sigma(\omega,\bm k_{\perp})&=\int du \int d\bm x_{\perp} \Sigma(u,\bm x_{\perp})e^{i\omega u-i\bm k_{\perp}\cdot\bm x_{\perp}},\\
\Sigma^+(\omega,\bm k_{\perp})&=\int dv \int d\bm x_{\perp} \Sigma^+(v,\bm x_{\perp})e^{i\omega v-i\bm k_{\perp}\cdot\bm x_{\perp}}.
\end{align}\es 
\section{Zeta-function regularizations}\label{zetaregu}
Usually, there are two kinds of zeta functions to proceed with this regularization, namely the Riemann zeta function $\zeta(s)$  and the Hurwitz zeta function $\zeta_{\rm H}(s)$ which are defined in the following \cite{elizalde2012ten}.
\subsection{Definitions}
1. Riemann zeta function $\zeta (s)$ is
\be
\zeta \left( s \right) =\sum_{n=1}^{\infty}{n^{-s}},\quad s\in \mathbf{C} ,\,\,\mathrm{Re}\,s>1.
\ee
We can use the integral representation 
\be\label{defriemannzeta}
\zeta \left( s \right) =\frac{1}{\Gamma \left( s \right)}\int_0^{\infty}{dt}\,\,\frac{t^{s-1}}{e^t-1}
\ee to analytically continue it to  the complex $s$-plane except the single pole $s=1$. In the neighbourhood of this pole, one may easily do the  Laurent expansion as follows
\bea
\zeta \left( s \right) =\frac{1}{s-1}+\gamma +\gamma _1\left( s-1 \right) +\gamma _2\left( s-1 \right) ^2+\cdots, \eea where $\gamma$ is Euler constant and
\bea 
\gamma _k=\underset{n\rightarrow \infty}{\lim}\left[ \sum_{\nu =1}^{\infty}{\frac{\left( \log \nu \right) ^k}{\nu}-\frac{1}{k+1}\left( \log n \right) ^{k+1}} \right],\quad k\ge 1.
\eea

Some special values may be  useful:
\be
\zeta \left( 0 \right) =-\frac{1}{2},\quad \zeta \left( -2n \right) =0,\quad \zeta \left( 1-2n \right) =-\frac{B_{2n}}{2n},\quad n=1,2,3,\cdots ,
\ee
where $B_n$ are Bernoulli numbers.\par 

2. Hurwitz zeta function $\zeta_{\rm H}(s)$ is also known as the generalized Riemann zeta function:
\be
\zeta _{\rm H}\left( s,a \right) =\sum_{n=0}^{\infty}{\left( n+a \right) ^{-s}},\quad \mathrm{Re}(s)>1,\,\,a\ne 0,-1,-2,\cdots .
\ee
One of its integral representations reads
\be\label{hzetaint}
\zeta _{\rm H}\left( s,a \right) =\frac{1}{\Gamma \left( s \right)}\int_0^{\infty}{dt}\,t^{s-1}\frac{e^{-ta}}{1-e^{-t}},\quad \mathrm{Re}(s)>1,\,\,\mathrm{Re}(a)>0.
\ee There is a  single pole at $s=1$ in the complex $s$ plane and we may expand it near this pole as 
\be\label{hzetalaurant}
\zeta _{\rm H}\left( 1+\varepsilon ,a \right) =\frac{1}{\varepsilon}-\psi \left( a \right) +\mathcal{O} \left( \varepsilon \right) ,
\ee
where $\psi (a)$ is the Digamma function.\par

Useful special values are
\be
\zeta _{\rm H}\left( -n,a \right) =-\frac{B_{n+1}\left( a \right)}{n+1},\quad n\in \mathbf{N} ,
\ee
where $B_n(a)$ are the Bernoulli polynomials.

\subsection{Regularization}
When trying to calculate the spectral zeta function, one may analytically continue the expression as follows
\be\label{fofzeta}
f\left( s;a,b,c \right) =\sum_{\ell =1}^{\infty}{\ell ^{-s+b}\left( \ell +a \right) ^{-s+c}},
\ee
where $a,b,c$ are arbitrary real numbers. In our work, they are natural numbers. We factor out $\ell$, separating the series into a finite summation and another series, and perform binomial expansion on it:
\begin{equation}
    \begin{split}
f\left( s;a,b,c \right) &=\sum_{\ell =1}^{\left[ a \right]}{\ell ^{-s+b}\left( \ell +a \right) ^{-s+c}}+\sum_{\ell =\left[ a \right] +1}^{\infty}{\ell ^{-2s+b+c}\left( 1+a\ell ^{-1} \right) ^{-s+c}}
\\
&=\sum_{\ell =1}^{\left[ a \right]}{\ell ^{-s+b}\left( \ell +a \right) ^{-s+c}}+\sum_{k=0}^{\infty}{\frac{a^k\Gamma \left( 1-s+c \right)}{k!\,\Gamma \left( 1-s-k+c \right)}}\sum_{\ell =\left[ a \right] +1}^{\infty}{\ell ^{-2s-k+b+c}},
    \end{split}
\end{equation}
where $[a]$ is the integer part of $a$. The second series is absolutely convergent due to $a\ell^{-1}<1$, and the first summation is an analytic function of $s$. Then we may write 
\be\label{regresspeczeta}
\begin{split}
f\left( s;a,b,c \right) &=\sum_{\ell =1}^{\left[ a \right]}{\ell ^{-s+b}\left( \ell +a \right) ^{-s+c}}+\sum_{k=0}^{\infty}{\frac{a^k\Gamma \left( 1-s+c \right)}{k!\,\Gamma \left( 1-s-k+c \right)}}
\\
&\times \left[ \zeta \left( 2s+k-b-c \right) -\sum_{\ell =1}^{\left[ a \right]}{\ell ^{-2s-k+b+c}} \right] .
\end{split}
\ee
After analysing the poles of $\Gamma \left( 1-s-k+c \right)$ and $\zeta \left( 2s+k-b-c \right)$, one could substitute the value of $a,b,c$ into \eqref{regresspeczeta} to finish the spectral zeta regularization.\par 
For example, let us consider $d=4$ and $d=6$, respectively. When $d=4$, we may write the corresponding spectral zeta function $\zeta _{\Delta _{S^2}}\left( s \right) $ as below
\be \label{zetad=4}
\zeta _{\Delta _{S^2}}\left( s \right) =\sum_{\ell =1}^{\infty}{\left( 2\ell +1 \right) \left[ \ell \left( \ell +1 \right) \right] ^{-s}},
\ee
and we have two different decomposition schemes at least:
\begin{equation}
    \begin{split}
\zeta _{\Delta _{S^2}}\left( s \right) &=2f\left( s;1,1,0 \right) +f\left( s;1,0,0 \right) ,
\\
\zeta _{\Delta _{S^2}}\left( s \right) &=f\left( s;1,1,0 \right) +f\left( s;1,0,1 \right) .
    \end{split}
\end{equation}
Note that choosing any of the schemes does not affect the final regularization result, since  the series are absolutely convergent. After fixing the constants $a,b,c$ , we can take the limit $s\rightarrow 0$. For the first scheme, we have
\be
\zeta _{\Delta _{S^2}}\left( s\rightarrow 0 \right) =2\times \left( 1-\frac{5}{6} \right) +\left( 1-2 \right) =-\frac{2}{3}.
\ee
For the second scheme, we have 
\be
\zeta _{\Delta _{S^2}}\left( s\rightarrow 0 \right) =\left( 1-\frac{5}{6} \right) +\left( 2-\frac{17}{6} \right) =-\frac{2}{3}.
\ee
As expected, they give the same results.\par 
When $d=6$, we may write the corresponding spectral zeta function $\zeta _{\Delta _{S^4}}\left( s \right) $ as follows
\be\label{zetad=6}
\zeta _{\Delta _{S^4}}\left( s \right) =\sum_{\ell =1}^{\infty}{\frac{1}{6}\left( 2\ell +3 \right) \left( \ell +2 \right) \left( \ell +1 \right) \left[ \ell \left( \ell +3 \right) \right] ^{-s}}.
\ee
At this time, we can write
\be
\zeta _{\Delta _{S^4}}\left( s \right) =\frac{1}{3}f\left( s;3,3,0 \right) +\frac{3}{2}f\left( s;3,2,0 \right) +\frac{13}{6}f\left( s;3,1,0 \right) +f\left( s;3,0,0 \right) .
\ee
Similarly, by taking the limit $s\rightarrow 0$, we can obtain
\be
\zeta _{\Delta _{S^4}}\left( s\rightarrow 0 \right) =\frac{1}{3}\times \frac{152}{15}+\frac{3}{2}\times \left( -\frac{9}{2} \right) +\frac{13}{6}\times \frac{13}{6}+\left( -2 \right) =-\frac{61}{90}.
\ee
Moreover, we can perform a substitution such as $t=\ell +\frac{3}{2}$, then \eqref{zetad=6} takes the form 
\begin{equation}
    \begin{split}
\zeta _{\Delta _{S^4}}\left( s \right) &=\sum_{\ell =1}^{\infty}{\frac{1}{3}t\left( t^2-\frac{1}{4} \right) \left( t^2-\frac{9}{4} \right) ^{-s}}
\\
&=\frac{1}{3}\sum_{k=0}^{\infty}{\frac{\left( -1 \right) ^k\Gamma \left( 1-s \right)}{k!\,\Gamma \left( 1-s-k \right)}}\left( \frac{9}{4} \right) ^k\left[ \zeta _{\mathrm{H}}\left( 2s+2k-3,\frac{5}{2} \right) -\frac{1}{4}\zeta _{\mathrm{H}}\left( 2s+2k-1,\frac{5}{2} \right) \right] ,
    \end{split}
\end{equation}
which can be regularized by the Hurwitz zeta functions. For this scheme, we obtain
\be
\zeta _{\Delta _{S^4}}\left( s\rightarrow 0 \right) =\frac{1}{3}\times\left(-\frac{61}{30}\right)=-\frac{61}{90}.
\ee
Once again, we find the same result. In summary, one may always get the regularization function with the form \eqref{regresspeczeta} where  $a,b,c$ are constant nature numbers.  After analytically continuing the expression, we can obtain an absolutely convergent series in the neighborhood of $s=0$. 

\bibliography{refs}

\end{document}